\let\includefigures=\iftrue
%
\let\useblackboard=\iftrue
%
%
\newfam\black
\input harvmac.tex
\includefigures
\message{If you do not have epsf.tex (to include figures),}
\message{change the option at the top of the tex file.}
\input epsf
\def\figin{\epsfcheck\figin}\def\figins{\epsfcheck\figins}
\def\epsfcheck{\ifx\epsfbox\UnDeFiNeD
\message{(NO epsf.tex, FIGURES WILL BE IGNORED)}
\gdef\figin##1{\vskip2in}\gdef\figins##1{\hskip.5in}
\else\message{(FIGURES WILL BE INCLUDED)}%
\gdef\figin##1{##1}\gdef\figins##1{##1}\fi}
\def\DefWarn#1{}
\def\figinsert{\goodbreak\midinsert}
\def\ifig#1#2#3{\DefWarn#1\xdef#1{fig.~\the\figno}
\writedef{#1\leftbracket fig.\noexpand~\the\figno}%
\figinsert\figin{\centerline{#3}}\medskip\centerline{\vbox{\baselineskip12pt
\advance\hsize by -1truein\noindent\footnotefont{\bf Fig.~\the\figno:} #2}}
\bigskip\endinsert\global\advance\figno by1}
\else
\def\ifig#1#2#3{\xdef#1{fig.~\the\figno}
\writedef{#1\leftbracket fig.\noexpand~\the\figno}%
\global\advance\figno by1}
\fi
\useblackboard
\message{If you do not have msbm (blackboard bold) fonts,}
\message{change the option at the top of the tex file.}
\font\blackboard=msbm10 scaled \magstep1
\font\blackboards=msbm7
\font\blackboardss=msbm5
\textfont\black=\blackboard
\scriptfont\black=\blackboards
\scriptscriptfont\black=\blackboardss

\else

\fi
%
\def\yboxit#1#2{\vbox{\hrule height #1 \hbox{\vrule width #1
\vbox{#2}\vrule width #1 }\hrule height #1 }}
\def\fillbox#1{\hbox to #1{\vbox to #1{\vfil}\hfil}}
\def\ybox{{\lower 1.3pt \yboxit{0.4pt}{\fillbox{8pt}}\hskip-0.2pt}}

\def\comments#1{}

\def\p{\partial}

\def\half{{1\over 2}}
\def\Tr{{{\rm Tr~ }}}
\def\tr{{\rm tr\ }}
\def\Re{{\rm Re\hskip0.1em}}
\def\Im{{\rm Im\hskip0.1em}}
\def\even{{\rm even}}
\def\odd{{\rm odd}}
\def\lcm{{\rm lcm}}
\def\diag{{\rm diag}}

\def\ket#1{|#1\rangle}
\def\bbra#1{{\langle\langle}#1|}
\def\kket#1{|#1\rangle\rangle}
\def\vev#1{\langle{#1}\rangle}
\def\Dslash{\rlap{\hskip0.2em/}D}

\def\CC{{\cal C}}

\def\CF{{\cal F}}

\def\CM{{\cal M}}
\def\CN{{\cal N}}
\def\CO{{\cal O}}


\def\ap{\alpha'}

\def\sgn{{\rm sgn\ }}

\def\II{\relax{I\kern-.10em I}}
\def\IIa{{\II}a}
\def\IIb{{\II}b}

\def\IZ{\relax\ifmmode\mathchoice
{\hbox{\cmss Z\kern-.4em Z}}{\hbox{\cmss Z\kern-.4em Z}}
{\lower.9pt\hbox{\cmsss Z\kern-.4em Z}}
{\lower1.2pt\hbox{\cmsss Z\kern-.4em Z}}\else{\cmss Z\kern-.4em
Z}\fi}
\def\IB{\relax{\rm I\kern-.18em B}}
\def\IC{{\relax\hbox{$\inbar\kern-.3em{\rm C}$}}}
\def\ID{\relax{\rm I\kern-.18em D}}
\def\IE{\relax{\rm I\kern-.18em E}}
\def\IF{\relax{\rm I\kern-.18em F}}
\def\IG{\relax\hbox{$\inbar\kern-.3em{\rm G}$}}
\def\IGa{\relax\hbox{${\rm I}\kern-.18em\Gamma$}}
\def\IH{\relax{\rm I\kern-.18em H}}
\def\II{\relax{\rm I\kern-.18em I}}
\def\IK{\relax{\rm I\kern-.18em K}}
\def\IP{\relax{\rm I\kern-.18em P}}

%
\def\barom{\overline{\Omega}}
\def\barA{\bar{A}}
\def\jb{{\bar \jmath}}

\def\inbar{\,\vrule height1.5ex width.4pt depth0pt}
\def\mod{{\rm\; mod\;}}

\def\p{\partial}

\def\pb{{\bar \p}}

\font\cmss=cmss10 \font\cmsss=cmss10 at 7pt
\def\IR{\relax{\rm I\kern-.18em R}}

\def\ker{{\rm ker\ }}

\def\im{{\rm im\ }}
\def\ind{{\rm ind\ }}

\def\one {{\bf 1}}
\def\dim{{\rm dim}}
\def\BR{\IR}
\def\BZ{Z} 
\def\BP{\IP}
\def\BR{\IR}
\def\BC{\IC}

\def\lp10{l_P^{10}}
\def\lp11{l_P^{11}}
\def\R11{R_{11}}

%

\def\delb{\delta_{{\rm B}}}

\def\frac#1#2{{#1 \over #2}}
\def\epb{\bar{\epsilon}}

\def\ie{{\it i.e.}}
\def\cf{{\it c.f.}}
\def\etal{{\it et. al.}}
\hyphenation{Di-men-sion-al}
\def\np{{\it Nucl. Phys.}}
\def\mpl{{\it Mod. Phys. Lett.}}
\def\cmp{{\it Comm. Math. Phys.}}
\def\ch{{\rm ch}}
\def\td{{\rm Td}}
\def\ahat{\hat{A}}

\def\unlockat{\catcode`\@=11}
\def\lockat{\catcode`\@=12}
\unlockat
\def\newsec#1{\global\advance\secno by1%
{\toks0{#1}\message{(\the\secno. \the\toks0)}}%
\global\subsecno=0\eqnres@t\let\s@csym\secsym\xdef\secn@m{\the\secno}\noindent
{\bf\hyperdef\hypernoname{section}{\the\secno}{\the\secno.} #1}%
\writetoca{{\string\hyperref{}{section}{\the\secno}{\the\secno.}} {#1}}%
\par\nobreak\medskip\nobreak}
\def\eqnres@t{\xdef\secsym{\the\secno.}\global\meqno=1\bigbreak\bigskip}
\def\sequentialequations{\def\eqnres@t{\bigbreak}}\xdef\secsym{}
%
\global\newcount\subsecno \global\subsecno=0
\def\subsec#1{\global\advance\subsecno by1%
\global\subsubsecno=0%
{\toks0{#1}\message{(\s@csym\the\subsecno. \the\toks0)}}%
\ifnum\lastpenalty>9000\else\bigbreak\fi
\noindent{\bf\hyperdef\hypernoname{subsection}{\secn@m.\the\subsecno}%
{\secn@m.\the\subsecno.} #1}\writetoca{\string\quad
{\string\hyperref{}{subsection}{\secn@m.\the\subsecno}{\secn@m.\the\subsecno.}}
{#1}}\par\nobreak\medskip\nobreak}
\global\newcount\subsubsecno \global\subsubsecno=0
\def\subsubsec#1{\global\advance\subsubsecno by1
\message{(\secsym\the\subsecno.\the\subsubsecno. #1)}
\ifnum\lastpenalty>9000\else\bigbreak\fi
\noindent{\it \quad{\secsym\the\subsecno.\the\subsubsecno.}\ {#1}}
\writetoca{\string\qquad{\secsym\the\subsecno.\the\subsubsecno.}{#1}}
\par\nobreak\medskip\nobreak}
\def\subsubseclab#1{\DefWarn#1\xdef
#1{\noexpand\hyperref{}{subsubsection}%
{\secsym\the\subsecno.\the\subsubsecno}%
{\secsym\the\subsecno.\the\subsubsecno}}%
\writedef{#1\leftbracket#1}\wrlabeL{#1=#1}}
\lockat
\Title{\vbox{\baselineskip12pt\hbox{hep-th/9906200}
\hbox{RU-99-25}\hbox{HUTP-99/A011}}}
{\vbox{
\centerline{D-branes on the Quintic} }}
\smallskip
\centerline{Ilka Brunner$^1$, Michael R. Douglas$^{1,2}$,}
\centerline{Albion Lawrence$^{3}$  {\it and} Christian R\"omelsberger$^1$.}
\medskip
\centerline{$^1$Department of Physics and Astronomy}
\centerline{Rutgers University }
\centerline{Piscataway, NJ 08855--0849}
\medskip
\centerline{$^2$ I.H.E.S., Le Bois-Marie, Bures-sur-Yvette, 91440 France}
\medskip
\centerline{$^3$Department of Physics}
\centerline{Harvard University }
\centerline{Cambridge, MA 02138}
\medskip
\bigskip
\noindent
We study D-branes on the quintic CY by combining results from several
directions: general results on holomorphic curves and vector bundles,
stringy geometry and mirror symmetry, and the boundary states in Gepner
models recently constructed by Recknagel and Schomerus, to begin
sketching a picture of D-branes in the stringy regime.  We also make
first steps towards computing superpotentials on the D-brane world-volumes.

\Date{June 1999}
%

\def\ib{\bar \imath}
\def\np{{\it Nucl. Phys.}}
\def\prl{{\it Phys. Rev. Lett.}}
\def\pr{{\it Phys. Rev.}}
\def\pl{{\it Phys. Lett.}}
\def\atamp{{\it Adv. Theor. Math. Phys.}}
\def\cqg{{\it Class. Quant. Grav.}}
%
%
\nref\dlp{J.~Dai, R.~G.~Leigh and J.~Polchinski, ``New
Connections Between String Theories'', Mod. Phys. Lett. 
{\bf A4}, 2073 (1989);
J.~Polchinski, ``Dirichlet Branes and
Ramond-Ramond Charges'' Phys.~Rev.~Lett.~75, 4724 (1995),
hep-th/9510017;
J. Polchinski, TASI Lectures on D-branes, hep-th/9611050.}
\nref\leigh{R. G. Leigh, ``Dirac-Born-Infeld
action from Dirichlet sigma model'', Mod. Phys. Lett. {\bf A4}, 2767
(1989).}
\nref\bebestro{K.~Becker, M.~Becker and A.~Strominger,
``Five-branes, membranes and nonperturbative string theory,"
\np\ {\bf B456}, 130 (1995) hep-th/9507158.}
\nref\bsvtop{M. Bershadsky, C. Vafa and V. Sadov,
``D-branes and topological field theories'',
\np\ {\bf B463}, 420 (1996) hep-th/9511222.}
\nref\quivers{M. R. Douglas and G. Moore, ``D-branes,
Quivers, and ALE Instantons'', hep-th/9603167.}
\nref\ooy{H. Ooguri, Y. Oz and Z. Yin, ``D-branes
on Calabi-Yau spaces and their mirrors'',
\np\ {\bf B477}, 407 (1996), hep-th/9606112.}
\nref\reckone{A.~Recknagel and V.~Schomerus, ``D-branes in Gepner
models," \np\ {\bf B531}, 185 (1998), hep-th/9712186.}
\lref\bwithinb{M. R. Douglas, ``Branes Within
Branes'', in the proceedings
of the NATO ASI on Strings, Branes and Dualities,
L. Baulieu \etal (eds), Kluwer Academic (1999), Dordrecht;
hep-th/9512077.}
\lref\mqcd{E. Witten, ``Branes and the Dynamics of QCD'',
\np\ {\bf B507}, 658 (1997) hep-th/9706109.}
\lref\ksorb{S. Kachru and E. Silverstein, ``4-d conformal
theories and strings on orbifolds'', \prl\ {\bf 80}, 2996
(1998) hep-th/9802183; A. Lawrence, N. Nekrasov and C. Vafa,
``On Conformal Field Theories in Four-Dimensions'',
\np\ {\bf B533}, 199 (1998), hep-th/9803015.}
\lref\fracbranes{J. Polchinski, ``Tensors From
K3 Orientifolds'', \pr\ {\bf D55}, 6423 (1997) 
hep-th/9606165; M.R. Douglas, ``Enhanced
Gauge Symmetry in M(atrix) Theory'', {\bf JHEP 9707:004}
(1997), hep-th/9612126; D.-E. Diaconsescu, M.R. Douglas
and J. Gomis, ``Fractional Branes and Wrapped Branes'',
{\bf JHEP:9802:013}\ (1998) hep-th/9712230.}
\lref\callanmalda{C. Callan and J. Maldacena,
``Brane Dynamics from the Born-Infeld Action'',
\np\ {\bf B513}, 198 (1998) hep-th/9708147.}
\lref\katzcurve{S. Katz, ``On the finiteness
of rational curves on quintic threefolds'',
{\it Comp. Math.}\ {\bf 60}, 151 (1986).}
\lref\reid{M. Reid, pp. 131-180, 
in Advanced Studies in Pure Mathematics 1, ed. S. Iitaka,
Kinokuniya (1983).}
\lref\thomas{R.P. Thomas, ``An obstructed bundle
on a Calabi-Yau threefold'', math.AG/9903034.}
\lref\dko{M.R. Douglas, A. Kato and H. Ooguri,
``D-brane actions on K\"ahler manifolds'',
\atamp\ {\bf 1}, 237 (1998) hep-th/9708012.}
\lref\wessbagger{J. Wess and J. Bagger, {\it
Supersymmetry and Supergravity}, Princeton Univ.
Press (1992).}
\lref\banksetal{T. Banks, L.J. Dixon, D. Friedan and
E. Martinec, ``Phenomenology and
conformal field theory: or, can
string theory predict the weak mixing angle?'',
\np\ {\bf B299}, 613 (1988).}
\lref\banksdixon{T. Banks and L.J. Dixon,
``Constraints on string vacua with spacetime
supersymmetry'', \np\ {\bf B307}, 93 (1988).}
\lref\wittnr{E. Witten, ``New issues in manifolds
of SU(3) holonomy'', \np\ {\bf B268}, 79 (1986).}
\lref\dinenp{M. Dine, N. Seiberg, X.-G. Wen and E. Witten,
``Nonperturbative effects on the string worldsheet'',
\np\ {\bf B278}, 769 (1986); and
``Nonperturbative effects on the string worldsheet(II)'',
\np\ {\bf B289}, 319 (1987).}
\lref\raysinger{D. Ray and I.M. Singer,
``Analytic Torsion and the Laplacian on Complex
Manifolds'', {\it Ann. Math.}\ {\bf 98}, 154 (1973).}
\lref\cardylewellen{J.L. Cardy and D.C. Lewellen,
``Bulk and boundary operators in conformal field theory'',
\pl\ {\bf B259}, 274 (1991).}
\lref\eguchiooguri{T. Eguchi and H. Ooguri,
``Conformal and current algebras on general Riemann
surfaces'', \np\ {\bf B282}, 308 (1987).}
\lref\holomanom{M. Bershadsky, S. Cecotti, H. Ooguri
and C. Vafa, ``Holomorphic anomalies in
topological field theories'', \np\ {\bf B405}, 279 (1993).}
\lref\bergetal{E. Bergsheoff, R. Kallosh, T. Ortin and
G. Papadopolous, ``$\kappa$-symmetry, Supersymmetry
and Intersecting Branes'', \np\ {\bf B502}, 149 (1997)
hep-th/9705040.}
\lref\bergtown{E. Bergsheoff and P. Townsend, ``Super D-branes'',
\np\ {\bf B490}, 145 (1997) hep-th/9611173.}
\lref\vijayrob{V. Balasubramanian and R.G. Leigh,
``D-branes, Moduli and Supersymmetry'', \pr\
{\bf D55}, 6415 (1997) hep-th/9611165.}
\lref\bpsalg{J.A. Harvey and G. Moore,
``On the algebras of BPS states'', \cmp\ {\bf 197}
489, (1998) hep-th/9609017.}
\lref\moorek{R. Minasian and G. Moore, ``K theory
and Ramond-Ramond charge'', {\bf JHEP 9711:002} (1997)
hep-th/9701230.}
\lref\wittenk{E. Witten, ``D-branes and K theory'',
{\bf JHEP 9812:019} (1998) hep-th/9810188.}
\lref\duy{K. Uhlenbeck and S.-T. Yau, ``On the
existence of Hermitian Yang-Mills connections
on stable vector bundles'', {\it Comm. Pure Appl. Math.}\
{\bf 39}, 257 (1986); and ``A note on our
previous paper: On the existence of
Hermitian Yang-Mills connections
on stable vector bundles'',
{\it Comm. Pure Appl. Math.}\ {\bf 42}, 703 (1989);
S.K. Donaldson, ``Infinite determinants,
stable bundles and curvature'', {\it Duke Math. J.}\
{\bf 54}, 231 (1987).}
\lref\friedman{R. Friedman, {\it Algebraic Surfaces
and Holomorphic Vector Bundles}, Springer-Verlag (1998),
New York.}
\lref\ghm{M. Green, J.A. Harvey and G. Moore, ``I-brane
inflow and anomalous couplings on D-branes'', \cqg\
{\bf 14}, 47 (1997) hep-th/9605033.}
\lref\cheungyin{Y.-K. E. Cheung and Z. Yin,
``Anomalies, branes and currents'', \np\ {\bf 517}, 69 (1998)
hep-th/9710206.}
\lref\morsheave{D. Morrison, ``The geometry
underlying mirror symmetry'', from
{\it Recent trends in algebraic geometry}, K. Hulek
\etal\ (eds), Cambridge Univ. Press (1999), 
alg-geom/9608006.}
\lref\fmw{R. Friedman, J. Morgan and E. Witten,
``Vector bundles and F theory'', \cmp\ {\bf 187},
679 (1997) hep-th/9701163.}
\lref\bjps{M. Bershadsky, A. Johansen,
T. Pantev and V. Sadov, ``On four-dimensional
compactifications of F theory'', \np\ {\bf 505},
165 (1997) hep-th/9701165.}
\lref\morII{D. Morrison, ``Mirror symmetry and
the type II string'', \np\ {\bf B}\ {\it Proc. Suppl.}\
{\bf 46}, 146 (1996), hep-th/9512016.}
\lref\katzrat{S. Katz, ``Rational Curves
on Calabi-Yau Threefolds'', in {\it Mirror
Symmetry I} (S.-T. Yau, ed.), American
Mathematical Society and International Press (1999);
alg-geom/9312009.}
\lref\hubsch{T. H\"ubsch, {\it Calabi-Yau Manifolds:
A Bestiary for Physicists}, World Scientific (1992),
River Edge, NJ.}
\lref\bhcy{J. Maldacena, A. Strominger and E. Witten,
``Black hole entropy in M-theory'', 
{\bf JHEP 9712:002}\ (1997) hep-th/9711053.}
\lref\nekschw{N. Nekrasov and A. Schwarz,
``Instantons on noncommutative $\IR^4$
and $(2,0)$ superconformal six-dimensional theory'',
\cmp\ {\bf 198}, 689 (1998) hep-th/9802068.}
\lref\cds{A. Connes, M.R. Douglas and A. Schwarz,
``Noncommutative geometry and Matrix theory: compactification
on tori'', {\bf JHEP 9802:003}\ (1998) hep-th/9711162.}
\lref\doughull{M.R. Douglas and C. Hull, ``D-branes
and noncommutative geometry'', {\bf JHEP 9802:008}\ (1998)
hep-th/9711165.}
\lref\ahbs{O. Aharony, M. Berkooz and N. Seiberg,
``Light cone description of $(2,0)$ superconformal
theories in six dimension'', \atamp\ {\bf 2}, 119 (1998)
hep-th/9712117.}
\lref\witten{E. Witten, ``String Theory
Dynamics in Various Dimension'',
\np\ {\bf B443}, 85 (1995), hep-th/9503124.}
\lref\dos{M. R. Douglas, H. Ooguri and S. H. Shenker,
hep-th/9702203.}
\lref\polgeo{J. Polchinski, hep-th/9611050.}
\lref\bfss{T. Banks, W. Fischler, S. H. Shenker and L. Susskind,
hep-th/9610043.}
\lref\DKPS{M. R. Douglas, D. Kabat, P. Pouliot and S. Shenker,
hep-th/9608024.}
\lref\tseytlin{A. Tseytlin, hep-th/9701125.}
\lref\dli{M. R. Douglas and M. Li, hep-th/9412203.}
\lref\connes{A. Connes, {\it Noncommutative Geometry,} Academic Press, 1994.}
\lref\hull{C. Hull, A. ii Karlhede, U. Lindstr\"om and M. Ro{\v c}ek,
\np\ B266 (1986) 1.}
\lref\bl{V. Balasubramanian and F. Larsen, hep-th/9703039.}
\lref\vafa{C. Vafa, hep-th/9602022.}
\lref\taylor{W. Taylor, hep-th/9611042.}
\lref\dewit{B. de Wit, J. Hoppe and H. Nicolai,
\np\ B 305 [FS 23] (1988) 545.}
\lref\dvv{R. Dijkgraaf, H. Verlinde and E. Verlinde,
``Notes on topological string theory and 2-d quantum
gravity''}
\lref\eguchiyang{T. Eguchi and S.-K. Yang, ``$N=2$
superconformal models as topological field theories'',
\mpl\ {\bf A5}, 1693 (1990).}
\lref\lvw{W. Lerche, C. Vafa and N.P. Warner,
``Chiral rings in $N=2$ superconformal theories'',
\np\ {\bf B324}, 427 (1989).}
\lref\topsig{E. Witten, ``Topological sigma
models'', \cmp\ {\bf 118}, 411 (1988).}
\lref\witmirr{E. Witten, ``Mirror manifolds and
topological field theory'', in {\it Mirror Symmetry I},
S.-T. Yau (ed.), American Mathematical Society (1998).}
\lref\vafamirr{C. Vafa, ``Topological mirrors
and quantum rings'', in in {\it Mirror Symmetry I},
S.-T. Yau (ed.), American Mathematical Society (1998).}
\lref\wittcs{E. Witten, ``Chern-Simons gauge theory
as a string theory'', in {\it The Floer Memorial
Volume}, H. Hofer \etal, eds., Birkhauser (1995), Boston.} 
\lref\petropen{P. Hor\v{a}va, ``Equivariant topological
sigma models'', \np\ {\bf B418}, 571 (1994), hep-th/9309124.}
\lref\gvone{R. Gopakumar and C. Vafa, ``On the gauge theory/geometry
correspondence'', hep-th/9811131, Harvard preprint HUTP-98/A078.}
\lref\cmr{S. Cordes, G. Moore and S. Ramgoolam,
``Lectures on 2-d Yang-Mills theory,
equivariant cohomology, and topological field theories'',
hep-th/9411210, in {\it Fluctuating Geometries
in Statistical Mechanics and Field Theory}, F. David
P. Ginsparg and J. Zinn-Justin, eds., Elsevier (1996), New York.} 
\lref\bcov{M. Bershadsky, S. Cecotti, H. Ooguri
and C. Vafa, ``Kodaira-Spencer theory of gravity
and exact results for quantum string amplitudes'',
\cmp\ {\bf 165}, 311 (1994), hep-th/9309140.}
\lref\antonetal{I. Antoniadis, E. Gava, K.S. Narain
and T.R. Taylor, ``Topological amplitudes in string theory'',
hep-th/9307158, \np\ {\bf B413} (1994) 162.}
\lref\joebook{J. Polchinski, {\it String Theory}
v. I-II, Cambridge Univ. Press (1998) NY.}
\lref\aticketal{J.J. Atick, L.J. Dixon and A. Sen,
``String calculation of Fayet-Iliopoulos D-terms
in arbitrary supersymmetric compactifications'',
\np\ {\bf B292}, 109 (1987).}
\lref\klt{Kawai, Lewellen and Tye}
\lref\lewellen{D.C.~Lewellen, ``Sewing constraints for conformal field
theories on surfaces with boundaries," \np\ {\bf B372}, 654 (1992).}
\lref\cand{P.~Candelas, X.C.~De La Ossa, P.S.~Green and L.~Parkes,
``A Pair of Calabi-Yau manifolds as an exactly soluble superconformal
theory, \np\ {\bf B359}, 21 (1991).}
\lref\cardy{J.L.~Cardy, ``Boundary Conditions, Fusion Rules And The
Verlinde Formula," \np\ {\bf B324}, 581 (1989).}
\lref\wittph{E.~Witten,``Phases of N=2 theories in two dimensions,"
\np\ {\bf B403}, 159 (1993), hep-th/9301042.}
\lref\gepone{D.~Gepner, ``Space-Time Supersymmetry In Compactified
String Theory And Superconformal Models," \np\ {\bf B296}, 757 (1988).}
\lref\geptwo{D.~Gepner, ``Lectures On N=2 String Theory,"{\it Lectures
at Spring School on Superstrings, Trieste, Italy, Apr 3-14, 1989}.}
\lref\recktwo{A.~Recknagel and V.~Schomerus, ``Boundary deformation
theory and moduli spaces of D-branes," \np\ {\bf B545}, 233
(1999) hep-th/9811237.}
\lref\bdone{M.~Berkooz and M.R.~Douglas,``Five-branes in M(atrix)
theory," \pl\ {\bf B395}, 196 (1997) hep-th/9610236.}
\lref\bdtwo{M.~Berkooz, M.R.~Douglas and R.G.~Leigh, ``Branes
intersecting at angles," \np\ {\bf B480}, 265 (1996) hep-th/9606139.}
\lref\gutpone{M.~Gutperle and Y.~Satoh, ``D-branes in Gepner models
and supersymmetry," \np\ {\bf B543}, 73 (1999) hep-th/9808080.}
\lref\gutptwo{M.~Gutperle and Y.~Satoh, ``D0-branes in Gepner models
and N=2 blackholes", Princeton preprint PUPT-1838, hep-th/9902120.}
\lref\zamo{V.A.~Fateev and A.B.~Zamolodchikov, ``Parafermionic
Currents In The Two- Dimensional Conformal Quantum Field Theory And
Selfdual Critical Points In Z(N) Invariant Statistical Systems,"
{\it Sov. Phys. JETP} {\bf 62}, 215 (1985).}  
\lref\greene{B.R.~Greene, ``String theory on Calabi-Yau manifolds",
in {\it Fields, Strings and Duality}, C. Efthimiou and B.R. Greene
(eds), World Scientific (1997) New Jersey,
hep-th/9702155.}
\lref\qiu{Z.~Qiu, ``Nonlocal Current Algebra And N=2 Superconformal
Field Theory In Two Dimensions," \pl\ {\bf 188B}, 207 (1987).}
\lref\mclean{R.~McLean,''Deformations of Calibrated Submanifolds,''
Duke preprint 96-01: www.math.duke.edu/preprints/1996.html.}
\lref\harlaw{F.R.~Harvey and H.B.~Lawson,''Calibrated
Geometries,'' {\it Acta Math.}\ {\bf 148}, 47 (1982).}
\lref\hitchin{N.J.~Hitchin,''The moduli space of special Lagrangian
submanifolds,'' dg-ga/9711002.}
\lref\bbs{K. Becker, M. Becker and A. Strominger}
\lref\syz{A.~Strominger, S.~Yau and E.~Zaslow, ``Mirror symmetry is T duality,"
\np\ {\bf B479}, 243 (1996) hep-th/9606040.}
\lref\dko{M.R.~Douglas, A.~Kato and H.~Ooguri, 
``D-brane actions on Kahler manifolds,"
{\it Adv. Theor. Math. Phys.}\ {\bf 1}, 237 (1998) hep-th/9708012.}
\lref\klm{A.~Karch, D.~L\"ust and A.~Miemiec,
``N=1 supersymmetric gauge theories and supersymmetric three cycles'',
Humboldt preprint HUB-EP-99-03, hep-th/9810254.}
\lref\dup{M. R. Douglas, work in progress.}
\lref\distler{J. Distler and S. Kachru, 
``$(0,2)$ Landau-Ginzburg Theory'', \np\ {\bf B413}, 213 (1994),
hep-th/9309110.}
\lref\sagnotti{A. Sagnotti, 
``Surprises in Open-String Perturbation Theory,''
{\it Nucl. Phys. Proc. Suppl.} {\bf 56B} 
332-343 (1997); hep-th/9702093.}
\lref\fusch{J.~Fuchs and C.~Schweigert,
``Branes: From free fields to general backgrounds,''
\np\ {\bf B530}, 99 (1998); hep-th/9712257.}
%
\lref\seibergmoduli{N. Seiberg, ``Observations
on the moduli space of superconformal 
field theories'', \np\ {\bf 303}, 286 (1988).}
\lref\bpsalg{J.A. Harvey and G. Moore,
``On the algebras of BPS states'', \cmp\ {\bf 197},
489 (1998).}
\lref\mukai{S. Mukai, ``Symplectic structure
of the moduli of sheaves on an abelian or K3 surface'',
{\it Invent. Math.} {\bf 77}, 101 (1984);
see also the review article
``Moduli of vector bundles on K3 surfaces,
and symplectic manifolds'', {\it Sugaku Expositions}
{\bf 1}, 139 (1988).}
\lref\eoty{T. Eguchi, H. Ooguri, A. Taormina and
S.-K. Yang, ``Superconformal algebras and string
compactification on manifolds with $SU(n)$
holonomy'', \np\ {\bf B315}, 193 (1989).}
\lref\egta{T. Eguchi and A. Taormina,
``Unitary representations of the $\CN=4$
superconformal algebra'', \pl\ {\bf B196},
75 (1986); ``Character formulas for the
$\CN=4$ superconformal algebra'', \pl\
{\bf B200}, 315 (1988); ``On the unitary 
representations of $\CN=2$ and $\CN=4$ 
superconformal algebras'', \pl\ {\bf B210},
125 (1988).}
\lref\fiol{M.R. Douglas and B. Fiol,
``D-branes and discrete torsion II'', Rutgers preprint
RU-99-11, hep-th/9903031.}
\lref\probe{M.R.~Douglas, ``Gauge fields and D-branes,"
J. Geom. Phys. {\bf 28}, 255 (1998) hep-th/9604198.}
\lref\asplect{P.S. Aspinwall, ``The Moduli Space
of $\CN=2$ Superconformal Field Theories'',
1994 Trieste lectures,
published in {\it 1994 Summer School in High Energy
Physics and Cosmology}, 
E. Gava \etal, eds., hep-th/9412115.}
\lref\smalldist{P.S. Aspinwall, B.R. Greene and
D.R. Morrison, ``Measuring Small Distances
in $\CN=2$ Sigma Models'', \np\ {\bf B420} 184 (1994),
hep-th/9311042.}
\lref\greenekanter{B.R. Greene and Y. Kanter,
``Small Volumes in Compactified String Theory'',
\np\ {\bf B497}, 127 (1997) hep-th/9612181.}
\lref\donald{S. Donaldson and P. Kronheimer,
{\it The Geometry of Four-Manifolds,} Clarendon 1990.}
\lref\okonek{C. Okonek, M. Schneider and H. Spindler,
{\it Vector Bundles on Complex Projective Spaces,} Birkhauser 1980.}
\lref\huybrechts{D. Huybrechts and M. Lehn,
{\it The Geometry of Moduli Spaces of Sheaves,} Vieweg 1997.}
\lref\gimonpol{E. Gimon and J. Polchinski, 
``Consistency Conditions for Orientifolds
and D-Manifolds'', \pr\ D54, 1667 (1996)
hep-th/9601038.}
\lref\lutytaylor{M. Luty and W. Taylor IV, \pr\ D53 3399-3405 (1996);
hep-th/9506098.}
\lref\greple{B.R. Greene and M.R. Plesser,
``Duality in Calabi-Yau Moduli SPace'',
\np\ {\bf B338}, 15 (1990).}
\lref\polstrom{J. Polchinski and A. Strominger, ``New vacua
for type \II\ string theory'', \pl\ {\bf B388}, 736 (1996),
hep-th/9510227.}
\lref\conifold{A. Strominger, ``Massless
black holes and conifolds in string theory'',
\np\ {\bf B451}, 96 (1995) hep-th/9504090;
B.R. Greene, D.R. Morrison and A. Strominger,
``Black hole condensation and the unification
of string vacua'', \np\ {\bf 451}, 109 (1995)
hep-th/9504145.}
\lref\ceretal{A. Ceresole, R. D'Auria, S. Ferrara and
A. Van Proeyen, ``Duality transformations in
Supersymmetric Yang-Mills Theory Coupled to Gravity'',
\np\ {\bf B444}, 92 (1995) hep-th/9502072.}
\lref\dgm{M. R. Douglas, B. R. Greene and D. R. Morrison,
\np\ B506 84-106 (1997),  hep-th/9704151.}
\def\ijmp{{\it Int. J. Mod. Phys.}}
\lref\clny{C. Callan, C. Lovelace, C.R. Nappi and S.A. Yost,
``Adding holes and crosscaps to the superstring'',
\np\ {\bf B293}, 83 (1987).}
\lref\polcai{J. Polchinski and Y. Cai, ``Consistency
of open superstring theories'', \np\ {\bf 296}, 91 (1988).}
\lref\ishone{N. Ishibashi, ``The boundary and crosscap states
in conformal field theories'', \mpl\ {\bf A4}, 251 (1989).}
\lref\ishtwo{T. Onogi and N. Ishibashi,
``Conformal field theories on surfaces with
boundaries and crosscaps'', \mpl\ {\bf A4}, 161 (1989);
erratum ibid. {\bf A4}, 885 (1989).}
\lref\bfk{W. Boucher, D. Friedan and A. Kent,
``Determinant Formulae and Unitarity for
$\CN=2$ Superconformal Algebras in Two Dimension
or Exact Results on String Compactification'',
\pl\ {\bf B172}, 316 (1986).}
\lref\snam{S. Nam, ``The
Kac formula for $\CN=1$ and $\CN=2$ superconformal
algebras'', \pl\ {\bf 172B}, 323 (1986).}
\lref\dpyz{P. Di Vecchia, J.L. Petersen, M. Yu
and H.B. Zheng, ``Explicit construction of unitary representations
of the $\CN=2$ superconformal algebra'', 
\pl\ {\bf B174}, 280 (1986).}
\lref\zamtwo{A.B. Zamolodchikov and V.A. Fateev, ``Disorder fields
in two- dimensional conformal quantum field theory and
$\CN=2$ extended supersymmetry'', {\it Sov. Phys. JETP}, 913 (1986).}
\lref\ravyang{F. Ravanini and S.-K. Yang, ``Modular
invariance in $\CN=2$ superconformal field theories'',
\pl\ {\bf 195B}, 202 (1987).}
\lref\qiumod{Z. Qiu, ``Modular invariant partition
functions for $\CN=2$ superconformal field theories'',
\pl\ {\bf 198B}, 497 (1987).}
\lref\kastms{D. Kastor, E.J. Martinec and S. Shenker, ``RG
flow in $\CN=1$ discrete series'', \np\ {\bf B316}, 590 (1989).}
\lref\emilalg{E.J. Martinec, ``Algebraic geometry and effective
Lagrangians'', \pl\ {\bf 217B} 431 (1989).}
\lref\vafawarner{C. Vafa and N. Warner, ``Catastrophes
and the classification of conformal field theories'',
\pl\ {\bf 218B}, 51 (1989).}
\lref\emilcatas{E.J. Martinec, ``Criticality, catastrophes
and compactifications'', in {\it Physics and Mathematics
of Strings}, eds. L. Brink, D. Friedan and A.M. Polyakov,
World Scientific (River Edge, NJ, 1990).}
\lref\muss{G. Mussardo, G. Sotkov and M. STashnikov,
``$\CN=2$ superconformal minimal models'', \ijmp\ {\bf A4},
1135 (1989).}
\lref\vafamirror{C. Vafa,
``Extending Mirror Conjecture to Calabi-Yau with Bundles,'' hep-th/9804131.}
\lref\RSmarg{A. Recknagel and V. Schomerus
``Boundary deformation theory and moduli spaces of D-branes,"
\np\ {\bf B545}, 233 (1999)
hep-th/9811237.}
\lref\runkel{I. Runkel, 
``Boundary structure constants for the A series Virasoro minimal models,"
\np\ {\bf B549}, 563 (1999)
hep-th/9811178.}
\lref\seibmoore{
G.~Moore and N.~Seiberg,
``Classical And Quantum Conformal Field Theory,"
\cmp\ {\bf 123}, 177 (1989).}
\lref\seibergsen{A. Sen, ``D0-branes
on $T^n$ and matrix theory'', \atamp\ {\bf 2}, 51 (1998),
hep-th/9709220;
N. Seiberg, ``Why is the
Matrix Model Correct?'', \prl\ {\bf 79}, 3577 (1987),
9710009.}
\lref\kls{S. Kachru, A. Lawrence and E. Silverstein,
``On the Matrix Description of Calabi-Yau Compactifications'',
\prl\ {\bf 80}, 2996 (1998), hep-th/9712223.}
\lref\hkt{S.~Hosono, A.~Klemm and S.~Theisen,
``Lectures on mirror symmetry,'' hep-th/9403096.}
\lref\bog{F. Bogomolov, ``Holomorphic tensors
and vector bundles on projective varieties'',
{\it Math. USSR. Izv.}\ {\bf 13}, 499 (1979).}
\lref\zffusion{A.B. Zamolodchikov and V.A. Fateev,
``Operator algebra and correlation functions in
the two-dimensional $SU(2)\times SU(2)$ chiral 
Wess-Zumino model'', {\it Sov. J. Nucl. Phys.}\
{\bf 43}, 657 (1986).}
\lref\albano{A. Albano and S. Katz, ``Lines on the
Fermat quintic threefold and the infinitesimal generalized
Hodge conjecture'', {\it Trans. Am. Math. Soc.}\ {\bf 324},
353 (1991).}
%
%
\newsec{Introduction}

In this work we study D-branes on the quintic
Calabi-Yau, historically the first CY to be intensively studied.
Our guiding question will be: to
classify all supersymmetry-preserving
D-branes at each point in CY moduli space, 
and find their world-volume moduli spaces.
As is well known, results of this type are quite relevant
for phenomenological applications of M/string theory, because the
world-volume theories we will obtain include a wide variety of
four-dimensional theories with $\CN=1$ space-time supersymmetry.
The problem includes the classification of holomorphic vector
bundles (which are ground states for wrapped six-branes); and almost all
M/string compactifications which lead to $d=4$, $\CN=1$
supersymmetry
(such as $(0,2)$ heterotic string compactifications and F theory
constructions) have a choice of bundle as one of the inputs.
Thus, many works have addressed this subject explicitly or implicitly.

As usual in string compactification this geometric data is only an input
and one would really like to answer the same questions with stringy
corrections included.  The primary question along these lines is:
is the effect of stringy corrections just quantitative -- affecting
masses and couplings in the effective Lagrangian but preserving the
spectrum and moduli spaces -- or is it qualitative?
If the latter, we might imagine that geometric branes undergo
radical changes of their moduli space or are even destabilized
in the stringy regime, with new branes which were unstable in the
large volume limit taking their place.  It should be realized that
at present very little is known about this question; 
for example it has not been
ruled out that the D$0$-brane becomes unstable in the
stringy regime or has moduli space dimension different from $3$.

Clearly these questions are of great importance for the string phenomenology
mentioned above and were asked long ago
in the context of $(0,2)$ models.
No simple answer has been proposed; we will return to this in the
conclusions.

A concrete framework which allows an exact CFT
study of the stringy regime is provided by the
Gepner models.  
The main lesson from the original study of Gepner models for type \II\
and heterotic strings was that these CFT
compactifications are continuously connected to CY compactification.
Mirror symmetry is manifest in 
the 2d superconformal field theory, and this connection was one of the
earliest arguments for it in the CY context.

The first detailed study of
D-branes in Gepner models was made by Recknagel and Schomerus \reckone\ 
who (following the general approach of Cardy) constructed a large set of
examples; further work appears in \refs{\gutpone,\gutptwo}.
So far no geometric
interpretation or contact with the large volume limit has been made.
We will do so in this work.  The main tool we will use is the 
(symplectic) intersection form for three-cycles in the large volume limit.
This form governs Dirac quantization in the effective $d=4$ theory and
as such must be invariant under any variation of the moduli.
As argued in \refs{\bdone,\fiol}\ it is given by the index
$\Tr_{ab} (-1)^F$ in open string CFT and thus is easily computed for
the Gepner boundary states.  The detailed study
of K\"ahler moduli space by Candelas \etal\ \cand\ then allows
relating this to the large volume basis for $2p$-branes.  We can also make
the large volume identification for the $3$-branes, aided by the large
discrete symmetry group.

The detailed outline of the paper is as follows.
In section 2 we review the quintic, its homology and moduli space,
and give a general overview of D-branes on the quintic in
the large volume limit.
In section 3 we review
the stringy geometry of its K\"ahler moduli space
and the monodromy group acting on B branes.
In section 4 we review Gepner models and Cardy's theory of boundary
states, which will allow us to 
review the boundary states constructed by Recknagel and Schomerus.
We briefly discuss the theory for K3 compactifications,
and show that the results agree with geometric expectations; in
particular that the dimension of a brane moduli space on K3 is given by the
Mukai formula.
In section 5 we compute the large volume charges for the quintic boundary
states, and compute the number of marginal operators.
This will allow us to propose candidate geometric identifications.
In section 6 we discuss the computation of world-volume superpotentials.
We begin by presenting evidence that the superpotential is
``topological'' in a sense that we explain.  If true, an important
consequence would be that the superpotential for B-type branes has
relatively
trivial K\"ahler dependence and can thus be computed in the large volume
limit.  This would imply general agreement between stringy and geometric 
results, analogous to the case of the prepotential.  
In section 7 we discuss
superpotential computations in the Gepner model and derive selection
rules.  Besides charge conservation rules similar to those in the closed
string sector, additional boundary selection rules appear, and we 
illustrate these with the examples of the $A_1$ and $A_2$ minimal models.
The selection rules will allow us to establish that certain branes have
non-trivial moduli spaces.  
The exact superpotentials should be calculable given the solutions
of the consistency conditions of boundary CFT 
\refs{\cardy,\cardylewellen}; this is work in progress.
In section 8 we summarize our results and draw conclusions.

A point of notation: in labeling a $p$-brane,
we always ignore its Minkowski space-filling dimensions
(for example, a D$4$ wraps four dimensions of the CY), but we describe
its world-volume Lagrangian in $d=4$, $N=1$ terms (appropriate if
the brane filled all $3+1$ Minkowski dimensions).

\newsec{Large volume limit of the quintic}

\subsec{General discussion of D-branes on large volume CY}

We are interested in BPS states in type II string theory
described by  collections of D-branes at points on
or wrapping some cycle in a Calabi-Yau manifold $M$.
A configuration for $N$ coincident D-branes
with worldvolume $\Sigma$ wrapped on such a cycle
is specified by an embedding 
$X:\Sigma\rightarrow M$ and a $U(N)$ gauge field $A$
on $\Sigma$, with field strength $F = dA + [A,A]$.
The $U(1)$ part of $U(N)$ appears in combination
with the B-field, $\CF = F - X^\ast B$, where $X^\ast B$ is the
pullback of the NS B-field onto the worldvolume. 

The conditions for supersymmetric embeddings with
nonabelian fields turned on has not been given,
but they have been worked out for single D-branes
in refs. \bebestro\bergetal, for which the
action of spacetime supersymmetry and worldvolume
$\kappa$-symmetry is known \bergtown.  A compactification
preserving supersymmetry will occur if there
are constant spinors $\eta^i$ on $M$ for each of the
spacetime SUSYs.  These supersymmetries transform
the embedding coordinates (and their superpartners)
on the D-brane worldvolumes; they are preserved
if one can find a $\kappa$-symmetry transformation
which cancels the SUSY transformation.
This condition can be written as
\eqn\dsusy{
	(1 - \Gamma) \eta^i = 0
}
and those $\eta^i$ which are solutions form
the unbroken SUSYs.  $\Gamma$ is defined as 
follows \bergetal.  Let $E_\mu^m$ be 
the vielbein connecting frame indices $m$ and
spacetime indices $\mu$.  We can pull this
back to the worldvolume, defining
\eqn\pbviel{
	E_{\alpha}^m = \p_\alpha X^\mu E_\mu^m (X)\ ,
}
where $\alpha$ is a worldvolume index for the $p$-brane.  
With this we can pull back the 10D $\gamma$-matrices $\Gamma_m$:
\eqn\pbgamma{
	\Gamma_\alpha = E_\alpha^m \Gamma_m\ .
}
Define
\eqn\gammawz{
	\Gamma_{(p+1)} = \frac{1}{(p+1)! \sqrt{g}}
	\epsilon^{\alpha_1\ldots \alpha_{p+1}}
	\Gamma_{\alpha_1 \ldots \alpha_{p+1}}\ ,
}
where
\eqn\induced{
	g_{\alpha\beta} = \eta_{mn}E^m_\alpha E^n_\beta
}
is the induced metric on the $Dp$-brane.  We can now write:
\eqn\project{
	\Gamma = \frac{\sqrt{g}}{\sqrt{g + \CF}}
	\sum_{\ell = 0}^{\infty}
		\frac{1}{2^\ell \ell!}
	\Gamma^{\alpha_1 \beta_1 \ldots \alpha_n \beta_n}
		\CF_{\alpha_1 \beta_1} \ldots
		\CF_{\alpha_n \beta_n} \Gamma_{(11)}^{n + (p-2)/2}
		\Gamma_{(p+1)}
}
When $\CF = 0$ this can be written in the simpler, more familiar form:
\eqn\nofproject{
	\Gamma = \epsilon^{\alpha_1 \ldots \alpha_{p+1}}
	\p_{\alpha_1}X^{\mu_1}\ldots\p_{\alpha_{p+1}}X^{\mu_{p+1}}
	\Gamma_{\mu_1} \ldots \Gamma_{\mu_{p+1}}
}
where $\Gamma_\mu = E_\mu^m \Gamma_m$.  The conditions in this
latter case have been worked out in some detail, as we 
will describe below.  These conditions match those
in refs. \bdtwo\vijayrob\ for boundary states of BPS
D-branes in flat space with constant background fields.

Solutions to Eq. \dsusy\ in the presence of nonzero $\CF$
have been worked out for flat, intersecting branes in refs.
\bdtwo\vijayrob\bergetal.
In the case of BPS D-branes
in Calabi-Yau 3-fold compactifications the
geometric conditions implied by \dsusy\ 
(and the analog for boundary states) have
been worked out in \bebestro\ooy.
These solutions fall into two classes:
``A-type'' branes wrapping special Lagrangian
submanifolds and ``B-type'' branes
wrapping holomorphic cycles.  Let us describe
each of these in turn.

\subsubsec{B branes}

``B-type'' BPS branes wrap even-dimensional,
holomorphic cycles in the Calabi-Yau \bebestro\ooy.
For B (even-dimensional) branes, \dsusy\ is solved by holomorphically
embedded curves ($2$-branes) and surfaces ($4$-branes), 
as well as by $0$ and $6$-branes with the
obvious (trivial) embeddings.
We may also have gauge fields on these
branes.  In general the gauge field may
change the definition of a supersymmetric
cycle via Eq. \dsusy.  However, if the brane
is wrapped around a holomorphic cycle,
we can find conditions for the gauge field to
preserve the supersymmetries.
In the case of
$N$ coincident $D6$-branes wrapping the entire CY threefold,
if we assume that the gauge fields live only
in the threefold then the SUSY-preserving gauge field
must satisfy the ``Hermitian Yang-Mills equations''~\bpsalg:
\eqn\genhitchin{
\eqalign{
	F_{ij} &= 0 \cr
	\omega^{2}\wedge \tr F &= c \omega^3 \ ,
}}
where $(i,j)$ and ${\bar i}, \jb$ are holomorphic and antiholomorphic
indices, respectively, on the CY.  These equations define
a ``Hermitian-Einstein'' connection $A$ with curvature $F$.  
The first equation tells us that the vector bundle is holomorphic.
The second equation tells us that the vector bundle is 
``$\omega$-stable''; conversely, $\omega$-stability guarantees
a solution to these equations \duy\
(\cf\ chapter 4 of \friedman\ for a discussion and definitions.)

For branes wrapped around holomorphic submanifolds of $M$,
these equations must be altered.  The
gauge fields polarized transverse to the cycle
are replaced by ``twisted'' scalars $\Phi$ which are
one-forms in the normal bundle to the embedding
\bsvtop, and Eq. \genhitchin\ becomes
a generalization of the Hitchin equations 
for $\Phi$ and $F$ \bpsalg.

It is believed that all topological invariants
of a D-brane configuration are given
by an element of a particular K-theory group
on $M$ \moorek\wittenk.  When the K-theory
group and/or the cohomology of $M$ has
torsion the K-theory interpretation is important;
one may have objects charged under the torsion.
The charge can be written \moorek\ as a generalization
of the results of \ghm\cheungyin:
\eqn\kcharge{
	v(E) = {\rm ch}(f_! E) \sqrt{\hat{A}(M)}
}
Here $E$ is a vector bundle on $\Sigma$;
remember that we must extend the $U(1)$
part of the gauge field $F$ by the NS B-field,
so properly the vector bundle $E$ is a polynomial
in $\CF$.  Let
$\pi: M\to\Sigma$ be the projection onto the
worldvolume and $N$ be the normal bundle of
$\Sigma\hookrightarrow M$.  There is a K-theory
element $\delta(N)$ which is roughly a delta function
on the worldvolume and depends on $N$; we can thus define
$f_! (E) = \pi^\ast E \otimes \delta(N)$.  The moduli
space of D-branes will not just be the
moduli space of vector bundles in this K-theory class
but rather the moduli space of coherent semistable sheaves
in this class \morsheave\bpsalg.
Some advantages of this
definition through K-theory and sheaves,
besides the fact that it seems to be correct, are that
it places configurations with D6-branes
(gauge field configurations on $M$) on an
equal footing with configurations without D6-branes,
and that it can describe certain singularities which lead to sensible
string compactifications.

In examples without torsion, such as the quintic,
one may describe the D-brane charge in a less esoteric fashion.  Assuming 
the branes give rise
to particles in the macroscopic directions,
for a $2n$-dimensional worldvolume $\Sigma$
we can write the D-brane coupling to the
RR gauge fields via the
``Wess-Zumino term'' as \refs{\bwithinb,\ghm,\cheungyin}:
\eqn\branecharge{
	\int_\Sigma C \wedge
	{\rm ch} (F-B) \sqrt{\frac{\hat{A}(M)}{\hat{A}(N)}}
}
where
$$ C = C^{(2n + 1)} + C^{(2n-1)}+\ldots + C^{(1)}
$$
is a sum over the $(k)$-form RR potentials that couple
to the $2n$-brane.

These RR charges reduce to conventional electric and magnetic
charges in the four noncompact dimensions.
Given two D-branes which reduce to particles, the most basic
observable we can study is the Dirac-Schwinger-Zwanziger
symplectic inner product on their charges,
\eqn\dszip{
	I(a,b) = Q_{Ea} \cdot Q_{Mb} - Q_{Ma} \cdot Q_{Eb}\ .
}
We will refer to this as the ``intersection form'' as it is closely
related to the topological intersection form for two- and four-branes.
For two six-branes, from the formulas above it is
\eqn\braneinter{
	I(a,b) = \int\ \ch (F_a)\ \ch (-F_b)\ \ahat(M)\ .
}

Finally, we quote a general theorem regarding stability
(Bogomolov's inequality \bog; \cf\ \refs{\huybrechts, \friedman}):
given a variety $X$ of dimension $n$ and $\omega$ an ample divisor on $X$,
then a $\omega$-semistable torsion free sheaf $E$ of rank $r$
and Chern classes $c_i$ will satisfy
\eqn\bogomolov{
	\int_S \left( 2 r\ c_2 - (r-1) c_1^2 \right)\wedge \omega^{n-2} \geq 0\ .
}
The parenthesized combination is called the ``discriminant'' of the
sheaf and is equal to $c_2(End(E))$.
In the special case $c_1(E)=0$ this amounts to requiring $c_2(E)\ge 0$.

\subsubsec{A branes}

An ``A-type'' BPS brane wraps a three-dimensional 
special Lagrangian submanifold $\Sigma$ \bebestro:\foot{
There is some evidence that the special Lagrangian condition
receives $\alpha'$ corrections \dup.}
\eqn\spLag{
\eqalign{
 	\omega|_\Sigma &= 0 \cr
	\Re e^{i\theta} \Omega|_\Sigma &= 
		0 \ .
}}
Here $\Omega$ is the holomorphic 3-form of the Calabi-Yau
and $\theta$ is an arbitrary phase.  Equivalently to the second
equation, we can require that $\Omega$ pulls back to a 
constant multiple of the volume element on $\Sigma$.
Furthermore the gauge field on this manifold
must be flat.
A nice introduction to the general theory of these is \hitchin.
It is shown there (and in the references therein)
that the moduli space has complex dimension $b^1(\Sigma)$.  The
space of flat $U(1)$ connections has real dimension
$b^1(\Sigma)$, and $\omega^{ij}$ can be used
to get an isomorphism between $T^*\Sigma$ and $N\Sigma$;
thus the deformations of $\Sigma$ pair up with the
Wilson lines to form $b^1(\Sigma)$ complex moduli.

For three-branes,
the DSZ inner product \dszip\ is precisely the geometric intersection form.

One application of these branes is the
Strominger-Yau-Zaslow formulation of mirror symmetry, a precise
formulation of the idea that ``mirror symmetry is T-duality'' \syz.
Since mirror symmetry exchanges the sets of A and B branes, an
appropriately chosen moduli space of A branes on $M$ 
will be the moduli space of D$0$-branes on the mirror $W$.  
Clearly $b^1=3$ for such A-branes, and SYZ argue that
$\Sigma$ will be a $T^3$ in this case.
A similar proposal was made for general B branes with bundles in \vafamirror.

Another application is the construction of $N=1$ gauge theories
with the help of brane configurations. Supersymmetric
three-cycles have been used to explore the strong
coupling limit by lifting the brane configurations to M-theory
in \klm.

Not too many explicit constructions 
of special Lagrangian submanifolds are known
and it appears (e.g. see \hitchin) that the problem is of the same
order of difficulty as writing explicit Ricci-flat metrics on a CY.
A general construction we will use below
is as the fixed point set of a real involution.

\subsec{General world-volume considerations}

Given a system $X$ of A or B D-branes, we can consider the system
which is identical except that it extends in the flat $3+1$ dimensions
transverse to $M$.  This system will have a $d=4$, $\CN=1$ supersymmetric
gauge theory as its low-energy world-volume theory, whose data is a
gauge group $G_X$; a complex manifold $C_X$ parameterized by 
chiral superfields $\phi^i$; a K\"ahler potential $K$ on $C_X$;
an action by holomorphic isometries of $G_X$ on $C_X$
(linearizing around a solution this corresponds to the usual choice of
representation $R$ of the gauge group), and
a superpotential $W$ (a holomorphic function on $C_X$ invariant under
the action of $G_X$).  If $G_X$ contains $U(1)$ factors, each of these
can have an associated real constant $\zeta_a$ (the ``Fayet-Iliopoulos
terms'').

In the classical
($g_s\rightarrow 0$) limit, the moduli space of this theory
is the solutions of $F_i=\p W/\p \phi^i=0$ (the ``F-terms'')
and $D^a=\zeta^a$ (the ``D-terms'')
modulo gauge transformations, where $D^a$ is the moment map generating
the associated gauge transformation (and $\zeta^a\equiv 0$ in the
non-abelian parts of the gauge group).

We review this well-known material for a number of reasons.  First,
we remind the reader that
although some of our later discussion will use other realizations of
this D-brane system (for example as particles in $3+1$ dimensions),
the world-volume theories for these other realizations are all obtained
by naive dimensional reduction from the $3+1$ theory (if $g_s\sim 0$),
while the $3+1$ language makes it easy to impose supersymmetry.

Second, it is known that
the study of bundles and sheaves on CY three-folds is 
much more complicated than that for K3; this complication has a direct
physical counterpart in the reduced constraints of $\CN=1$ supersymmetry.
The most basic example of this is the fact that -- unlike the
case for K3 -- there is no formula
for the dimension of the moduli space of $E$ given $c(E)$.
The main reason for this is that this dimension is not necessarily
constant -- the moduli space can have branches of different dimension,
and can depend on the moduli of the CY as well.

Physically, this corresponds to the possibility of a fairly
arbitrary superpotential in the low energy theory.  Indeed, the
language of superpotentials and $\CN=1$ effective Lagrangians might
be the best one for these problems, much as hyperkahler geometry and
hyperkahler quotient is for instanton problems in four dimensions.
Just as the self-dual Yang-Mills 
equations can be regarded as an infinite-dimensional
hyperkahler quotient, we might pose the problem of rephrasing the YM
equations under discussion as the problem of finding the moduli space
of an $\CN=1$ effective theory with an infinite number of fields.

The basic outlines of part of this treatment are known
(see \donald, ch. 6 for a very clear discussion
of the four-dimensional case).
The two equations \genhitchin\ will correspond directly to the F-term
(superpotential) constraints and the D-term constraints, respectively.
Indeed, the problem of solving $F_{ij}=0$ is a purely holomorphic problem,
while it is not hard to see that the expression $F^a\wedge \omega^{n-1}$ is
the moment map generating conventional gauge transformations.
The stability condition on the bundle is exactly the infinite-dimensional
counterpart of the usual condition in supersymmetric gauge theory
for an orbit of the complexified gauge group to contain a solution of the
D-flatness conditions (e.g. see \lutytaylor).
Donaldson's theorem proving the existence of such solutions proceeds exactly
by considering the flow generated by $i$ times the moment map to a minimum;
the Uhlenbeck-Yau generalization is quite similar (for technical reasons a
different equation is used).

The other part of the story -- translating the problem of finding holomorphic
vector bundles into solving constraints on a finite-dimensional configuration
space, which can be derived from a superpotential -- does not seem to have been
addressed in as systematic a manner; clearly this could be useful.

In a sense the six-dimensional problem is the ``universal''
one which also describes the lower-dimensional branes.  Not only can their charges
be reproduced, but gauge field singularities will correspond to specific lower
dimensional branes (e.g. the small instanton).  Furthermore, there is a sense in
which even the
lower-dimensional brane world-volume theories are six-dimensional if we include
``winding strings'' (by analogy to tori and orbifolds, although 
this idea has not yet been made precise).  Treating a system of N D$0$'s as
quantum mechanics requires neglecting these strings, which one expects to be
problematic once the separation between branes approaches the size of the space.

We now turn from these abstract ideas to our concrete example.

\subsec{D-branes on the quintic}

Perhaps the best-studied family of Calabi-Yau manifolds
is the quintic hypersurfaces in $\IP^4$.  A relatively
thorough discussion of these is contained in the
classic paper \cand.  The moduli space of these
manifolds is locally the product of $b_{2,1}=101$
complex structure deformations and $b_{1,1}=1$
deformations of the complexified K\"ahler forms
$B + i J$ (where $B$ is the flux of the NS-NS B-field).
We will be particularly interested in the Fermat quintic
\eqn\fermateq{
	P = \sum_{i=1}^5 z_i^5 = 0
}
where $z_i$ are the homogeneous coordinates on $\IP^4$.
Note that this equation has a $S_5 \times Z_5^4$ 
discrete symmetry; the $Z^5$ generators are
$g_i:z_i\rightarrow \omega z_i$ and satisfy the relation 
$\prod_{i=1}^5 g_i = 1$, while the $S_5$ permutes the
coordinates in the obvious way.

\subsubsec{B branes on the quintic}

As we have discussed, D-branes on the quintic can
be described by vector bundles or sheaves on this space.
Let us denote the charge carried by a single D$2p$-brane wrapped
about a generator of $H^{2p}$ as $Q_{2p}=1$.

Transporting a D-brane configuration about closed,
nontrivial cycles of the moduli space of K\"ahler structures
will induce an associated $Sp(4,\BZ)$ monodromy 
on the B branes.  We will discuss the
monodromy more completely in the next section,
but there is already one cycle in the moduli space
which can be understood in the large 
volume limit: $B\rightarrow B+1$, where $B$ is the NS 2-form.
The action on the charge $Q$ can be seen from 
Eqs.~\kcharge,\branecharge\ \morII.
Mathematically this corresponds to the possibility to tensor the
vector bundle $V_{2p}$ with a $U(1)$ bundle of $c_1=1$.  
This preserves stability and the dimension of the moduli space.  
Given a bundle $V$ this operation and its inverse can be used to produce
a related bundle with $-r< c_1\le 0$: this is referred to as
a ``normalized'' vector bundle.

There is no classification of
vector bundles and coherent sheaves on the quintic,
but we can write down a few examples in order
to orient ourselves when discussing specific
boundary states at the Gepner point.

BPS D2-branes wrap holomorphic 2-cycles of the Calabi-Yau,
the same cycles as appear in worldsheet instanton corrections.
Such cycles can have arbitrary genus and arbitrary degree.
Degree one rational curves
are generically rigid on the quintic \katzcurve.
Nontheless for special quintics, families
may exist; for example, in the case
of the Fermat quintic \fermateq, there
are 50 one-parameter families essentially
identical to the family \albano\katzrat:
\eqn\ratfam{
\eqalign{
	&(z_1, z_2, z_3, z_4, z_5) = (u, -u, av, bv, cv)\cr
	&a^5 + b^5 + c^5 = 0;\ \ \ a,b,c \in \IC\ ,
}}
where $(u,v)$ are homogeneous coordinates in $\IP^1$.
Once we perturb away from the Fermat point, these
moduli are lifted and a finite number of
rational curves remain \albano.
This could be described in the world-volume
theory by a superpotential of the general form
$$
W = \phi \psi^2
$$
where $\phi$ are complex structure moduli; $\phi=0$ is the Fermat point;
$\psi$ are curve moduli, and $\psi=0$ a curve which exists for generic
quintics.

The infinitesimal description of deformations of such cycles
is as sections
of the normal bundle, which by the Calabi-Yau condition will be
$\CO(a)\oplus \CO(b)$ with $a+b=-2$ for a rational curve.  One
might think that all one needs to find examples of families is to
find examples with $a\ge 0$ or $b\ge 0$, but this is not true as
deformations can be obstructed.  The canonical example is given
by resolving the singularity in $\BC^4$
\eqn\obsing{
xy = z^2 - t^{2n} \ .
}
For $n=1$ this is the conifold singularity and the ``small'' resolution
contains a rigid $\IP^1$, parameterized by $x/(z-t) = (z+t)/y$.
It can be shown \reid\ that for 
$n>1$ the resolution also contains a $\IP^1$, now
with normal bundle $\CO\oplus \CO(-2)$, 
but the deformation is obstructed at $n$'th
order, as could be described by the superpotential
\eqn\obssuper{
W = \psi^{n+1} .
}
Intuitively this can be seen 
by deforming \obsing\ by a generic polynomial in $t^2$,
which splits the singularity into 
$n$ conifold singularities, each admitting a rigid
$\IP^1$.  If we then tune the parameters 
to make these $\IP^1$'s coincide, a superpotential
describing the $n$ vacua will degenerate to \obssuper.
Such singularities do appear in large families of quintic CY's 
\katzcurve.\foot{
(Note added in v2):
The idea that the moduli space of such a curve
can always be described as the critical 
points $W'=0$ of a single holomorphic function
was apparently not known to mathematicians.
We thank S. Katz for a discussion on this point.}

It turns out that
the curves in \ratfam\ provide another example of
obstructed deformations \albano.\foot{(Note added in v3):
We would like to thank S. Katz for explaining this
example, pointing out a mistake in our earlier draft,
and suggesting the superpotential discussed here.}  
The normal bundle of these
curves is $\CN = \CO(1)\oplus\CO(-3)$;
as $\dim H^0(\CN) = 2$, there must be another
obstructed deformation; call it $\rho$.
The correct counting of curves upon deforming away from the Fermat point
can be reproduced by a superpotential $\rho^3$.
The modulus $\rho$ is also connected to the fact that pairs of the
$50$ families in \ratfam\ intersect (e.g. take \ratfam\ and
the family $(av,bv,u,-u,cv)$ with $c=0$); it describes deformations
into the second family.
All of this structure can be summarized in the superpotential
$$ W(\rho, \psi) = \rho^3\psi^3 + \phi F(\rho,\psi) + \ldots\ ; $$
where $\phi F$ generalizes the $\phi\psi^2$ term discussed above.

Higher genus curves can generically come in families and examples
can be found as complete intersections of hypersurfaces in $\IP^4$
with the quintic.  A particular example is the intersection
of two hyperplanes with the quintic \katzrat:
\eqn\hyperpl{
	\sum_{k=1}^5 a_k z_k = \sum_{k=1}^5 b_k z_k = 0\ ,
	a_k, b_k \in \IC\ .
}
It is easy to see that there are six independent 
complex parameters after rescaling the equations.
The curve is genus 6, and the area of the
curve $C$ is $\int_C J = 5$, where $J$ is the unit
normalized K\"ahler form of $\IP^4$,
i.e. 
\eqn\jnorm{
\eqalign{
	&\int_{\IP^4} J\wedge J \wedge J \wedge J = 1\cr
	&\int_{{\rm Quintic}} J\wedge J\wedge J = 5
}}
Thus this brane
has $Q_2=5$.\foot{See 
ref. \hubsch, chapters 1 and 2 for a nice 
description of complete intersections in projective
spaces, and of techniques for performing the
calculations we allude to here.}
There will be six additional complex moduli coming
from Wilson lines of the $U(1)$ gauge field around the
12 cycles of the curve.

Similarly, four-branes can be obtained as the intersection with another
hypersurface in $\BP^4$.  For example, the intersection
of the quintic with a single hyperplane
$$ \sum_k a_k z_k = 0 $$
produces a four-parameter family of four-cycles $S$.
Their volume is $\int_S J = 5$ and so $Q_4=5$.
In addition $c_2 (TS) = 11 J^2$; so that the coupling
of $C^{(1)}$ to $p_1/48$ in eqs. \kcharge,\branecharge\ 
leads to an induced $0$-brane charge of $55/24$.
The four-brane generically may support nontrivial
gauge field flux over two-cycles, corresponding
to D2-brane charge, or instanton solutions,
corresponding to zero-brane charge.
Some discussion of the moduli space of
four-branes in a Calabi-Yau can be found in
\bhcy. By \bogomolov,
stability of the vector bundle on the four-brane
requires $Q_0 > 0$.

Finally we can look at the case of D6-branes wrapping the
entire Calabi-Yau manifold.  In fact we will find that all
of the boundary states we examine at the Gepner point
will have non-trivial six-brane charge.  A single
six-brane by itself will have no moduli.
The $U(1)$ gauge
field on a single 6-brane can support flux with 
first Chern class $c_1 = n$ corresponding to $Q_4=n$.
We can get the relevant bundles by restriction from
$U(1)$ bundles on $\IP^4$.  The latter have no moduli, and we will not gain
any upon restriction.  

We can also imagine binding D2-branes to the D6-brane, by analogy
to $2-6$ (or $0-4$) configurations in flat space.
For $Q_6=1$ and $Q_4=0$ this appears singular;
$U(1)$ gauge fields do not support smooth instanton solutions.
The brane counterpart to this is that the $2-6$
strings cannot be given vevs which bind the branes
and give mass to the relative $U(1)$s.  
This might lead us to predict that such states, if they exist at all,
exist only as quantum-mechanical bound states.
Such a state should be easily identifiable because it appears at the
junction of Coulomb and Higgs branches of the moduli space;
a small perturbation should put it on the Coulomb branch and
produce two $U(1)$
gauge fields in the macroscopic direction.  In the classical considerations
of this paper, it should not show up at all.

For $Q_6 > 1$, we require information
about vector bundles on the Calabi-Yau.
A well-known example with $Q_6=3$ is deformations of
the tangent bundle.
This has vanishing $c_1$ and $c_2 (E) = 10 J$
giving us $Q_2=50$.
The dimension of the moduli space is $224$.
This example can be generalized
as follows. 
(Such generalizations
are due to for example \refs{\wittnr,\distler} in the physics
literature, and were previously known as 
``monads'' in the math literature).
We consider a complex of holomorphic vector bundles
$$
0 \rightarrow A \rightarrow^a B \rightarrow^b C \rightarrow 0
$$
such that $\ker a=0$, $\im a$ is a subbundle of $B$, $\im b = C$
and define our new bundle as
$$
E = \ker b/\im a.
$$
For a hypersurface $M$ in $\IP^n$, simple bundles to start with are
direct sums of the line bundles $\CO(n)$ restricted to $M$, as in
$$
0 \rightarrow \oplus \CO \rightarrow \oplus_{i=1}^m \CO(q_i)
\rightarrow \CO(\sum_{i=1}^m q_i) \rightarrow 0
$$
This data allows computing the Chern classes: 
$$
c_n = (\sum_{i=1}^m q_i)^n - \sum_{i=1}^m q_i^n .
$$
The dimension of the moduli space can also be computed,
but this is not as easy.

A physical realization of this construction is to start with fields $\lambda^i$
parameterizing sections of $B$ (e.g. the world-sheet fermions of
a heterotic string theory), include a superpotential
enforcing the constraints $b^a_i \lambda^i =0$, 
and gauge invariances identifying $\lambda^i \sim \lambda^i + a^i$.
Although it is not the only place this construction 
appears (e.g. see \probe), 
the most relevant version for present 
purposes is in linear $(0,2)$ models \distler.  
These constructions have the advantage that they can be studied
with conventional world-sheet techniques; a disadvantage is that one
requires the anomaly cancellation conditions $c_1=0$ and $c_2(V)=c_2(T)$
to get a sensible model, so only a subset of possible $V$ can be obtained.

The anomaly cancellation conditions also appear in D-brane constructions
of the dual type I theories
as the consistency condition that the total RR charge vanishes \gimonpol.
However in this
context we need not consider branes which fill the noncompact dimensions
but can instead consider lower dimensional branes, for which these
consistency conditions are not required (a point emphasized in \quivers).
It seems likely that this additional freedom will lead to a simpler theory.

Another construction of vector bundles on a CY is the Serre construction.
Given a holomorphic curve (satisfying certain conditions),
this produces a rank $2$ vector bundle with a section having
its zeroes on the curve.  In \thomas\ this is used to produce an example
of a vector bundle with an obstructed deformation (on a different CICY).

Finally, to conclude this section, there are a few explicit constructions
of bundles on $\IP^4$ in the literature using monads, such as the
Horrocks-Mumford bundle ($r=2, c_1=5, c_2=10$) and the bundle of Tango
($r=3, c_1=3, c_2=5, c_3=5$) \okonek, which can be restricted to
the hypersurface $P=0$ to produce new examples. 

\subsubsec{A branes on the quintic}

The simplest example of supersymmetric $3$-cycles on the quintic
are the real surfaces $\Im\omega_jz_j=0$ with $\omega_j^5=1$;
this was described in \bebestro\ for $\omega=1$. 
These cycles are determined by the five phases
$(\omega_1,\omega_2,\omega_3,\omega_4,\omega_5)$ up to the 
diagonal $\BZ_5$ action $\omega_i\rightarrow \omega\omega_i$
(which is just a remnant of the equivalence of
homogeneous coordinates under complex multiplication), 
so they come in a $625$-dimensional 
irrep of the discrete symmetry $S_5\times Z_5^4$.

The equation $\sum(\omega_jx_j)^5=0$, 
where $\omega_i x_i \in \IR$, always has a unique
solution for $x_k$ in terms of the other real coordinates; thus
the cycle is the real projective space $\BR P^3$. 
The first homotopy group is $\pi_1 (\IR P^3) =\BZ_2$;
by the discussion above (\cf\ \hitchin) the wrapped 3-branes
cannot have any continuous moduli, but they
can support a discrete $Z_2$-valued Wilson line.

To compare these cycles with Gepner boundary states it will be useful
to find their intersection matrix. 
Let us choose the coordinate system $z_1=1$ on $\IP^4$,
so that $\omega_1 = 1$.
Regard the cycle $(1,1,1,1,1)$ as an embedding
of the coordinates $x_2$,$x_3$ and $x_4$ into the quintic
with positive orientation.
The other surfaces are obtained by $Z_5^4$ rotation
from this one, $\prod_{i=1}^5 g_i^{k_i} (1,1,1,1,1)$. 
Since the intersection matrix must respect the $Z_5^4$
symmetry, it can be written as a polynomial in the generators $g_i$
and is determined by the matrix elements
\eqn\rpthreeintersect{
\vev{(1,1,1,1,1)|(1,\omega_2,\omega_3,\omega_4,\omega_5)}
	= \vev{(1,1,1,1,1)|g_2^{k_2} g_3^{k_3} g_4^{k_4}
		g_5^{k_5}|(1,1,1,1,1)}
}
where $g_i^{k_i}:z \to \omega^{k_i} z$.
$S_5$ symmetry also constrains the problem in an obvious way.

There are different possibilities for intersections with the surface
$(1,1,1,1,1)$ in this coordinate system.  If $\omega_2$, $\omega_3$,
$\omega_4$ and $\omega_5$ are all different from $1$ there is no
intersection in this coordinate patch. 
If only three of them are
different from $1$ there is exactly one intersection in this
coordinate patch and the intersection has the signature
$\sgn\Im\omega_2\Im\omega_3\Im\omega_4$ assuming that
$\omega_5=1$. If the two surfaces intersect on a higher dimensional locus the
intersection number has to be calculated by a small deformation of one
of the two surfaces. This deformation has to be normal to both
surfaces. Because of the special Lagrangian property of the undeformed
surfaces this ``normal bundle'' of the intersection locus can be
identified with its tangent space. The intersection number is then
given by the number of zeros of a section of the tangent bundle of the
intersection locus.

For example,
in the case that exactly two $\omega_j$'s are not $1$ the intersection
locus is a circle. A circle can have a nowhere vanishing section of
its tangent bundle and the intersection number in this coordinate
patch is $0$. As another example, let precisely one $\omega_j\ne 1$.
The intersection locus is then an $\BR P^2$. A section of its tangent bundle
has one zero, as can be seen by modding out the 'hedgehog
configuration' of an $S^2$ by $\BZ_2$. The orientation of this
intersection is given by the intersection in the remaining complex
dimension, i.e. by $\Im\omega_j$.

In order to compute the full intersection
we must look at all possible patches.  This can be done
by using the constraint $\prod_{i=1}^5 g_i = 1$
to rewrite $(1,\omega_2,\omega_3,\omega_4,\omega_5)$
as $(\omega_2^{-1},1,\omega_2^{-1}\omega_3,
\omega_2^{-1}\omega_4,\omega_2^{-1}\omega_5)$
and so on.  We then add all of the intersection numbers
for all of these patches.  Thus, although we find that
$\vev{(1,1,1,1,1)|(1,\omega,\omega,\omega,\omega)}=0$
in the $z_1 = 1$ coordinate patch, the total intersection
number -- the coefficient of $\prod_{i=2}^5 g_i$ in
the intersection matrix -- is $1$.  Another example is the
intersection of $(1,1,1,1,1)$ with $(1,\omega,\omega,\omega,1)$
which gives a circle in the patch $z_2 = 1$ and a point
in the patch $z_1 = 1$.

A simple general formula that matches all of these results is
\eqn\intersectguess{
	I_{\BR P^3} = \prod_{i=1}^5 (g_i+g_i^2-g_i^3-g_i^4) .
}

\newsec{Stringy geometry}

Type \IIb\ string compactification on 
a general CY threefold $M$ leads to an $\CN=2$, $d=4$
supergravity with $b_{2,1}+1$ vector fields 
($b_{2,1}$ vector multiplets plus the graviphoton)
and $b_{1,1}+1$ hypermultiplets (including the 4d dilaton); 
in \IIa\ these identifications are reversed.
The most basic physical observables which reflect the structure of $M$
are those described by the special 
geometry of the vector multiplets. This geometry is determined by a
prepotential $F_K$ of K\"ahler deformations
in the \IIa\ case, and by the prepotential $F_c$ for complex
structure deformations in the \IIb\ case.

A fundamental result from the study of the worldsheet 
sigma model is that $F_c$ can be determined entirely from classical 
target space geometry;
it receives no worldsheet quantum ($\ap$) corrections.  
Let us then discuss the complex structure moduli space.
Choose a basis for the $3$-cycles $\Sigma^i\in H_3(M,\IZ)$
(where
$i=0,\ldots,b_{2,1}, b_{2,1}+1,\ldots,2 b_{2,1}+2$),
so that the intersection form $\eta^{ij} = \Sigma^i \cdot \Sigma^j$
takes the canonical form
$\eta^{i,j}=\delta_{j,i+b_{2,1}+1}$ for $i=0,\ldots,b_{2,1}$ 
(an $a$ cycle with a $b$ cycle).
The $b_{2,1}+1$ vector fields come from reducing the RR
potential $C^{(4)}$ on the $a$ cycles, while the $b$ cycles produce their
$d=4$ electromagnetic duals.  Thus a three-brane wrapped about the
cycle $\Sigma = \sum_i Q_i \Sigma^i$ has (electric,magnetic) charge
vector $Q_i$.  Note that $H_3(X)$ forms a nontrivial
vector bundle over the moduli space $\CM_c$ of
complex structures; a given basis
in $H_3(X,\IZ)$ will have monodromy in $Sp(b_3,\IZ)$
as it is transported around singularities in $\CM_c$.

The primary observables are the periods of the holomorphic three-form,
$$
\Pi^i = \int_{\Sigma^i} \Omega .
$$
In $\CN=2$ language these are the vevs of the scalar fields in the
corresponding vector multiplets.  The $a$-cycle $\Pi^i$'s can be used
as projective coordinates on the moduli space; the $b$-cycle periods
then satisfy the relations 
$\Pi^j = \eta^{ij} \p \CF/\p \Pi^i$.  If we fix (for example)
$\Pi^0 = 1$ to pass to inhomogeneous coordinates, the
related vector field is the graviphoton.
These periods determine the 
central charge of a three-brane wrapped about the cycle 
$\Sigma = \sum_i Q_i [\Sigma^i ]$:
$$
Z = \int_\Sigma \Omega = Q_i \Pi^i .
$$
Thus the mass of a BPS three-brane is \ceretal:
\eqn\bpsmass{
m_Q = c |Z| = c |Q\cdot \Pi|
}
where $c$ is independent of $Q$.
If we use four-dimensional Einstein units for $m$, 
it is $c = 1/g_s (\int \Omega\wedge\bar\Omega)^{1/2}$.

In contrast to $F_c$, $F_K$ receives world-sheet instanton corrections
to the classical computation.
The exact worldsheet result can be 
obtained by mirror symmetry: $F_K$ for \IIa\ on $M$
is equal to $F_c$ for \IIb\ on the mirror $W$ to $M$.
Of course this requires a map between the 
periods of $M$ and $W$.
This analysis has been carried out for the quintic
in \cand\ (see \asplect\ for a summary) and we will
quote the result in this case.

The mirror $W$ to the quintic threefold $M$ can be 
obtained \greple\ as a $Z_5^3$ quotient of a special quintic
$$
0 = \sum_{i=1}^5 z_i^5 - 5\psi z_1z_2z_3z_4z_5\ .
$$  
The transformation $\psi\rightarrow \alpha\psi$ with $\alpha^5=1$ can
be undone by the coordinate transformation $z_1\rightarrow
\alpha^{-1}z_1$
and thus the complex moduli space of $W$'s can be
parameterized by $\psi^5$.  This is an ``algebraic''
coordinate, which although not directly observable, does
appear naturally in the world-sheet formulations  \refs{\wittph,\smalldist}.

The moduli space $\CM$ has three singularities, about which
the three-cycles in $W$ will undergo monodromy.
Each singularity has physical significance.
First, $\psi^5\rightarrow\infty$ is the ``large complex structure limit''
mirror to the large volume limit.  In this limit \asplect\
\eqn\asymptmap{
	(5\psi)^{-5}\to e^{2\pi i (B + i J)}\ ,
}
where $B$ is the NS B-field flux around the 2-cycle forming
a basis of $H_2(M)$, and $J$ is the size of that 2-cycle.
Next, $\psi^5\rightarrow 1$ is a conifold singularity; here a
wrapped three-brane becomes massless \conifold.  
This turns out to be mirror to the ``pure''
six-brane \refs{\polstrom,\greenekanter}.
Finally, at $\psi^5=0$ the model obtains an additional $Z_5$ 
global symmetry; this is an orbifold singularity of moduli space.
The Gepner model $(3)^5$ lives at this point
in K\"ahler moduli space of $M$ \wittph.

Each singularity in $\CM$ gives a noncontractible
loop, which is associated with a monodromy on the basis of
$3$-cycles in $W$ (or even homology in $M$)
and thus on the periods.  We let $A$ be the monodromy
induced by $\psi\rightarrow \alpha\psi$ around $\psi=0$; clearly $A^5=1$.
$T$ will be the monodromy induced by going once around the conifold
point, and $B$ will be the monodromy induced by taking
$\psi\rightarrow\alpha^{-1}\psi$ around infinity.  These satisfy the
relation $B=AT$.  One may make the physics associated
with a given singularity manifest by choosing variables 
(the periods) for which the associated monodromy is simple.

In our case the periods $\Pi_i$ satisfy a 
Picard-Fuchs differential equation
of hypergeometric type.  Since $b^3=4$ it is fourth order and quite
tractable.  There will be four independent
solutions and as per the discussion above,
we generally want to choose a basis making one of
the monodromies simple.  Two such bases are particularly
natural.  The first is the large volume basis which we will denote 
$(\Pi_6,\Pi_4,\Pi_2,\Pi_0)^t$.  Up to an upper triangular transformation
this is determined by the asymptotics as $\psi^5\rightarrow \infty$
\eqn\largevol{
	\left(\matrix{\Pi_6\cr \Pi_4\cr 
		\Pi_2\cr \Pi_0}\right) \rightarrow
	\left(\matrix{-{5\over 6}(B+iJ)^3
		\cr -{5\over 2}(B+iJ)^2\cr B+iJ\cr 1}\right)\ .
}
The coefficients correspond to the classical volumes of the cycles.
The signs were chosen so that the supersymmetric brane configurations
have positive relative charges.  We will derive the monodromy
below.

The other natural basis for us makes
the monodromy $A$ simple, and is appropriate
for describing the Gepner point.
If we choose a solution $\Pi^G_{0}(\psi)$
analytic near $\psi=0$, the set of solutions
\eqn\gepnerbasis{
	\Pi^G_{i}(\psi) = \Pi^G_{0}(\alpha^i\psi)
}
will provide a basis with the single linear relation
$0=\sum_{i=0}^4 \Pi^G_{i}$.  
It turns out that the $0$-brane period $\Pi_0$
(the solution $\tilde\omega_0$ of \cand, equation (3.15))
is analytic near $\psi=0$ and thus we can set $\Pi^G_{0}=\Pi_0$
and define the others using \gepnerbasis.
We then (as in \cand) choose the period vector
$(\Pi^G_{2}, \Pi^G_{1}, \Pi^G_{0}, \Pi^G_{4})^t$.
In this basis, the three monodromy matrices are \foot{
There is a typo in table I in \cand\ as published in Nuclear Physics B.}
\eqn\Bgbasis{
\eqalign{
A^G &= \left(\matrix{
	-1&-1&-1&-1 \cr
	1&0&0&0 \cr
	0&1&0&0 \cr
	0&0&1&0 }\right) \cr
T^G &= \left(\matrix{
	1& 4& -4& 0 \cr
	0& 0& 1& 0 \cr
	0& -1& 2& 0 \cr
	0& 4& -4& 1 }\right)\cr
B^G &= \left(\matrix{
	-1& -7& 5& -1 \cr
	1&  4& -4& 0  \cr
	0& 0& 1& 0 \cr
	0&  -1& 2& 0 }\right)
}}

In \cand, the relation between the large volume and Gepner bases
proceeds through a third basis which we will call $\Pi^3$,
which is naturally described by a particular basis of 3-cycles
in $W$.  The intersection form in this basis has the canonical form
$\eta_{13}=\eta_{24}=-1$, and the $T$ monodromy is simple:
$\Pi^3_i \rightarrow \Pi^3_i + \delta_{i,2} \Pi^3_4$.
Thus $\Pi^3_4$ is the vanishing cycle at the conifold
and $\Pi^3_2$ is its dual.
This turns out to be enough information to relate it to the Gepner
basis uniquely up to a remaining $SL(2,Z)$ acting on $\Pi^3_1$ and
$\Pi^3_3$, which we may fix arbitrarily.  
One then finds a transformation of $\Pi^3$
to a basis satisfying \largevol.
This is an $SL(2,Z)$ transformation of the type which was unfixed
in the previous step; so the $\Pi^3$ basis has no
significance intrinsic to our problem of relating
$\Pi^G$ to the large-volume basis.  Thus we will merely quote the
final result for this change of basis, which is:
\eqn\geptobig{
\eqalign{
	&\Pi = M \Pi^G \qquad 
		Q = Q^G M^{-1} \qquad A = M A^G M^{-1} \ldots \cr
	&M = L \left(\matrix{
	0& -1& 1& 0 \cr
	-{3\over 5}& -{1\over 5}& {21\over 5}& {8\over 5} \cr
	{1\over 5}& {2\over 5}& -{2\over 5}& -{1\over 5} \cr
 	0& 0& 1& 0 }\right)
}}
Here $Q$ and $Q^G$ are the charge vectors in the large-radius
and Gepner basis respectively.
(In the notation of \cand, $M=KNm$: with $K$ a matrix taking
the vector $(Q_4,-Q_6,Q_2,Q_0)$ of their conventions to our conventions;
and $N$ taken with $a'=b'=c'=0$.)
The matrix $L$ is an as-yet undetermined $Sp(4,Z)$ 
ambiguity in the $Q_2$ and $Q_0$ charges of the 
six- and four-branes:
$$
L =  \left(\matrix{
1&0&-b&-c\cr 0&1&a&b \cr 0&0&1&0 \cr 0&0&0& 1
}\right)
$$
with $(a,b,c)$ integers (the $(a',b',c')$ of \cand).

Given the classical
intersection form $\eta$ in the large-radius limit,
we can now determine the intersection form in the Gepner basis:
\eqn\gepint{
	\eta_g = M^{-1} \eta (M^{-1})^t =
	\left(\matrix{
	0& -1& 3& -3 \cr
	1& 0& -1& 3 \cr
	-3& 1& 0& -1 \cr
	3& -3& 1& 0 }\right)\ ,
}
where
$\eta_{14}=-\eta_{41}=-\eta_{23}=\eta_{32}=1$ from \cand.\foot{
The signs 
$\Sigma_6 \cdot \Sigma_0 = +1$ and $\Sigma_4 \cdot \Sigma_2 = -1$
in the large
volume intersection form $\eta$ follow from the definition \braneinter.}
$L$ does not enter since it is symplectic, and so preserves $\eta$.
$\eta_g$ has determinant $25$ and thus the Gepner basis is not
canonically normalized; this point will not be important for us.

We want to better understand the ambiguity $L$.
We can start by comparing the monodromy $B$
with our expectations from the large volume 
limit.  One may define a basis of charges
such that $\Gamma^{RR}_k$ is the charge under
the RR potential $C^{(k+1)}$, with the switch in four- and
six-brane charge as in \largevol.
In this basis the effect of the shift 
$B\rightarrow B+1$ follows from
Eq. \branecharge:
\eqn\Bmonopre{
	B_L = \left(\matrix{
	1&1&-{5\over 2}&-{5\over 6} \cr
	0&1&-5&-{5\over 2} \cr
	0&0&1&1 \cr
	0&0&0&1 }\right)\ .
}
The factors $1/2$ and $1/6$ in this expression come from expanding the
exponential (they can also be seen in \largevol)
and indicate that in this basis
the charges are not integers.

The $B$ monodromy in the $\Pi$ basis \largevol\ is
\eqn\Blvol{
	B = \left(\matrix{
	1& 1& 3- a& -5 - 2 b \cr
	0& 1& -5& -8+ a \cr
	0& 0& 1& 1 \cr
	0& 0& 0& 1 }\right)\ .
}
Eqs. \Bmonopre\ and \Blvol\ agree if $a=11/2$ and $b=-25/12$,
\ie\ if we make a non-integral redefinition of the charge lattice.
The explanation of this is that the intersection form in the
conventions leading to \Bmonopre\ is actually not canonical, because
it includes the other terms in \branecharge.
If we act on the basis \largevol\ with the matrix
$L(a=11/2,b=-25/12,c)$, we can see that the charges
are modified in precisely this way.
The modification due to $b$
comes from the $\hat A$ term in \braneinter
(as $c_2=50$ for the quintic).
$a$ induces a two-brane charge on the four-brane 
and might come
from $c_1$ of its normal bundle.  These effects were
referred to in \bpsalg\ as the ``geometric Witten effect''.

The most interesting ambiguity comes from $c$ which
induces zero-brane charge on the six-brane.  
In \cand\ this was attributed to the
sigma model four-loop $R^4$ correction in the bulk Lagrangian.
In the D-brane context, one possibility is that this comes from an as yet
unknown term at this order in the D-brane world-volume Lagrangians.
We should also keep in mind that the intersection form we are computing
involves the bulk propagation of the RR fields between the branes, so
another possibility is that
it comes from a partner to the $R^4$ term in the bulk Lagrangian which
affects the RR kinetic term in a curved background.

In \cand, the redefinition $L$ was used to make the charge basis integral,
but an overall $Sp(4,Z)$ ambiguity was left over.  It is in general more
useful to have an integer charge basis so we will follow this procedure
(this was already done implicitly as we took integer coefficients
in the change of basis).
We can resolve most of the $Sp(4,Z)$ ambiguity by calling 
the state which becomes massless at the mirror of the
conifold point a ``pure''
six-brane with large volume charges $(1\ 0\ 0\ 0)$, following 
\refs{\polstrom,\greenekanter}.
This determines $b=c=0$.  A geometrical argument for this is that any fluxes
on the six-brane would produce additional contributions to its energy.
If there is a line from the large volume limit to the conifold point
along which the six-brane becomes massless with no marginal stability
issues, this argument will presumably be valid.  Another argument
is that we will find 
this state as a Gepner model boundary state with no moduli,
as is appropriate for a pure six-brane.  Finally, this choice simplifies
the charge assignments for the other boundary states.

We still have the ambiguity in $a$ to fix.  As it happens this does not
enter into the results we discuss, so we have no principled way to do this.
We will simply set it to zero.  

\newsec{Boundary states in CFT}

\subsec{Some results from boundary conformal field theory}

A CFT on a Riemann surface with boundary requires specifying 
boundary conditions on the operators. 
For sigma models 
these conditions can be derived
by imposing Dirichlet and/or Neumann boundary conditions
directly on the sigma model fields. 
For more general CFTs we do not have a nice
Lagrangian description; so the construction,
classification, and interpretation of boundary
conditions is not as straightforward.
(See \refs{\sagnotti,\reckone,\fusch,\gutpone}\ and references there for recent work
in this direction.)

If the CFT has a chiral symmetry algebra one may
simplify the problem by demanding that the boundary conditions
are invariant under the symmetry. 
We can start with the Virasoro
algebra which must be preserved (particularly in string
theory where the symmetry is gauged).  
Let the boundary be at $z=\bar{z}$ in some local coordinates.
Reparameterizations should leave the boundary fixed, so
we must impose $T=\bar T$. If the remaining
symmetry algebra is
generated by chiral currents $W^{(r)}$ with spin $s_r$, then the
boundary conditions are
\eqn\chiralbc{
	W^{(r)}=\Omega\bar W^{(r)}\Omega^\dagger\ ,
}
where $\Omega$ is an automorphism of the symmetry algebra.

We are interested in describing BPS D-branes
which preserve $\CN=1$ spacetime SUSY.
The closed-string sector will have at least
$\CN=(2,2)$ worldsheet SUSY and
the boundary conditions must preserve
a diagonal $\CN=2$ part \refs{\banksetal,\banksdixon}.
Eq. \chiralbc\ leads to two classes of boundary conditions
\ooy: the ``A-type'' boundary conditions
\eqn\atypebc{
	T=\bar T, \;\;
	J=-\bar J, \;\;
	G^+=\pm\bar G^-\ ,
}
and the ``B-type'' boundary conditions
\eqn\btypebc{
	T=\bar T, \;\;
	J=\bar J, \;\;
	G^+=\pm\bar G^+\ .
}
These conventions correspond to the open-string
channel where the boundary propagates in worldsheet time.
For Calabi-Yau 
compactification at large volume,
A-type boundary conditions correspond to D-branes
wrapped around middle-dimensional
supersymmetric cycles;
and B-type boundary conditions to D-branes wrapped around
even-dimensional supersymmetric cycles \ooy.

A CFT on an annulus can also be studied in
the closed-string channel where time flows
from the one boundary to the other.  The
boundaries appear as initial and final
conditions on the path integral and are
described in the operator formalism
by ``coherent'' boundary states \refs{\clny,\polcai}.
The boundary conditions \chiralbc\ can
be rewritten in the closed-string channel
as operator conditions 
on these boundary states; for example
$$\eqalign{
J_n &= ~ \bar J_{-n} \qquad\qquad {\rm A\ type} \cr
J_n &= -\bar J_{-n} \qquad\qquad {\rm B\ type} .
}$$
The relative sign change from \atypebc,\btypebc\ can be understood as the result
of a $\pi/2$ rotation on the components of the spin one current;
it means that the A-type states are charged under $(c,c)$ operators and
the B-type under $(c,a)$ operators.

The solution to these conditions \refs{\ishone,\ishtwo} are linear combinations
of the ``Ishibashi states'':
\eqn\ishstate{
	\kket{i}_\Omega=
	\sum_N|i,N\rangle\otimes U\Omega\overline{|i,N\rangle}\ .
}
Here $|i\rangle$ is a highest weight state of the
extended chiral algebra; the sum is over all descendants 
of $|i\rangle$; and $U$ is an
anti-unitary map with
$U\overline{|i,0\rangle}=\overline{|i,0\rangle}^\star$ and 
$U\bar W^{(r)}_n U^\dagger=(-1)^{s_r}\bar W^{(r)}_n$.

Modular invariance requires that
calculations in either channel have the same result. This
gives powerful restrictions on possible boundary states.
In particular one requires that a transition
amplitude between different boundary states
can be written as a sensible open-string
partition function, via a modular transformation.  
For rational CFTs with certain
restrictions, Cardy \cardy\ showed that
the allowed linear combinations of Ishibashi
states \ishstate\ are:
\eqn\cardybs{
	\kket{I}_\Omega=\sum_j B_I^j\kket{j}_\Omega=
	\sum_j {{S_I^j}\over{\sqrt{S_0^j}}} \kket{j}_\Omega\ .}
If $\chi_j$ is a character of the extended chiral algebra,
then $S_i^j$ is the matrix representation of the modular
transformation $\tau\to -1/\tau$.
In this notation capital and lower-case letters denote the
same representation; we use capital letters to denote this
particular linear combination of Ishibashi states.
We may also associate
a bra state to the representation $I^{\vee}$ conjugate to $I$:
\eqn\cardybra{
	{}_{\Omega}~\bbra{I^{\vee}} = 
	\sum_j {}_{\Omega}~\bbra{j} B^j_I\ .
}
These boundary states are in one-to-one correspondence
with open-string boundary conditions which we will label the same way.
Cardy argued that the open-string partition function
was determined by the fusion rule coefficients.
Let worldsheet time and space be labeled by $\tau$ and $\sigma$
respectively; and let the boundary run from $\sigma=0$ to $\sigma=\pi$,
and the boundary conditions be $I^{\vee}$ and $J$, respectively.
Then the number of times that the representation $k$ appears
in the open-string spectrum is precisely the
fusion rule coefficient $N^k_{IJ}$; in other words,
the open-string partition function will be
\eqn\openpart{
	Z_{I^{\vee} J}
		= \sum_k N^k_{IJ} \chi_k \ .
}

\subsec{The Gepner model in the bulk}

Gepner models \refs{\gepone,\geptwo} (see also \greene\ for a quick
review) are exactly solvable CFTs which correspond to Calabi-Yau
compactifications at small radius \wittph.  They are
tensor products of $r$ $\CN=2$ minimal models
together with an orbifold-like projection 
that couples the spin structures and allows
only odd-integer $U(1)$ charge.  We will
review their construction here.  For simplicity
we will discuss theories with $d+r=\even$, where $d$ is
the number of complex, transverse, external 
dimensions in light cone gauge.

Our building blocks are the 
$\CN=2$ minimal models at level $k$;  these are
SCFTs with central charge
$c={3k\over{k+2}}<3$ \refs{\bfk,\zamtwo,\dpyz,\snam}.  
The superconformal primaries are 
labelled by 3 integers, $(l,m,s)$
with
\eqn\standardone{
	0\leq l\leq k;\ \ |m-s|\leq l;\ \ s\in\{-1,0,1\};\ \ 
	l+m+s = 0\ {\rm mod}\  2\ .
}
The integers $l$ and $m$ are familiar from the 
$SU(2)_k$ WZW model and can
be understood from the parafermionic construction of the minimal
models \refs{\zamo,\qiu}. 
$s$ determines the spin structure:
$s=0$ in the NS sector;
and $s=\pm 1$ are the two chiralities in the $R$
sector.\foot{The variable $m$ in \qiu, in sec. 2.1 of \gepone,
and sec. 4 of \geptwo, is what we are calling $m-s$.}
The conformal weights and $U(1)$
charges of these primary fields are:
\eqn\ntwoweights{
\eqalign{
	h^l_{m,s}&={{l(l+2)-m^2}\over{4(k+2)}}+
		{{s^2}\over 8}, \cr
	q^l_{m,s}&={m\over{k+2}}-{s\over 2}\ .
}}

The $\CN=2$ chiral primaries are clearly $(l,\pm l,0)$
in the NS sector.  The related Ramond sector
states $(l,\pm l, \pm 1)$ can be reached by spectral flow.
The minimal models
can also be described by a Landau-Ginzburg model
of a single superfield with superpotential
$X^{k+2}$ \refs{\kastms,\emilalg,
\vafawarner,\emilcatas,\lvw}.  At the conformal
point $X^l = (l,l,0)$ and the Landau-Ginzburg fields
provide a simple representation of the chiral ring.

The $\CN=2$ characters and their modular properties
are described in \refs{\ravyang,\qiumod,\gepone,\geptwo};  
we will follow the notation in \refs{\gepone,\geptwo}.
One extends the $s$ variable to take values in $Z_4$.
The NS characters are labelled by $s=0,2$ and the different values of $s$
denote opposite $Z_2$ fermion number.  The contribution
from the NS primary is in $\chi_{l,m,0}$.  
Similarly, in $R$ sector $s=\pm 1$ denotes contributions
from opposite fermion number: the $s=1$($s=3$)
character includes the contribution from the $s=1$($s=-1$)
Ramond-sector primary.
These characters are actually defined in the range
$l\in\{0,\cdots,k\}$, $m\in \BZ_{2k+4}$ and $s\in\BZ_4$, 
where $l+m+s=\even$. They obey the identification
$\chi^l_{m,s}=\chi^{k-l}_{m+k+2,s+2}$ by which the fields can be
brought into the range \standardone.

Not every $c=9$ tensor product of minimal models will give
a consistent string compactification with 4d spacetime SUSY.
We must find a reasonable GSO projection, and we must
project onto states with odd integer $U(1)$ charges \banksetal.
We must then add ``twisted'' sectors in order to maintain
modular invariance.  The resulting spectrum is most
easily represented by the partition function, for which
we require some notation.  We will tensor $r$ minimal models
at level $k_j$ with the CFT of flat spacetime.  The latter also has a
$\CN=2$ worldsheet SUSY in our case, and we denote the characters by the 
indices $i$.
The vector $\lambda=(l_1,\cdots,l_r)$ gives the $l_j$ quantum numbers
and the vector
$\mu=(m_1,\cdots,m_r;s_1,\cdots,s_r)$, the charges and spin structures. 
Now define $\beta_{j=1,\ldots,r}$ to be the charge
vector with a two at the position of $s_j$, 
and all other entries zero;
and define $\beta_0$ to be the charge vector with all entries one.
The modular invariant partition function in light cone gauge can be
written as \refs{\gepone,\geptwo}:
\eqn\clpartfn{
	Z=\sum_{(i,\bar i),\lambda,\mu}\sum_{b_0,b_j}
	\delta_\beta(-1)^{b_0}\chi_{i,\lambda,\mu}(q)
	\chi_{\bar i,\lambda,\mu+b_0\beta_0+
	\sum_j b_j\beta_j}(\bar q)\ ,
}
Here $\chi_{i,\lambda,\mu}$ is the character for the $r$
minimal models specificed by $\lambda,\mu$ and for the
character of the flat transverse spacetime coordinates
(labelled by $i$).
In the sum, $b_0=0,\cdots,2K-1$, $b_j=0,1$ and $K=\lcm\{2,k_j+2\}$.
$\delta_\beta$ is a Kronecker delta function enforcing both
odd integral $U(1)$ charge and the condition
that all factors of the tensor product have the same spin structure.

The $k$th minimal model has a $\BZ_{k+2}\times \BZ_2$ symmetry \refs{\gepone,\muss}\ 
which acts as:
\eqn\discrete{
	\eqalign{
	g\phi^l_{m,s}&=e^{2\pi i{m\over{k+2}}}\phi^l_{m,s},\cr
	h\phi^l_{m,s}&=(-1)^s\phi^l_{m,s}\ .
}}
With the above projection, all $\BZ_2$ symmetries have
the same action on a given state and are identified. The
remaining $\BZ_2$ symmetry acts only on R states by reversing their
sign.  The $Z_{k+2}$ symmetry is
correlated with the $U(1)$ charge.  In particular,
the diagonal generator $G = \prod_j g_j$ is the identity for
integral $U(1)$ charges.  The Gepner model is
an orbifold theory; the orbifold group $H$ is the group
generated by $G$.  The remaining discrete
symmetry is $\otimes_{i=1}^r Z_{k_r+2}/H$.  For
example, the $(k=3)^5$ model is an orbifold by the
diagonal $Z_5$ of $(Z_5)^{\otimes 5}$\ .

\subsec{Boundary states in the Gepner model}

It is difficult to construct the most general boundary state for the Gepner
model, because the Gepner model is not rational. 
Following \reckone, 
we will consider states which respect the $\CN=2$ world-sheet algebras
of each minimal model factor of the Gepner model separately,
and can be found by Cardy's techniques.
These might be called ``rational boundary states.''
They are labeled according to Cardy's notation by $\alpha=(L_j,M_j,S_j)$
and an automorphism $\Omega$ of the chiral symmetry algebra. 
In our case there are
two choices of $\Omega$ giving either A- or B-type boundary conditions;
$\Omega$ must have the same action on every factor of the
tensor product.

Recknagel and Schomerus \reckone\ proved the modular invariance of A-
and B-type boundary states with internal part:
\eqn\rsstate{
	\kket{\alpha}={1\over\kappa^\Omega_\alpha}\sum_{\lambda,\mu}\delta_{\beta}\delta_{\Omega}
	B^{\lambda,\mu}_\alpha\kket{\lambda,\mu}_\Omega\ .
}
The coefficients are:
\eqn\rscoeff{
	B^{\lambda,\mu}_{\alpha}=
	\prod_{j=1}^r{1\over{\sqrt{\sqrt{2}(k_j+2)}}}
		{{\sin(l_j,L_j)_{k_j}}\over
	{\sqrt{\sin(l_j,0)_{k_j}}}}e^{i\pi{{m_j M_j}\over{k_j+2}}}
	e^{-i\pi{{s_j S_j}\over{2}}}\ ,
}
a result of eq. \cardybs\ for the minimal models and the extra
coefficient $\kappa^\Omega_\alpha$ described in the appendix.
Here
$$ (l,l')_k = \pi \frac{(l+1)(l'+1)}{k+2}\ . $$
$\delta_\Omega$ denotes the constraint that the Ishibashi state
$\kket{\lambda,\mu}_\Omega$ must appear in the closed string
partition function \clpartfn. For A-type boundary states this
is no constraint as the Ishibashi states are already built on
diagonal primary states and $\delta_\beta$ already
enforces that total $U(1)$ charge is integral.
However, the B-type Ishibashi states have opposite $U(1)$ charge
in the holomorphic and antiholomorphic sector,
and these only appear as a consequence of the GSO projection;
so the $\delta_B$ constraint requires
that all the $m_j$ are the same modulo $k_j+2$.  Finally,
an integer normalization constant $C$ has to be included in
$\kappa^\Omega_\alpha$ to get the correct normalization for
the open-string partition function.

It is easy to see from eqs. \rsstate,\rscoeff\
that the action of the $\BZ_{k_j+2}$ ($\BZ_2$)
symmetries is $M_j\rightarrow M_j+2$ ($S_j\rightarrow S_j+2$).
As a result of the $\delta_\beta$ constraint,
the two physically inequivalent choices for $S_j$
are $S=\sum S_j=0,2\mod 4$. The $S_j=\odd$ case
seems to be inconsistent because 
their RR-charges do not fit into a charge 
lattice together with the $S=\even$
states; thus they will violate the charge
quantization conditions\foot{The amplitude between a $S=\odd$ boundary state 
and a $\tilde S=\even$ boundary state also has interchanged roles of R- and
NS-states in the open string sector.}. In the end,
due to the $Z_2$ symmetry, it 
is enough to consider only boundary states
with $S=0$. A boundary state can be written as
$$\eqalign{
	& g_1^{M_1\over 2}\cdots g_r^{M_r\over 2}h^{S\over 2}
	\ket{L_1\cdots L_r}_\Omega :=
	\ket{L_1\cdots L_r;M_1\cdots M_r;S}_\Omega = \cr
	& g_1^{{M_1-L_1}\over 2}\cdots g_r^{{M_r-L_r}\over 2}
		h^{S\over 2}
	\ket{L_1\cdots L_r;M_1'=L_1\cdots M_r'=L_r;S'=0}_\Omega\ .}
$$
For B-type boundary states, the $\delta_\beta$ 
constraint in eq. \rsstate\
implies in addition that the physically inequivalent choices
of $M_j$ can be described by the quantity
$$
M=\sum_j{{K'M_j}\over{k_j+2}}\ ,
$$
where $K'=\lcm\{k_j+2\}$.

We will be interested in counting the number of
moduli for a D-brane state; these
will be the massless bosonic (\ie\ NS) open-string states.
To find their contribution to the open-string partition
function, it is enough to
examine the NS-NS part of a transition amplitude in the
internal dimensions.  The reason is that
the (open-string) NS characters arising from the modular
transformations of the RR part of the transition amplitude
come with an insertion of $(-1)^F$ \refs{\ravyang,\qiumod}.
With this in mind, a calculation similar to that in \reckone\ leads
to\foot{
$N^l_{L,\tilde L}$ are the $SU(2)_k$ fusion rule coefficients \zffusion: 
they are one if $|L-\tilde L|\le l\le\min\{L+\tilde L,2k-L-\tilde L\}$
and $l+L+\tilde L=\even$, and zero otherwise; note that
our indices thus differ from those in \zffusion\ by a factor of two.}
\eqn\apart{
	Z_{\alpha\tilde\alpha}^A(q)={1\over C}\sum^{NS}_{\lambda',\mu'}
	\sum_{\nu_0=0}^{K-1}\prod_{j=1}^rN^{l_j'}_{L_j,\tilde L_j}
	\delta^{(2k_j+4)}_{2\nu_0+M_j-\tilde
	M_j+m_j'}\chi^{\lambda'}_{\mu'}(q)\ ,
}
and
\eqn\bpart{
	Z_{\alpha\tilde\alpha}^B(q)={1\over C}
	\sum^{NS}_{\lambda',\mu'}
	\delta^{(K')}_{{{M-\tilde M}\over 2}+\sum{K'\over {2k_j+4}}m_j'}
	\prod_{j=1}^rN^{l_j'}_{L_j,\tilde L_j}
	\chi^{\lambda'}_{\mu'}(q)\ .
}
(Here $\delta_x^{(n)}$ is one when $x=0 \mod n$ and zero otherwise.)
This shows that only a $U(1)$ projection and the $SU(2)_k$ fusion rule
coefficients constrain the open string spectrum of B-type boundary
states; these states are much richer as a consequence.

The condition that two D-brane boundary states
$\kket{\alpha}$ and $\kket{\tilde\alpha}$, with the same external
part, preserve the same supersymmetries is \reckone:
\eqn\pressusy{
	Q(\alpha-\tilde\alpha):=
	-\frac{S-\tilde S}{2}
	+\sum\limits_{j=1}^r\frac{M_j-\tilde M_j}{k_j+2}
	=\even\ .
}

To explore the charge lattice of the boundary states, and
to find the geometric interpretation of given boundary
states, we wish to calculate the intersection \dszip\braneinter\
of our branes.  The CFT quantity which computes this is
$I_\Omega=\tr_R(-1)^F$ in the
open string sector \fiol. The best way to do this is to start in the
closed string sector and to do a modular transformation to the open
string sector. In the closed string sector this trace corresponds to
the amplitude between the RR parts of the boundary states with a
$(-1)^{F_L}$ inserted. The calculation is done in the Appendix
and the result for A-type boundary states is:
\eqn\atrace{
	I_A={1\over C}
	(-1)^{{{S-\tilde S}\over 2}}
		\sum_{\nu_0=0}^{K-1}\prod_{j=1}^r
	N^{2\nu_0+M_j-\tilde M_j}_{L_j,\tilde L_j}\ .
}
For B-type boundary states,
\eqn\btrace{
	I_B={1\over C}(-1)^{{{S-\tilde S}\over 2}}\sum_{m_j'}
	\delta^{(K')}_{{{M-\tilde M}\over 2}+
		\sum{K'\over{2k_j+4}}(m_j'+1)}
	\prod_{j=1}^r N^{m_j'-1}_{L_j,\tilde L_j}\ .
}
The intersection matrix depends only on
the differences $M-\tilde M$ as was required by the discrete symmetry. 
We also see that the $\BZ_2$ action $S\rightarrow S+2$ changes
the orientation of a brane. 

In the next section we will rewrite these formulas in a more compact notation and
use them to identify the charges of the boundary states.

\subsec{D-branes on K3 and the Mukai formula}

For compactifications with $\CN=4$ worldsheet supersymmetry, the
index in the Ramond sector is directly related 
to the number of marginal operators in the NS sector.
We now use this to give a CFT proof of Mukai's formula \refs{\mukai,\bpsalg}\
for the dimension of the moduli space of
$1/2$-BPS D-brane states.

K3 compactifications are geometric
throughout their moduli space \seibergmoduli.
The BPS D-brane states in these compactifications
are described by coherent semistable sheaves $E$ \bpsalg\
which can be labelled by the Mukai 
vector \refs{\mukai,\bpsalg}.  In terms of the rank $r$ and Chern classes
$c_i$ of $E$, this is
\eqn\mukaivector{
\eqalign{
	v(E) &= \left(r, c_1, \half c_1^2-c_2+r\right)\cr
	&\in H^0(M,\IZ)\oplus H^2(X,\IZ) \oplus H^4(M,\IZ)
}}
There is a natural inner product
on the space of Mukai vectors:
\eqn\mukaiip{
	\langle (r,s,\ell),(r',s',\ell') \rangle =
		s\cdot s' - r\ell' - \ell r'
}
where $s\cdot s'$ is defined by the natural
intersection pairing of 2-cycles on $M$.
In fact this is just (minus) the intersection form \braneinter.

Mukai's theorem~\mukai\ states that the complex
dimension of the moduli space of an irreducible
coherent sheaf $E$ is:
\eqn\mukaitheorem{
	{\rm dimension} = \langle v(E),v(E) \rangle + 2.
}

We now argue that this follows from the relation
\eqn\indextoip{
	\tr_{a,a} (-1)^F = \langle v(E_a),v(E_a) \rangle
}
and general properties of supersymmetry.
First, only two $d=2$, $\CN=4$ representations have nonvanishing
Witten indices \refs{\egta,\eoty}.  We list them below together with
the NS weights related by spectral flow:
\eqn\nfourreps{
\eqalign{
	&{\rm identity\ rep.}:\ (h=0,\ell = 0)_{{\rm NS}} 
	\longrightarrow (h=1/4, \ell =  1/2)_{{\rm R}} 
	\ \ \ \ \ {\rm tr} (-1)^F = -2 \cr
	&{\rm ``massless''\ rep.}:\ 
	(h=1/2, \ell = 1/2)_{{\rm NS}} \longrightarrow
	(h=1/4, \ell = 0)_{{\rm R}} \ \ \ \ \ {\rm tr} (-1)^F = 1\ ,
}}
where $\ell$ is the $SU(2)_R$ isospin.  
The identity representations lead to world-volume $d=6$,
$\CN=1$ (or $d=4$, $\CN=2$) gauge multiplets,
while the massless representations lead to world-volume 
half-hypermultiplets, so there will
be one complex scalar in the open-string sector
for each massless multiplet.

Let there be $N_g$ identity and $N_m$ massless multiplets; then
the Witten index is
\eqn\wittenindex{
	\tr(-1)^F = N_m  - 2 N_g .
}
Using \indextoip\ we find that \mukaitheorem\ will be true if
the world-volume theory has a (Higgs branch) moduli
space of complex dimension $N_m - 2 N_g + 2$.
This moduli space is essentially determined by the $d=6$, $\CN=1$ 
world-volume supersymmetry: it is
the hyperk\"ahler quotient of the configuration space by the subgroup
$G$ of the gauge group which acts non-trivially on the
hypermultiplets.  The resulting space has complex dimension
$N_m - 2 \dim G$.

Now, any brane configuration will have an overall $U(1)$ acting trivially
whose partners in the vector multiplet are the center of mass position
of the brane; if more $U(1)$s act trivially we will have more center
of mass moduli, so such a configuration must correspond to a reducible
bundle.  Therefore $\dim G = N_g-1$ for an irreducible bundle and
we have proven \mukaitheorem.

\subsec{Generalizations}

Mukai's theorem used the Hirzebruch-Riemann-Roch formula
together with special properties of K3 surfaces; these properties
allowed one to extract the dimension of the moduli space of a bundle
directly from the holomorphic Euler characteristic.
We have a similar statement for CY threefolds if we keep track of
both chiralities separately.
The self-intersection
number of a brane on a threefold is of course zero, 
but we can get non-trivial 
statements if we consider the intersection 
of two different branes.

For example, consider the index of the Dirac operator on the bundle $E$.
Since the world-volume is K\"ahler this is
$$
\ind \Dslash = \sum_{i=0}^3 (-1)^i \dim H^i(M,E) = \chi(E)
$$
which is the holomorphic Euler characteristic.
By the Hirzebruch-Riemann-Roch formula,
\eqn\hrr{
	\chi(E) = \int_M \ch(E) \td(TM)\ .
}
Here 
$$
	\ch(E) = r + c_1 (E) + 
	\half\left( c_1^2(E) - 2 c_2(E)\right)
	+ \frac{1}{6}
	\left(c_1^3(E) -3 c_1(E)c_2(E) + 3c_3(E)\right)+\ldots\ ,
$$	
and
$$
	\td(TM) = 1 + \frac{c_2(TM)}{12} + \ldots
	= 1 - \frac{p_1(TM)}{24} +
		+ \ldots 
$$
Thus on a threefold, $\td(M)=\ahat(TM)$, and combining eqs. \hrr\ 
and \braneinter, we find:
\eqn\indextointer{
	\ind \Dslash = \vev{D6,D(E)} = \tr_{D6,D(E)}(-1)^F\ ,
}
where $D(E)$ is the D-brane representation or generalized
Mukai vector for $E$.

On the other hand, the Ramond ground states which contribute to the 
open string index are exactly the fermion zero modes which contribute
to the index of $\Dslash$.  In the type I case where $E$ is a gauge
bundle with vevs entirely in an $SU(3)$ subgroup and with
the gauge connection equal to the spin connection,
$c_1(E)=0$; this gives a brane picture of the
standard result
$$ 
	N_g = \hbox{(\# of generations)} = 
	\int_M \frac{c_3}{2}
$$
for this case.  If we are interested not in the
bulk gauge theory on $9$-branes in type I but in a gauge
theory on a brane $B$ intersecting another brane $A$, 
the generalization is that
the number of generations (with respect to the $B$ gauge group) associated
with the brane $A$ is the intersection form $\vev{A,B}$.
For B-type branes this follows from eq. \braneinter\ and the 
Hirzebruch-Riemann-Roch theorem
for the bundle $E(A)^* \otimes E(B)$; for A-type branes each intersection contributes
a chiral multiplet with chirality given by the sign of the intersection \bdtwo.

\newsec{Discussion of the $3^5$ model}

Let us apply these results to the example to model $(k=3)^5$,
the Gepner point in the moduli space of the quintic. 
We will consider boundary states labelled by
$L_j~\in~\{0,1\}$,  $0\le~M_j~<~(2k+4)~=~10$, and $S~=~0$.
Let the $\BZ_5^4$ symmetry be generated by the operators
$g_j$ taking $M_j\rightarrow M_j+2$, and satisfying
$g_1\cdots g_5=1$.  Note that $g_j^{1/2}$ 
which takes $M_j\rightarrow M_j+1$
is well-defined for these states
(using the identifications on $LMS$, 
it relates branes to antibranes).

We will be particularly interested in computing the
intersection forms \atrace\ and \btrace,
as we will be able to use them to
extract the charges and open string spectrum for a given brane.
The main advantage of considering these quantities over the charges themselves
is that they are canonically normalized, as already noted in \dlp.

We can consider the intersection form as a matrix $I$ acting on the space of
boundary states; since it commutes with 
$\BZ_5^4$ it can be written as a function
of the generators $g_i$.  The main content of formulae \atrace\ and
\btrace\ is contained in the $SU(2)$ fusion rule coefficients.
In these equations the labels $M_j$, $\widetilde{M}_j$
can be thought of as indices of a matrix acting on the states.
The particular fusion coefficients we will need are:\foot{The
coefficients for $m>l$ are defined in the Appendix.} 
\eqn\fusionc{\eqalign{
N_{00}^{M_j-\widetilde M_j}&\rightarrow (1-g_j^4),\cr
N_{01}^{M_j-\widetilde M_j}&\rightarrow g_j^\half(1-g_j^3) 
  = N_{00}\ g_j^\half(1+g_j^4); \cr
N_{11}^{M_j-\widetilde M_j}&\rightarrow (1+g_j-g_j^3-g_j^4)
  = N_{01}\ g_j^\half(1+g_j^4).\cr
}}
These various fusion matrices are related by successive multiplication
with $g_j^\half(1+g_j^4)$, so we can express the RR charges
of all our boundary states in 
terms of those for $Q(\ket{00000}_\Omega)$.

By eq. \pressusy\ there are two cases of pairs of branes preserving a common susy.
If the total $\Delta L$ is even (so integral powers of $g$ appear),
a pair with $\Delta M =\Delta S=0$ (brane and brane) will preserve susy.
If the total $\Delta L$ is odd (powers $g^{5k+5/2}$ appear),
a pair with $\Delta M=5$ and $\Delta S=2$ (brane and anti-brane) will preserve susy.

In the case that the two D-branes are both A-type or B-type, 
the massless open string spectrum can also 
be expressed in terms of the fusion
coefficients. It is easy to 
see from \apart\ and \bpart\ that if the two boundary
states are the same, there is exactly one vacuum and one spectral flow
operator in the open string channel; if they are not the same,
neither state  propagates. This means
that the unbroken worldvolume gauge group
is (the center-of-mass) $U(1)$, and
the brane can be viewed as a single object 
(a priori, it still might be a bound state).

The SUSY-preserving moduli of the D-branes are constructed
from chiral vertex operators.  The Witten index
counts these operators albeit with a sign depending on
their chirality.  In our explicit CFT calculation we can
remove this sign by hand, and thus the total number of chiral
fields can be calculated using \atrace\ and \btrace\ with
the fusion matrices replaced by their absolute values.\foot{
In other words, we define $N_{L{\tilde L}}^m = + N_{L{\tilde L}}^{-m-2}$,
rather than the opposite sign in the Appendix.}
We can again write this ``modified'' matrix as
a polynomial $P_\Omega(g_j)$ in the shift matrices $g_j$. 
For example, the matrix for boundary states
$\ket{11111}_B$ is:
\eqn\specmat{
P_B(g)=(1+g+g^3+g^4)^5\ .
}
If spacetime supersymmetry is preserved, the chiral fields have 
integer $U(1)$ charges,
and are related to antichiral fields by spectral flow.
In particular charge-$2$ 
chiral fields in $Z_{\alpha\tilde\alpha}^\Omega$,
are related to charge-$-1$ antichiral fields in
$Z_{\alpha\tilde\alpha}^\Omega$; the latter
are the hermitian conjugate of charge-$1$ chiral
fields in $Z_{\tilde\alpha\alpha}^\Omega$.
Thus $\sum_k m_k$ in the open-string channel
will be a multiple of 5 for marginal, chiral vertex
operators.  Examination of the fusion coefficients in \atrace\
and \btrace\ reveals that
the number of massless chiral 
superfields is given by counting terms in 
$$
\half(P_\Omega(g_j)-2)
$$ 
with the total power of $g$ being a multiple of $5\over 2$. 

Applying these statements to
eq. \specmat\ shows that the D-brane described by
$\ket{11111}_B$ has 
$101$ marginal operators.  This particular
case can also be worked out by checking that the fusion 
rules lead to all possible
$L$ values, so for every operator in the $(c,c)$ 
ring of the model there is a corresponding
chiral open string operator.

\subsec{A boundary states}

The intersection matrix \atrace\ for the A-type 
boundary states with $L_j=0$ is
\eqn\Ainter{
I_A=(1-g_1^4)(1-g_2^4)(1-g_3^4)(1-g_4^4)(1-g_1g_2g_3g_4).}
To determine the rank of the intersection matrix we can count
the number of nonzero eigenvalues. The $g_j$ can be diagonalized as
$g_j=\diag(1,e^{{{2\pi i}\over 5}},e^{{{4\pi i}\over 5}},
e^{{{6\pi i}\over 5}},e^{{{8\pi i}\over 5}})$. Zero eigenvalues
appear if a $g_j=1$ or if $g_1g_2g_3g_4=1$. The combinatorics
leads to $204$ nonzero eigenvalues, which is the number of independent
3-cycles on the quintic. 
Thus, the $L_j=0$ states provide a basis for the 
charge lattice.  So far as we can
tell they do not provide an integral basis of the charge lattice.
Furthermore, the charges of the other A-type
Gepner boundary states can be obtained from these by successive
multiplication by $g_j^\half(1+g_j^4)$; for example,
$Q(g_1^\half\ket{10000}_A)=Q(\ket{00000}_A)+Q(g_1\ket{00000}_A)$,
so these are even farther from an integral basis.

The intersection matrix for the $\ket{11111}_A$ states,
$$
\prod_{i=1}^5 (1+g_i-g_i^3-g_i^4)
$$
coincides with the intersection matrix 
\intersectguess\ for the three-cycles $\Im\omega_jz_j=0$,
and thus we identify these states with the $\BR P^3$'s.

This leads to a potential contradiction with the large volume limit in that
the $L=1$ states have one marginal operator, while the $\BR P^3$'s do
not.  Although it might be that this is indeed a contradiction,
from what we know at present
an equally likely resolution is that the $L=1$ marginal operator
is not strictly marginal; in other words the world-volume theory
has a superpotential for the corresponding field $\psi$, perhaps of the
form
$$W = \psi^3 + \psi \phi$$
where $\phi$ is the K\"ahler modulus ($\psi^5$ in the notation of
section 3).  Such a superpotential has two ground states and would also
fit the fact that the $\BR P^3$ has a $\BZ_2$ Wilson line in the large
volume limit.\foot{
(Note added in v2):
Actually, the two choices of Wilson line are topologically distinct bundles
so they would not be continuously connected in the large volume limit.
This would suggest that the potential should have a unique minimum.
On the other hand, it can be shown that any simply connected six-dimensional
manifold $X$ with $H^*(X)$ torsion-free (such as the quintic CY) has $K(X)\cong H^*(X)$,
and thus the K theory class distinguishing the two bundles becomes trivial when lifted
to the CY.
(We thank D. Freed and J. Morgan for explaining this to us.)
Thus there is no candidate for a space-time topological charge which could
distinguish the two D-branes,
and it is not ruled out that
transitions between the two choices of bundle are possible in the
full string theory.}

\subsec{B boundary states}

As we discussed in the previous section,
the B-type boundary
states at fixed $L_j$ are described 
by the single integer, $M=\sum M_j$ and the $g_j$ 
for different $j$ are
identified. The intersection matrix \btrace\ 
for $L=0$ states can be written as: 
\eqn\Binter{
I_B = (1-g^{-1})^5 = 5g - 10g^2 + 10g^3 - 5g^4.
}
We want to describe these boundary
states in the Gepner basis.
The Gepner intersection form \gepint\ in
the same notation is:
\eqn\gepintg{
I_g = -g + 3g^2 - 3g^3 + g^4\ .
}
A linear change of basis preserving the action of $\BZ_5$ can be written
as a polynomial in the operator $g$ as well and a transformation of the
form $I\rightarrow mIm^t$ will be $I\rightarrow I m(g)m(g^{-1})$.
The relation
$$
I_B = (1-g)(1-g^{-1})I_g
$$
provides this change of basis.

The results of section 3 allow us to write these
charges in the large volume basis. The Gepner charge vector
$Q_G$ is related to the large volume charge vector $Q$ as
$$
Q = Q_G M^{-1}\ .
$$
Thus 
$Q_G=\left(\matrix{0& 1& -1& 0}\right)$ becomes 
$Q=\left(\matrix{-1& 0& 0& 0}\right)$ which is a pure 
(anti)six-brane.
The other charges can be found by acting with the operator $A_L$.

One can now compute the charges for the $L\ne 0$ branes by 
using the multiplicative relation in \fusionc. For example, we have
$$
Q(g^{5\over 2}h\ket{10000}_B) = -Q(g^{2}\ket{00000}_B) - Q(g^{3}\ket{00000}_B).
$$
Starting with $M=0$ and successively applying this operation
produces a subset of branes which preserve the same supersymmetry.  
This can be checked by computing the central charges using the
periods at the Gepner point, which are simply the fifth roots of unity.
Thus the central charge for the $L$'th brane in this series is
$$Z(L) = (2 \cos {\pi\over 5})^L Z(0).$$

The charges in the
Gepner basis charges can written in large volume basis
viq eq. \geptobig.
Tabulating these results and the numbers 
of marginal operators, we have
(for the $Z_5$ representatives related to the six-brane)
\eqn\branelist{\matrix{
L\qquad&Q_6&Q_4&Q_2&Q_0\qquad&\dim \cr
00000& -1& 0& 0& 0& 0 \cr
10000& 2& 0& 5& 0& 4 \cr
11000& 1& 0& 5& 0& 11 \cr
11100& 3& 0& 10& 0& 24 \cr
11110& 4& 0& 15& 0& 50 \cr
11111& 7& 0& 25& 0& 101 \cr
}}
The simple pattern 
$Q_{L+1}=Q_{L}+Q_{L-1}$
follows from the identity 
$(-g^2-g^3)^2 = 1-g^2-g^3$.

It is also easy to compute the number of marginal operators between
pairs of distinct boundary states.  For example, $\ket{00000}_B$
and $\ket{(1\ldots)^L(0\ldots)}$ have (for $1\le L\le 5$)
$4, 3, 3, 4$ and $1$ (respectively) marginal operators.
Each corresponds to a chiral superfield of charge $(1,-1)$ and
its charge conjugate (since the mutual intersection numbers are zero,
none of these pairs has chiral spectra).  
The number of operators between two branes of
higher $L$ of course depends on which $L_i$ are non-zero.

\subsec{Comparison with geometrical results}

To what extent can we compare these results with the geometrical branes and
bundles we discussed in section 2?
The only clear match is the six-brane which indeed has no moduli
as expected.

Our states can plausibly be identified with vector bundles
since they obey the stability condition $c_2 > 0$.  
We were not able
to identify any of them
with the explicit constructions we mentioned in
section 2.
This may just reflect our lack of knowledge of vector
bundles on the quintic; thus 
we might regard our results as predictions
of the existence of new vector bundles.
We should note that the numbers of marginal operators we obtained
are only upper bounds for the dimension of the moduli space as in general
these theories will have potentials.

The problematic objects are 
the $\ket{11000}_B$ branes as an
object with these charges cannot 
be a classical line
bundle.  For reasons explained 
in section 2 we do not believe it is a quantum
bound state either, since we have found it at string tree level.
There is a piece of evidence that it is
some sort of bound state of the six-brane with the two-brane \hyperpl:
namely, they come in the same multiplet of the discrete symmetries.
Like all B branes, the $\ket{11000}_B$ branes are invariant under
$Z_5^4$, while $S_5$ acts by permuting the $L_i$ labels.
The two-brane construction \hyperpl\ also picks out two of the five
coordinates and thus comes in the same multiplet.
This identification creates a puzzle 
opposite to the one we faced for
the $\BR\BP^3$'s: the geometric object appears to 
have more moduli ($12$) than 
the boundary state.  Such a mismatch 
could not be fixed by a superpotential.
On the other hand, it could be that the 
(unknown) mechanism which binds the
two-brane to the six-brane removes moduli, 
so this is not a clear disagreement.

One candidate for such a bound state is the instanton in noncommutative
$U(1)$ gauge theory \nekschw.  Again by analogy with flat space,
(since noncommutative gauge theory
has not been formulated on curved spaces, this
is all we can say), at generic values of $B$ we might expect
the $D6$-brane gauge theory to be 
noncommutative \refs{\cds,\doughull,\ahbs}.  
The center-of-mass position of the instanton would then (presumably)
give the moduli of a two-brane and provide at least
some of the moduli we observe.
A potential problem with this idea is that we can continue to $B=0$ in the large
volume limit, and there is no sign that this bound state is unstable there.

One may ask why the D0-brane does not appear on our list.
One possible explanation is that the path from the
large volume limit to the Gepner point crosses 
a line of marginal stability, and the D0 does not exist at the Gepner
point.  To test this we found the periods for all the branes in
\branelist\ by numerically
integrating the Picard-Fuchs equations along the negative
real $\psi$ axis.  We found that the D0 is lighter than any brane from
the list along the whole trajectory, so we have no evidence for instability.
Our favored explanation is simply that all of the B branes by construction
are invariant under the $\BZ_5^4$ discrete symmetry, while any
location we might pick for the D0 would break some of this symmetry.
Thus, even if the D0 exists at the Gepner point, it cannot be
a rational boundary state, at least in this model.

\newsec{Superpotential and topological sigma models}

The calculations of the previous section describe
the field content of the D-brane world-volumes,
but not their dynamics.  
The primary question in this regard is to find the
world-volume potential and true moduli spaces for the brane theories.
In CFT language, the marginal boundary operators
operators we found might not be strictly marginal.

$\CN=1$, $d=4$ supersymmetry tells us 
that the world-volume potential
will be a sum of F-terms and Fayet-Iliopoulos D-terms.
The D-terms are simply
determined by the gauge group and charges of the matter fields.  In the
case of a single brane or $N$ identical branes we have checked 
in the models we are studying that the
gauge group is $U(N)$ with all matter uncharged under the diagonal
$U(1)$, so there is no possibility for a D-term.
More generally we must consider such terms,
for example in the
case of D0-branes near orbifold points.

However, we may expect a non-vanishing superpotential,
in general constrained only by holomorphy and the
symmetries of the problem.  These
conditions are often stronger than they might appear,
but in general the superpotential must be found by explicit computation.
It should eventually be possible to do
exact calculations at the Gepner point,
as we will discuss in the next section.  
In this section we will
try to make some general statements about the
superpotential in these models by showing that they
can be calculated as amplitudes in
some topologically twisted version of the open string theory.
In particular we will use this fact to describe the cubic
term in the superpotential, and to discuss to
what extent the superpotential couples to the background
CY geometry.

\subsec{Known examples of brane superpotentials}

In order to motivate the search for
superpotentials in these theories we will
start with a few examples where we know they arise.
The most obvious example is $N$ D3-branes in flat space;
one may write the $N=4$ Lagrangian in $N=1$ notation
so that there are 3 adjoint complex scalar
fields $Z^{i=1,2,3} = Z^i_a t^a$ with the superpotential
$\Tr Z^1[Z^2,Z^3]$ (here $t^a$ are adjoint matrices for $U(N)$).  
Of course this vanishes for $N=1$ but not for $N>1$.

A plausible generalization of this to weak curvature (still
preserving $N=1$ world-volume SUSY) is a function
$W$ written as a single trace of the adjoint chiral superfields and
with the property that
\eqn\weakcurve{
	\frac{\delta}{\delta Z^i_a} 
	\frac{\delta}{\delta Z^j_b}
	\frac{\delta}{\delta Z^k_c} W = f^{abc} \Omega_{ijk}(z)
}
for variations around the diagonal vevs $Z^i = z^i{\bf 1}$.
The assumption of this form of the superpotential
is a fairly weak constraint; see \dko\ for analysis along these
lines.  In the case of a large Calabi-Yau threefold, we will find
below that this assumption is correct, and that $\Omega$
is the holomorphic $(3,0)$ form of the threefold.

A well-studied genuinely stringy example is that of D-branes
near orbifold singularities, or near resolved orbifolds
with string-scale curvature.
In these examples a ``single'' brane (in the orbifold limit
they are described by Chan-Paton factors in the regular 
representation of the orbifold group) can have a superpotential, which
furthermore can have non-trivial dependence on the closed string moduli.
A ``single'' brane is described via Chan-Paton factors transforming
in the regular representation of the orbifold group \quivers.
The superpotential takes the general form
\eqn\orbifsp{
W = \Tr Z^1[Z^2,Z^3]|_{proj} + \zeta_i \Tr Z_i \ .
}
The spectrum of these models is obtained as a subset of the $\CN=4$ SYM
spectrum \refs{\quivers,\dgm,\ksorb}, 
and the notation ``$W|_{proj}$'' indicates that the $\CN=4$
superpotential is simply restricted to this subset. 
The $\zeta_i$ are closed string moduli.

This intrinsically stringy
background illustrates the important lesson that by 
varying both closed and open string moduli, 
it is possible to bring down new
massless open string states invisible in the weakly-curved
geometric limit described above.
For example, the $\BC^2/\BZ_2$ model of a single brane has $U(1)$ gauge
symmetry generically; but when both closed- and open-string moduli
are tuned to the orbifold point, 
the gauge symmetry is enhanced to $U(1)^2$.  Furthermore
a new branch of moduli space meets this point, where the
single brane breaks up into branes wrapping the
shrunken cycles of the orbifold \refs{\quivers,\fracbranes}.
This new branch is a transition from Higgs to Coulomb branch and
as usual in supersymmetric gauge theory it is almost impossible to predict
such transitions starting from the Higgs branch.
In the present context we see that we should be wary of arguments
that rely on the distinction between configurations involving ``one'' or
``several'' branes, or equivalently ``one'' or ``several'' distinct
world-sheet boundary conditions, as they can be continuously connected.
Other examples where this distinction is
questionable are a small instanton leaving a
D$p$-brane as a D$p-4$-brane, or an intersection of $2$-branes as described
by \callanmalda.

Another point we will return to is that the closed string moduli
$\zeta_i$ which appear in the superpotential in this example
are complex structure moduli.  Of course orbifold 
resolution also depends on K\"ahler moduli,
but these enter in the Fayet-Iliopoulos D-terms.

Our final example is the superpotential on a wrapped
two-brane.  Recall that a supersymmetric theory
arises when we wrap the 2-brane on a holomorphic
cycle.  The massless fields correspond to
infinitesimal deformations of this cycle into
a cycle close by in the D-brane moduli space.  
Witten \mqcd\ has argued that an M-theory two-brane (or 5-brane)
wrapped around a two-cycle $\Sigma$ has the superpotential
\eqn\trivtwo{
W(\Sigma) = \int_B \Omega
}
when $\Sigma$ is homologically trivial,
where $B$ is a three-manifold bounded by $\Sigma$.
Indeed, this is a holomorphic functional of the
embedding coordinates, which is stationary
by holomorphic curves.  When $\Sigma$ is
in a nontrivial homology class, the superpotential
is defined up to an additive constant as:
\eqn\gentwo{
	W(\Sigma) - W(\Sigma_0)
	= \int_B \Omega}
where $\Sigma_0$ is an arbitrarily chosen referent
holomorphic 2-cycle in the same homology class as $\Sigma$,
and $B$ has boundary $\Sigma - \Sigma_0$.
Here the additive constant depends both on $\Sigma_0$
and the homology class of $B$.  For purely 
classical, geometric deformations
these formulae should hold for D2-branes;
there may also be terms arising from the
gauge fields on the D2-brane worldvolume.

Before discussing the computation in general, we note that
in all of our examples, which are of B-type branes,
the superpotential depends on closed string moduli only through
the complex structure, not through the K\"ahler structure.
Could it be that this is a general statement?

We can see some potential problems with the statement by considering
the other branes on the list.  First of all, we need to describe not
just the embedding but also the gauge bundle on the branes.  To the
extent that this is determined by a choice of a holomorphic vector
bundle, this will fit into the same class of problems
depending only on complex structure data.
However, one might object that general four- and six-brane
configurations involve a gauge bundle with $c_2\ne 0$ and such
a holomorphic bundle will correspond to a solution of Yang-Mills
only if it is stable, a condition which depends on the K\"ahler class.
This condition indeed should enter into the potential but
as we discussed in section 2, it is more natural to expect that it appears
as a D-term, which would not contradict the decoupling statement.

Mirror symmetry for boundary states
\ooy\ and the statement we are considering would together
imply that the superpotential for
A-type branes depends on the K\"ahler structure of
the background, but not the complex structure.
But any formula for the potential on the brane analogous
to \trivtwo\ will necessarily involve both structures, since the
special Lagrangian condition 
cannot be stated without bringing in $\Omega$.

This situation can only be compatible with our decoupling statement
if the terms involving $\Omega$ are D-terms, an assertion not contradicted by
any existing results.\foot{Related questions are being considered by G. Tian.}
One might object that this possibility
would require charged matter under a gauge group which is not immediately
apparent, but the small instanton and orbifold examples show that such gauge groups
can be broken and become invisible in the large volume limit.
This would lead to the further interesting possibility that, at special
moduli points in the space of a ``single'' $3$-brane, enhanced gauge symmetry
could appear.  The simplest way this could happen is for the brane to split in
two at a self-intersection, leading to $U(1)^2$ gauge symmetry.

We conclude that we have several examples in which decoupling (before taking
stringy corrections into account) is clear, and no examples in which it is clearly false.
Thus we will consider this decoupling statement further below.

\subsec{CFT computation of superpotential}

Given the above examples, we have good reason to believe
that the Gepner model boundary states we have
constructed correspond to D-branes with worldvolume
superpotentials.  We want to know how to calculate
these in the models at hand.

We are interested in BPS D-branes in $\CN=2$ compactifications
of type II string theory; these lead to $\CN=1$
worldvolume theories.  Thus the open- (closed-) string
sectors will have $\CN=2$ ($\CN=(2,2)$) worldsheet
supersymmetry \refs{\banksetal,\banksdixon}.
To fix notation we write out the OPE algebra
for the holomorphic piece:
\eqn\ntwosusy{
\eqalign{
	T(z)T(w) &\sim \frac{\half c}{(z-w)^4} + 
		\frac{2 T(w)}{(z-w)^2} + 
		\frac{\partial T(w)}{(z-w)} + \cdots \cr
	T(z) G^{\pm}(w) &\sim \frac{\frac{3}{2} G^\pm(w)}{(z-w)^2}
		+ \frac{\partial G^\pm(w)}{(z-w)} + \cdots\cr
	G^\pm(z)G^\pm(w) &\sim \cdots\cr
	G^+(z)G^-(w) &\sim \frac{\frac{c}{6}}{(z-w)^3}
		+ \frac{\half J(w)}{(z-w)^2} +
		\half \frac{T(w) + \half \partial J(w)}{(z-w)}
		+ \cdots \cr
	J(z) G^\pm(w) &\sim \pm \frac{G^\pm(w)}{(z-w)} + \cdots \cr
	J(z) J(w) &\sim \frac{\frac{c}{3}}{(z-w)^2} + \cdots\ .
}}
For compactifications with $c=9$ 
$J(z)$ can be constructed from the internal part of the
spacetime SUSY current \banksetal.  It can be written in
terms of a single boson $H$,
\eqn\abelcur{
	J(z) = i\sqrt{3} \partial H
}
and operators with charge $q$ under this $U(1)$ R-symmetry
can be written as:
\eqn\chargedop{
	\CO_q = e^{i(qH/\sqrt{3})} \CO_0\ .
}

The spacetime SUSY currents can be constructed from the
macroscopic spin fields and the internal $U(1)$
current algebra.  In the $(\pm 1/2)$ picture,
the currents can be written as:
\eqn\stsusy{
	Q_{\pm \half,\alpha}(z) = e^{\pm \phi/2} S_\alpha
		\Sigma^\pm
}
where
\eqn\halfspectral{
	\Sigma^\pm = e^{\pm i \sqrt{3} H/2}
}
is the spectral flow operator of the $\CN=2$
worldsheet algebra, mapping the NS sector
to the R sector and vice-versa. $\phi$ is the bosonized
superconformal ghost.

On the 4d noncompact worldvolume, we can have
massless chiral superfields $\Phi^{i}_{IJ}$ 
with scalar components
$\phi^i$, fermionic components $\psi^i$ and auxiliary components
$F^i$.  $i$ will label the (complex) internal moduli of the D-brane
configuration on the CY threefold $M$; these
moduli correspond to
marginal boundary operators of the internal CFT.
$(IJ)$ label gauge indices,
which are described by Chan-Paton factors on the worldsheet.
These could be adjoint indices if there are
coincident branes, or bifundamental indices
if there are several types of (possibly intersecting)
branes.
More abstractly, the off-diagonal terms are
boundary condition-changing
operators \cardy\ and the diagonal terms
are boundary condition-preserving
operators.  

The superpotential can
be written via a holomorphic function $W(\Phi)$ and
it contributes the following terms to the Lagrangian:
\eqn\Fterms{
\eqalign{
	\int d^4 x & \left( d^2 \theta\ \tr W(\Phi)
		+ {\rm h.c.}\ \right)\cr
	& = \int d^4 x \left( \partial_i \partial_j
		\frac{\partial}{\partial\phi^i_{IJ}} 
		W(\phi) F^i_{IJ} - 
		\frac{\partial}{\partial\phi^i_{IJ}}
		\frac{\partial}{\partial\phi^j_{KL}}
		W(\phi) \psi^i_{IJ} \psi^j_{KL}
		+ {\rm h.c.} \right)
}}
where we use the superfield conventions in \wessbagger.
We are interested in small fluctuations about a reference
D-brane state, so we expand $W(\phi)$ in
a Taylor series in $\phi$.  All of the
terms of interest in Eq. \Fterms\ will be of the form
$$ \tr \left[ w_{i_1 i_2 i_3 \ldots i_n}\left( 
	F^{i_1} \phi^{i_2}\ldots \phi^{i_n} - \half
	\psi^{i_1}\psi^{i_2}
	\phi^{i_3}\ldots\phi^{i_n}\right)\right]\ .
$$
The coefficients $w$ may also depend on the closed-string
background.  We will examine small
fluctuations about some reference background
so that we can sensibly expand $w$ in a Taylor series
in fluctuations of the closed-string background.

The worldvolume fermions are represented in the open string theory
by  dimension $1$ boundary Ramond
vertex operators constructed from spin fields.
In the $(-1/2)$ picture they can be written
as:
\eqn\rvertex{
	V^{(-1/2)}_{R,IJ}
	= \zeta^\alpha_{i,a} e^{-\phi/2} S_\alpha \Sigma^i t^a_{IJ}\ ,
}
where $S_\alpha$ is the spacetime part of
the spin field and has dimension $1/4$;
$\Sigma^i$ is the internal part of the
boundary vertex operator and has dimension $3/8$;
$\zeta$ carries polarization and gauge
indices; and $t^a_{IJ}$ are the Chan-Paton matrices.
Since we are interested in computing a potential
term we are interested in zero-momentum
amplitudes, so we can omit spacetime momentum factors
$e^{ik\cdot X}$.  Note that the
spacetime directions will have standard Dirichlet or
Neumann conditions, so that $S_\alpha$ is easily
related to its bulk counterpart, for example by the
doubling trick.\foot{\cf\ \joebook\ for a nice discussion
of this method.}

Similarly, the worldvolume scalars
are represented by NS vertex operators.
In the $(-1)$ picture they can be written as 
\eqn\minusns{
	V^{(-1)}_{NS,IJ} = \zeta_{i,a} e^{-\phi} \psi^i t^a_{IJ}
}
where $e^{-\phi}$ has dimension $1/2$, and
$\psi$ is a dimension $1/2$ boundary operator arising from
the internal sector.  To find the $0$-picture operator,
we find the superpartner of $\psi$ under the
(gauged) worldsheet $\CN=1$ SUSY,
\eqn\wspartner{
	T_F(z) \psi^i (w) = \frac{\half}{(z-w)} \CO^i (w)\ ,
}
where $T_F = \frac{1}{\sqrt{2}}(G^+ + G^-)$ is the gauged $\CN=1$ part
of the $\CN=2$ superconformal currents.
Here $\CO^i$ has dimension $1$.  The vertex operator
is simply $\CO^i$; if it is exactly marginal
its integral over the boundary is a valid conformal
deformation of the worldsheet action. 
Note that if $\psi^i$
is a chiral primary,
$$ G^+(z) \psi^i(w) = ({\rm non-singular\ terms}) $$
as $z\to w$, and may write
$$ G^-(z) \psi^i(w) = \frac{\half}{(z-w)} \CO^i \ . $$

The internal part of the vertex operators 
for the auxiliary fields, in the $(0)$ picture,
can be constructed from the internal part of the
$(-1)$-picture scalar vertex operators,
via the spectral flow operator
mapping the NS sector back to itself \aticketal:
\eqn\specone{
	V_{{\rm aux},IJ} = \lim_{z\to w} \left[
		(z-w) e^{-i\sqrt{3} H} V^{(-1)}_{NS, IJ}\right]
}
Essentially this is because one gets the auxiliary
component by acting on the scalar
fields twice with the spacetime SUSY current.

For deformations preserving spacetime SUSY,
the internal part of the vertex
operators should be constructed from the chiral
ring of the $\CN=2$ algebra.\foot{Or antichiral
ring.  We will fix the overall sign ambiguity
of the $U(1)$ charges by demanding that the
boundary operators be chiral.}  This is because the
marginal $(0)$-picture operators will have vanishing
R-charge; thus they may be added to the
worldsheet Lagrangian while maintaining
$\CN=2$ worldsheet supersymmetry.
The $(-1)$-picture operators will
have charge $q=1$; the $(-1/2)$-picture
Ramond operators will have charge $q=-1/2$;
and the auxiliary fields will have charge
$q=-2$. 

We will also want to include bulk vertex
operators in order to measure the effects
of the closed-string background.  For
SUSY-preserving deformations,
we will be interested in marginal
operators in the $(c,c)$
or $(a,c)$ ring.  The
vertex operators for massless fields can
be constructed from dimension $(\half,\half)$
operators $\psi(z,\bar{z})$ in the
internal sector, with charge $(1,1)$ if they are
in the $(c,c)$ ring, $(-1,1)$ if they are in the
$(a,c)$ ring, and so on.  In the $(-1)$
picture these operators are:
\eqn\bulklocal{
	V_{{\rm bulk}}^{(-1)} = e^{-\phi(z)-\phi(\bar{z})}
		\psi(z,\bar{z})\ .
}
The $(0,0)$ picture operators can be constructed
via the $\CN=2$ supercurrents which cancel
the $U(1)$ charge, i.e.
\eqn\intcc{
	V_{{\rm bulk}}^{(0,0)}(w,\bar{w}) = \oint dz\oint d\bar{z}
		G^-(z)G^-(\bar{z}) \psi(w,\bar{w})
}
for $(c,c)$ operators, and
\eqn\intac{
	V_{{\rm bulk}}^{(0,0)}(w,\bar{w}) = \oint dz\oint d\bar{z}
		G^+(z)G^-(\bar{z}) \psi(w,\bar{w})
}
for $(a,c)$ operators.

Now we wish to calculate the
tree-level contribution to the $n$th order term of
the superpotential; we will expand
out the coefficients to $k$th order in the
closed string fields.  We are particularly interested
in the case $n>2$, as we are studying putative
moduli.  We will examine the contribution
of this term to the fermion bilinear part of the
action.  On the disc, we must fix 3 real moduli
due to the $SL(2,\IR)$ symmetry.  In addition, we must absorb
the superconformal ghost number violation on the
disc.  These requirements can be met by using the
$(-1/2)$ picture for the two fermionic vertex operators
and placing them at opposite sides of the disc, or
equivalently at $z=0$ and $z=\infty$ in the upper half plane.
Furthermore we will take one of the NS vertex operators
to be in the $(-1)$ picture and fix its location
between the two R vertex operators, i.e. at $z=1$ in the
upper half-plane.  The remaining open- and closed-string
vetex operators are in the $(0)$ and $(0,0)$ pictures, 
and are integrated respectively over the boundary and bulk
of the worldsheet.  The resulting amplitude on the disc is:
\eqn\nksuperpot{
\eqalign{
	A & = \lim_{\delta_i,\epsilon_i\to 0}\langle
	e^{-\phi/2}S_\alpha\Sigma^{i_1}_{I_1 I_2}(x_3)
		\int_{x_2 + \epsilon_2}^{x_3 - \delta_2} 
		dy_1\CO^{(i_2,(0))}_{I_2 I_3}(y_1) 
		\int^{y_2 - \delta_{3}}_{x_2 + \epsilon_3}
		dy_2\CO^{(i_3,(0))}_{I_3 I_4}(y_2)\ldots \cr
	&\ \ \ \ \ e^{-\phi}\psi^{i_{k}(-1)}_{I_k I_{k+1}}(x_2)\ldots
	e^{-\phi/2}S_\beta\Sigma^{i_\ell}_{I_\ell I_{\ell+1}}(x_1)\ldots
	\int^{y_{n-4}-\delta_{n-3}}_{x_3 + \epsilon_{n-3}}
	dy_{n-3} \CO^{(i_n,(0))}(y_{n-3})\times \cr
	&\ \ \ \ \ \int_D dz_1\ldots dz_k \CO^{1(0,0)}(z_1,\bar{z}_1)\ldots
	\CO^{k(0,0)}(z_k,\bar{z}_k) \rangle
}}
where $\epsilon_2 > \epsilon_3 > \ldots \epsilon_{k-1}$,
$\epsilon_{k+2} > \ldots > \epsilon_{\ell-1}$, 
$\epsilon_{\ell+1} > \ldots > \epsilon_{n-3}$;
this prescription of the limits of integration
ammounts to a point-splitting regularization on the boundary.
In addition we sum over all orderings consistent
with the Chan-Paton indices; amplitudes
with adjacent operators $\phi_{IJ}\phi_{KL}$
are only nonvanishing if $K=L$.  For gauge-invariant
amplitudes we would sum over all such indices and thus
over all orderings.

So far we have not specified which of the four
chiral rings the closed-string vertex operators
live in.  We will discuss below how operators
in the different rings may or may not couple
to these amplitudes for a given boundary condition.

We will also be interested in superpotential terms
which are linear in the superfields and contain couplings
to closed-string moduli, such as the last term in Eq.
\orbifsp.  This term will not show up as a fermion bilinear;
only the auxiliary field will in fact couple.
Such a term can be computed on the disc with a single
closed-string insertion and a single open-string
insertion.  $SL(2,\IR)$ invariance allows
us to fix the positions of both vertex
operators.  In addition, if we place the
closed-string vertex operator in the $(-1,-1)$
picture and the open-string operator in the 
$(0)$-picture we have absorbed the superconformal
ghost number violation (the left- and right-moving
ghost zero modes will be tied together
by the boundary condition.)  The relevant
tree-level amplitude is thus:
\eqn\oneone{
	\langle e^{-\phi-\bar{\phi}} \psi^{(-1,-1)}(z,\bar{z})
	V_{F,II}(x) \rangle
}

All of these prescriptions allow us
to perform tree-level calculations
for fixed boundary conditions in the
Gepner models.  In the rest of this section
we will discuss these amplitudes
in general compactifications as correlators in 
the topologically twisted version of the
internal CFTs.  In this language we can revisit our
question regarding K\"ahler decoupling from the
superpotential of B-type branes.

\subsec{Topological CFT with boundaries}

We begin by reviewing and generalizing
the discussions in refs. \refs{\wittcs,\petropen,\bcov}
of topological CFTs with boundary.

Topological CFTs can be constructed
from $\CN=2$ CFTs via ``twisting''
the stress tensor with the $U(1)$ current \eguchiyang;
that is, we define a new stress tensor:
\eqn\twistedstress{
	T^{{\rm top}}(z) = T(z) \pm \half \partial J(z)\ .
}
Note the sign ambiguity; as we will discuss, 
the overall sign
is physically unimportant but the relative
sign between left- and right-moving sectors
is physically meaningful.
This twisting
may be achieved by adding a charge
of $\pm c/3$ at infinity; the change
in the stress tensor is simply the
shift derived in the Feigin-Fuchs construction.
In closed string theories one can see this
most simply by adding to the action 
the coupling of the $U(1)$ current to
a background gauge field $A=\half\omega$
where $\omega$ is the worldsheet spin connection \bcov.
In the cases we are interested, where $J=i\sqrt{3}\partial H$,
the term
\eqn\spincouple{
	\int d^2 z \half \left(J\bar{\omega} + \bar{J}\omega\right)
}
can be integrated by parts to get a coupling of $H$ to
the Riemann curvature.  For amplitudes on
the sphere, one may use conformal invariance
to write the sphere as a flat cylinder with two hemispherical
caps.  The initial and final states are created
by the path integral on those caps with any
operator insertions one might have there; 
the curvature on these hemispherical caps means that
the above terms in the Lagrangian become the
half-unit spectral flow operators applied to the initial
and final states.

If one constructs the open-string case via
the doubling trick on the Riemann sphere,
one finds again that the topological
twisting is equivalent to an amplitude with half-units
of spectral flow applied to the initial and final states.
More generally, to derive the twisted theory
on a surface with boundary
via the above coupling to the background field, 
one must take the boundary contribution in 
$\int J\bar{\omega} + {\rm c.c.}$ into account.
If we rewrite the disc as a long strip with two caps,
the background charge will be concentrated
on the boundary of these caps
and the result will again be spectral
flow applied to the in- and out-states \bcov.
Care should be taken with any boundary operator
insertions on or near this part of the boundary, 
as they may have contact terms
with the charge insertion.

The relative sign of the twisting of the holomorphic
and anti-holomorphic parts of the stress tensor
comes from the relative sign of the background charge.
The ``A-model'' arises from an axial twisting
while ``B-model'' arises from a vector twisting \witmirr.
In the presence of D-branes, the twisting must be
compatible with the boundary conditions.
We can see easily that
A-type boundary conditions are compatible with the
A-model and B-type boundary conditions
are compatible with the B-model, as in each case
the ``twisted'' stress tensor satisfies
$$ T^{{\rm top}} = \bar{T}^{{\rm top}} $$
and thus satisfies sensible boundary conditions.

In these twisted models, the conformal dimensions
of $\CN=2$ primary operators
are shifted by half their $U(1)$ charge,
with the sign depending on the twisting.
In the B-model, $G^+$ and $\bar{G}^+$ become
dimension-zero ``scalar'' Grassman operators
and suitably define BRST currents.  The NS
$(c,c)$ operators are annihilated by them and
have dimension $0$ with respect to
$T^{{\rm top}}$.  These operators are
denoted the ``topological'' operators and
their correlators are independent of position,
as one can see using the conformal Ward identities
in the presence of the background charge.
We denote them as the ``0-form'' operators
$\CO^{i(0)}$.  In the closed string theory,
we can also define ``$(1,0)$-'' and ``$(0,1)$-form'' operators
\eqn\oneform{
\eqalign{
	\oint_{z\to w} dzG^-(z) \CO^{(0)}(w,\bar{w}) & = \CO^{(1,0)}
		(w,\bar{w})\cr
	\oint_{\bar{z}\to\bar{w}} 
	d\bar{z} \bar{G}^-(\bar{z}) \CO^{(0)}(w,\bar{w})
	& = \CO^{(0,1)}(w,\bar{w})\ ,
}}
and ``2-form'' (or ``$(1,1)$-form'') operators:
\eqn\twoform{
	\oint_{z\to w} \oint_{\bar{z}\to\bar{w}} 
	dz d\bar{z} G^-(z) \bar{G}^-(\bar{z})
	\CO^{(0)}(w,\bar{w}) 
}
We can see that the $(0)$-form operators
are simply the internal parts of the
$(-1,-1)$ operators of the untwisted
theory, while the $(1,1)$-form operators
are the internal parts of the $(0,0)$-picture
operators of the untwisted theory.
The operators
$$ \oint dz \CO^{(1,0)}\ ;\ \oint d\bar{z} \CO^{(0,1)}\ ;\ 
	\int d^2 z \CO^{(1,1)}\ ,
$$
are BRST-invariant on Riemann surfaces
without boundary.  On surfaces with boundary,
the integrated 2-form operator is only
BRST-invariant up to an integral
of the one-form operators along the boundary,
as one can see by integrating by parts;
similarly the BRST transformations of the one-form
operators pick up boundary terms if the
curve of integration ends on a boundary of $D$.

One may similarly construct topological
operators on the boundary from the chiral
primary boundary operators.  From these
one may also construct ``one-form''
operators from the commutators or anticommutators
of the operator with modes of the spin-2
operators $G$:
\eqn\boneform{
	\frac{1}{\sqrt{2}}\left\{ G_{-\half}^- + \bar{G}^-_{-\half},
	\CO^{(0)}\right\} 
}
In the cases that we can construct the boundary condition
and boundary operators via the doubling trick, these
can be written as closed-string one-form
operators via holomorphic contour integrals, as above.
Again, the integral of these operators along the
boundary are BRST-invariant, up to potential contact terms
with other boundary operators.

One may similarly construct BRST-invariant operators
in the A-model with A-type boundary conditions.
If the CFTs correspond to geometric Calabi-Yau
sigma-models,
then we can see following refs. \refs{\witmirr,\ooy}  
that the open-string A-model describes
D-branes wrapped around special Lagrangian
submanifolds and the topological closed-string
operators are the K\"ahler deformations of
the target space; while the B-model describes
D-branes wrapped around holomorphic cycles,
and the topological closed-string operators
correspond to complex structure deformations
of the target space.  Note that although
refs. \refs{\wittcs\bcov} discuss
only the case of purely Neumann boundary conditions
for the B-model, our general
discussion shows that we may couple
this topological theory to any supersymmetric,
even-dimensional brane. (Mirror symmetry requires
this, if we are
allowed to discuss any supersymmetric 
3-cycle in the mirror).  Note also that
in this geometric picture, the
almost-BRST-invariance of the integrated
2-form observables makes sense:  a change of
complex structure (K\"ahler class) will
change the definition of holomorphic (special
Lagrangian) submanifolds.  

\ifig\figone{Computing the disc contribution
to the D-brane superpotential.  {\bf A)}. CFT
contribution.  R vertex operators on ``caps''
can be written as R ground states.
{\bf B)}. Representation of R ground state as path
integral on half-disc.  Spectral flow maps
this to NS ground state created by insertion
of NS vertex operator.}{\epsfxsize5.0in
\epsfbox{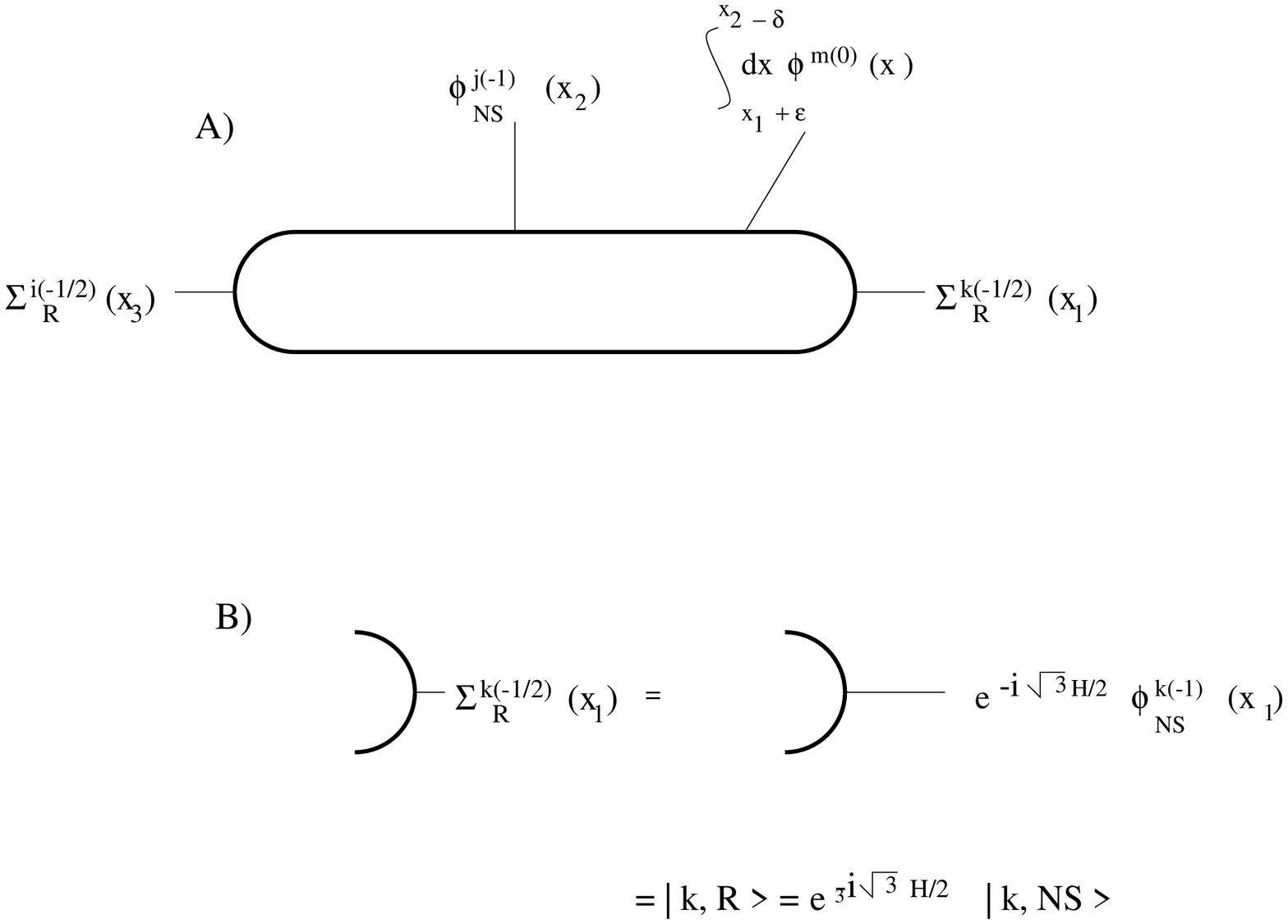}}

Now let us return to the fermion bilinear
part of the ($n>2$)th order superpotential.
By stretching the cylinder out into the capped
strip (\figone) we may write the amplitude
as the expectation value of some set of
NS vertex operators between Ramond states;
these states are created
by applying the Ramond vertex operators to the vacuum.
$\CN=2$ worldsheet supersymmetry allows us
to write the internal-CFT part of these states
as the spectral flow operator applied to
NS operators acting on the vacuum;
\eqn\RtoNS{
	\Sigma^i(0)|{\rm vac}\rangle =
	e^{-i\sqrt{3}H/2}\psi^i(0)|{\rm vac}\rangle\ .
}
The amplitude \nksuperpot\ factorizes into
three pieces.  The first is the superghost piece;
the second is the two-point function of the
spin fields polarized in the spacetime directions.
These give essentially universal answers
which we can expect from 4d Lorentz invariance.
The internal CFT amplitude is the interesting part.
It is an expectation value of $n$ chiral 
or antichiral NS boundary
operators and $k$ bulk NS-NS operators in one of the
four closed-string chiral rings, with two
additional half-unit spectral flow operators each mapping NS states
to R states.  The fixed boundary
operators become 0-form observables
and the integrated boundary operators,
1-form observables.  If the 
closed-string operators are
$(c,c)$ for the B-type twisting, 
corresponding to complex structure deformations,
they become (almost)-invariant topological observables.
If they are $(a,c)$ operators, corresponding to
K\"ahler deformations, they are exact with respect to
the left-moving BRST current and one might hope that
they decouple.  We will address this issue below.

The result is (up to the caveats above)
a correlator of topological operators
in the topologically twisted theory.  The fixed
operators become 0-form observables and the
integrated operators become 1-form and 2-form observables.
We may bring the techniques of topological
field theory to bear on this calculation, and
will do so below.

Similarly, the computation of Eq. \oneone\ is
topological (with the same caveats).  
Here the auxiliary field is
related by a full unit of spectral flow to the
$(0)$-form observable of the associated scalar field.
The superghost part of the amplitude merely
takes care of the relevant zero modes. The
internal CFT part is once again an amplitude
in the topologically twisted theory of a $(0)$-form
boundary observable and a $(0,0)$-form bulk observable.

\subsec{Computations in the geometric sigma model}

The topological symmetry of these correlators,
and the localization properties of the
topological path integrals \refs{\topsig,\witmirr},
make the above calculations relatively
straightforward.  To see this we will compute
the cubic part of the superpotential
for a D0-brane in a weakly curved background,
and discuss the linear part of the superpotential for
generic wrapped B-branes.

To begin with we need to construct the
relevant topological observables.  The closed-string
case has been described in ref. \witmirr\ and the open-string
case for fully Neumann boundary conditions
has been described in \wittcs.  We need to
generalize these results to arbitrary B-model
boundary conditions.  

In the untwisted sigma-model,
the propagating worldsheet 
fields are 3 complex scalars $\phi^i$,
and three complex fermions $\psi_\pm^i$.  $\psi^i$
has $U(1)$ charge $1$ and $\psi^{\ib}$ has charge $-1$.
Thus in the B-twisted theory $\psi^i$ have dimension
zero and become worldsheet scalars, while $\psi^{\ib}$
have dimension one and become worldsheet one-forms.
The BRST currents $G^+$, $\bar{G}^+$ give rise
to global symmetries parameterized by constant Grassman
scalars $\epsilon$,$\epb$ (since these are
scalars such constants are well-defined on any worldsheet).
In order to write these transformations
in the simplest form, it is convenient to rewrite
the fermions as:
\eqn\fermtoform{
\eqalign{
	\xi^i & = \psi^i_+ + \psi^i_- \cr
	\theta_{\jb} & = g_{i\jb}
	\left(\psi^i_- - \psi^i_+\right)\ .
}}
If we integrate out the auxiliary fields on the worldsheet,
the BRST transformations become
\eqn\bmodelbrst{
\eqalign{
	\delb\phi^i &= \frac{i}{2}
		\epsilon\left(\xi^i + 
		g^{i\jb}\theta_{\jb}\right)\cr
	\delb\phi^{\bar{i}} & = 0\cr
	\delb\xi^i & = i\Gamma^i_{jk}
	\left(\epsilon\psi^i_+\psi^k_-
	     + \epb\psi^j_-\psi^k_+\right)\cr
	\delb\theta_{\jb}
		& = ig_{i\jb}\Gamma^i_{jk}
		\left(\epsilon\psi^j_+\psi^k_-
		-\epb\psi^j_-\psi^k_+\right)
	+ g_{i\jb,k}g^{i\bar{\ell}}(i\epsilon
		\psi^i_+ + i\epb\psi^i_-)
	\theta_{\bar{\ell}}\cr
	\delb\psi^{\jb}_-&=-\epb\pb\phi^{\jb}\cr
	\delb\psi^{\jb}_+&=-\epsilon\p\phi^{\jb}\ .
}}
These do not necessarily close off-shell once
we have integrated out the auxiliary fields.
In the presence of a boundary we must set 
$\epsilon=\epb$.\foot{This is a sensible thing
to do in the closed-string sector as well, as
B-model path integrals localize onto constant maps
into the target space \witmirr.}
Then the transformations simplify: in particular, the
important transformations are:
\eqn\bmbrstmod{
\eqalign{
	\delb\phi^i&=i\epsilon\xi^i\cr
	\delb\xi^i &= 0\cr
	\delb\theta_{\jb} &= 0\ .
}}
Recall that in the Dirichlet directions of the
untwisted model, $\psi^i$ is fixed and
$\psi^i_+ = - \psi^i_-$; in the
Neumann directions (when
we have turned off the NS 2-form and boundary gauge
field), $\psi_+ = \psi_-$.  Thus along
Dirichlet directions, $\xi$ vanishes at the boundary;
while along Neumann directions, $\theta$ vanishes
at the boundary.  Of course, for curved boundaries,
whether a given polarization
is ``Dirichlet'' or ``Neumann'' will depend on $\phi$;
this can be defined by a projection matrix
$\IP^i_j(\phi(\CC)):TM\longrightarrow T\CC$.

The $(0)$-form topological observables in the bulk
were constructed in \witmirr. They are of the form:
\eqn\bmobserv{
	\Lambda_{i_1\ldots i_p}{}^{\jb_1\ldots\jb_q}
	\xi^{i_1}\ldots\xi^{i_p}
	\theta_{\jb_1}\theta_{\jb_q}\ ;
}
$\Lambda$ is a $\p$-closed $(0,p)$ form
with values in $\wedge^q T^{(0,1)}M$.

The boundary observables will live
on the appropriate holomorphic
submanifold $\CC \subset M$ and will take
values in the Chan-Paton algebra $\hat{g}$.
In the fully Neumann case,
the boundary observables are $\hat{g}$-valued
$(p,0)$-forms, while in the fully Dirichlet case
they will be antiholomorphic functions of the
position of the boundary, with values
in $\wedge^q T^{(0,1)}M\otimes \hat{g}$.
In the 2-brane and 4-brane cases, the
observables will be forms on $\CC$ valued in the
normal bundle times the gauge group
of the internal worldvolume, $N\CC|_M\otimes\hat{g}$.
The 0-form boundary observables
corresponding to the marginal, chiral
primaries of the untwisted model are
linear in the worldsheet fermions.
For all of these observables, open and closed,
it is easy to see that the BRST operator acts
as the holomorphic differential on $\CC$.  Thus
topological observables are $\p$-closed,
and trivial BRST-exact observables are $\p$-exact.

The construction of these topological amplitudes
makes it clear that we have some anomalous $U(1)$ charge.
This essentially counts fermion number in these models.
In the closed-string B-model, nonvanishing correlators
have fermion number $3$ for both $\theta$ and $\xi$
corresponding to zero modes for each of these fields.
On the disc, the boundary conditions
will kill the $\xi$ zero modes polarized along the
Dirichlet directions and the $\theta$ zero
modes polarized along the Neumann directions.
Thus for a holomorphic $p$-cycle, nonvanishing
correlators will have $\xi$ fermion number $p$
and $\theta$ fermion number $3-p$.

The other fact that makes these amplitudes straightforward
to calculate is that the topological path integral
localizes onto constant maps, restricted to the
submanifold defined by the boundary conditions.
The correlation functions are then
integrals of the appropriate forms over the
moduli space of constant maps;
in these cases they will be integrals
of the pullback of forms on $M$ onto
the submanifold $\CC$.

Let us start by computing the cubic
term in the superpotential for a 
brane sitting at a point in the CY.
The topological boundary observables
corresponding to the world-volume chiral fields
will be
\eqn\dzerovertex{
	\CO = \phi^{\ib}_{a} \theta_{\ib}t^a
}
where $t^a$ is a matrix in the adjoint of the Chan-Paton
group.  The correlation function is:
\eqn\cubiccorr{
	\phi^{\ib_1}_a \phi^{\ib_2}_b \phi^{\ib_3}_c
	\left(\langle \theta_{\ib_1}\theta_{\ib_2}
		\theta_{\ib_3}\rangle +
		\langle \theta_{\ib_2}\theta_{\ib_1}
		\theta_{\ib_3}\rangle\right)
	\tr (t^a t^b t^c)
}
Note that the expectation value is antisymmetric
in the fermions; thus this will vanish if there
is only one D-brane, after summing over the ordering.
The moduli space of constant maps is a point.
Chiral deformations of the location by boundary observables
live in $T^{0,1}M$ and anti-chiral deformations
are valued in $T^{1,0}M$.  The latter are BRST-exact in
our picture, so the above correlator is $\partial$-closed
as a function of the location of the point.\foot{For
multiple derivatives there is potentially a holomorphic anomaly.}
The correlators must be antiholomorphic functions
of the location of this point, 
and they live in $\wedge^3 T^{(0,1)}M$.
Serre duality implies that they are components of
a closed antiholomorphic $(0,3)$ form.  On
the Calabi-Yau manifold there is only one such form 
$\barom$
within cohomology.  Thus the superpotential is
\eqn\dzerosuper{
	W = \barom_{\ib_1 \ib_2 \ib_3}f^{abc}
	\Phi^{\ib_1}_a \Phi^{\ib_2}_b \Phi^{\ib_3}_c\ .
}

This superpotential is similar to the term coming from
fully Neumann boundary conditions.  The
topological string theory in this latter case
has been argued to be a holomorphic six-dimensional
version of Chern-Simons theory \wittcs; the
vertex operators which describe
chiral fields in the spacetime Lagrangian
describe antiholomorphic gauge fields $\barA$
on the Calabi-Yau.
The low-energy Lagrangian has been argued
to be
\eqn\sixdlag{
	S = \int_M \Omega\wedge
	\left( \barA\wedge \p \barA + \frac{2}{3} \barA\wedge\barA 
		\wedge \barA
	\right)
}
This second term is the superpotential.

Finally, we can look for linear terms in the superpotential
coming from a coupling to closed strings.  Again,
the boundary topological operator will be linear
in the worldsheet scalar fermions; for the
correlator to have the right fermion number, the
closed-string operator must be quadratic.
We can analyze these couplings for 0-, 2-, 4- and 6-cycles
separately and we find that some of these amplitudes
vanish automatically.

For boundaries living on points, the open-string operator
is written in Eq. \dzerovertex.  The closed-string operator
must be quadratic in $\theta$ and have no $\xi$ charge
by $U(1)$ charge conservation.  This latter
operator is an element of $H^0(M,\wedge^2 T^{(1,0)}M)$.
By Serre duality this group is equivalent to the Dolbeaux
cohomology group $H^{(0,1)}(M)$ which vanishes for
Calabi-Yau compactifications.  Thus there
is no closed-string operator which couples
to a single open-string operator on the D0-brane.
The argument is almost identical for D6-branes
and rests on the fact that $H^{(2,0)}(M)$ is
trivial on a Calabi-Yau manifold.

For 2-cycles the story is a bit richer.  If the open-string
vertex operator is polarized along a Neumann direction,
\eqn\neumannvertex{
	V_N = A_{i,a} \xi^i t^a\ ,
}
(we work in a coordinate patch where
the tangent-normal split is trivial),
then fermion number conservation requires that the closed-string
operator be quadratic in $\theta$ and there are no such nontrivial
operators as we have just argued.  But for
vertex operators polarized in a Dirichlet direction,
the closed-string operator must be bilinear in $\xi$ 
and $\theta$, making it a one-form valued in $T^{(0,1)}M$.
Serre duality relates this to an element of $H^{(2,1)}(M)$
and this group is certainly nontrivial,
so these open-closed correlators are allowed.  
Similarly we can have nontrivial linear terms for
chiral fields coming from Neumann directions along
a 4-cycle.

Indeed, these results should not surprise us.  If
we change the complex structure of the manifold,
the holomorphic 2- and 4-cycles, and the
homomorphic bundles on them, will change.
We should generically find that the reference cycle is no
longer a stable, supersymmetric configuration.  On
the other hands, the 0- and 6-cycles
are holomorphic regardless
of the complex structure, so we expect them
to be supersymmetric so long as the closed-string background
maintains $\CN=2$ spacetime SUSY.

\subsec{Decoupling of non-topological moduli}

One of the more powerful statements one can make in
topological closed string theory is that the
K\"ahler (complex structure) deformations decouple
from topological amplitudes in the B (A) model.
This is related to the fact that the
spacetime of the theory has $\CN=2$ supersymmetry
and the vector multiplets and hypermultiplets decouple
(away from singular points in the moduli space).
One can show in the topologically twisted B (A) models
that insertions of integrated $(c,a)$ ($(a,c)$) operators,
which one would get by taking derivatives
of the amplitudes with respect to the K\"ahler (complex structure)
moduli, lead to vanishing amplitudes.  In the open
string case the status of this decoupling is less clear.
To start with, the spacetime SUSY is only $\CN=1$ and
the Lagrangian is far less constrained.
For example, recall the analogous $E_8 \times E_8$ heterotic
string, compactified on a CY 3-fold.  
There the charged multiplets
arising from the K\"ahler and complex structure
decouple from each other at finite
order in $\alpha'$ \wittnr\ 
but couple due to worldsheet instantons \dinenp.
Furthermore, in the generic $(0,2)$ model
it may not make sense to identify deformations
with K\"ahler or complex structure deformations.

Actually, a total decoupling is not to be expected, 
even from geometric considerations.
For example, if we consider the theory of D$9$-branes wrapped on the CY,
the four-dimensional action will come with the prefactor $V_6/g_s$ where $V_6$ is
the volume of the CY.  On the other hand this is a B-brane so the topological
amplitudes naturally depend on complex moduli.
Thus the strongest conjecture we could make is that the superpotential (for a B
brane) takes the form
\eqn\decoupling{
W = m(\phi_K) W(\phi_c,\psi)
}
where $m(\phi_K)$ is proportional to 
the brane tension \bpsmass.

A known example which illustrates this is the
one-loop topological open string amplitude.  For the D$6$-brane this is
the Ray-Singer torsion $I(V)$ associated
to the Chan-Paton bundle $V$ on $M$ \bcov.  In general
$I(V)$ is not independent of the K\"ahler moduli,
but ratios $\ln (I(V_1)/I(V_2))$ are, where $V_{1,2}$
are two different bundles on $M$ \raysinger.
This is consistent with \decoupling; furthermore this amplitude
also corresponds to a chiral ($\int d^2\theta$) 
term in the effective action,
the one-loop correction to the gauge coupling.

In a system involving several different branes,
\decoupling\ does not even predict a universal multiplicative dependence
of the total superpotential on the K\"ahler moduli.  At the very least it will be the sum
of several terms of this form but with different $m(\phi_K)$. There will also be
terms involving strings stretched between different branes.  Geometrically these
would be expected to come with $m(\phi_K)$ for the surface of intersection;
it would be quite interesting to make a more general proposal along these lines.

In any case, it is preferable to have string world-sheet arguments for decoupling.
Thus we proceed to consider the 
the cubic and quartic terms in the superpotential as computed on the disk,
to see if derivatives with respect to K\"ahler moduli $\phi_K$
are consistent with \decoupling.
We will work in the untwisted theory in order to
ensure that we are not avoiding any subtleties;
our statements can be carried over to the twisted theory.

\ifig\figtwo{Perturbation of
cubic part of superpotential by K\"ahler
deformation.  The superconformal Ward 
identities allow us to pull $\oint dz G^+(z)$
to the contour $C$.  This can be deformed to
a contour integral around each
of the boundary operators and
an integral over $C$ of $d\bar{z}\bar{G}^+(\bar{z})$
which can be deformed back to the insertion of $\bar{G}^- \psi$.
The result is an integral over the insertion of $\pb_w \psi$
which can be integrated by parts to an integral
of $\psi$ over $C$.}{\epsfxsize5.0in
\epsfbox{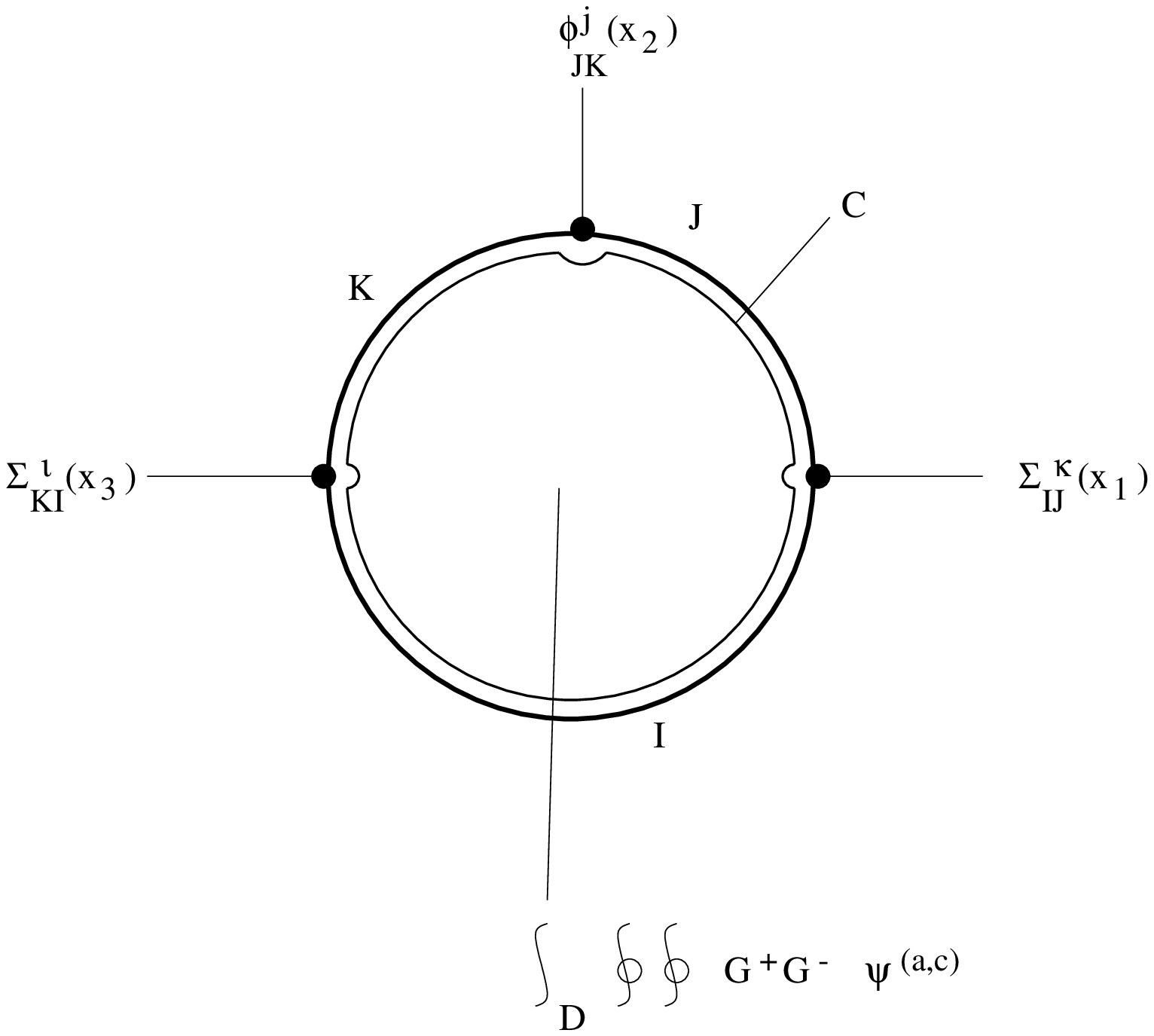}}

Recall that the cubic term in the superpotential
for B-type boundary conditions
is calculated via the 3-point disc amplitude.
The part arising from the internal CFT is:
\eqn\threepoint{
	\langle \Sigma^{i_1}_{IJ} \CO^{i_2}_{JK} \Sigma^{i_3}_{KI}
	\rangle
}
plus a sum over any orderings consistent with the Chan-Paton
factors.  We may fix the ordering by picking
suitable Chan-Paton factors, which we will do here.
Now the first derivative of this amplitude with respect to
some K\"ahler deformation will lead to the above
amplitude with the insertion of an integrated
$(0,0)$-picture vertex operator constructed from the $(a,c)$
(or $(c,a)$) ring:
\eqn\baddef{
	V = \int_D d^2 w \oint_{z\to w}\oint_{\bar{z}\to\bar{w}}
	dz d\bar{z}G^+\bar{G}^-(\bar{z}) \psi^i(w,\bar{w})\ ,
}
and the complete amplitude is show in \figtwo. 
Conformal invariance allows us to deform the
integral of $G^+$ out to the boundary.  This
amounts to using the superconformal Ward 
identities.  Let us
concentrate on the case where the doubling
trick allows us to describe amplitudes on the
upper-half plane via amplitudes on the full complex
plane.  The contour may be deformed to a sum of
integrals of $G^+$ around each boundary
operator\foot{Taking some care with the
branch cuts created by the spin fields.}, 
plus a contour integral around the image of (see \figtwo)
\eqn\badoneform{
	V_{(a,c)}^{(0,1)} = 
	\int d^2 w \oint d\bar{z} \bar{G}^-(\bar{z}) 
	\psi^i(w,\bar{w})
}
In the end, the contour
integrals around the boundary operators
will vanish as the operators are chiral.
The contour integral around the image of the bulk operator
in the lower half-plane
may be expressed in the
upper half-plane as an integral
$$\oint_{\bar{z}\to\bar{w}} \bar{G}^+(\bar{z}) $$
around $V_{(a,c)}^{(0,1)}$.  Using the superconformal
algebra, this term becomes:
\eqn\badopbound{
	\int d^2 w \pb_w \psi^i = \oint_{\p D} \psi^i\ .
}
In this integral over the boundary, we must take some
care when the contour passes near one
of the boundary operator insertions.
The result is the correlation function
\eqn\threepointbad{
	\langle \Sigma^{\alpha_1}_{IJ} \CO^{\alpha_2}_{JK} 
	\Sigma^{\alpha_3}_{KI}
	\oint_C \psi^i \rangle
}
where the contour $C$ is shown in \figtwo.  We get two potential
contact terms from this correlator.  One arises from
the operator product of the $(a,c)$ operator with the
boundary \cardylewellen:
\eqn\btobope{
	\lim_{y\to 0}\psi^i(x+iy,x-iy) \sim
	C^{I}_{\psi^i \CO^\alpha} 
	\frac{1}{y^{1-\delta_{\CO^\alpha}}} \CO^\alpha(x)
}
(here $I$ lables the boundary condition in the region
of the contact term);
the other from the operator product of $\psi^i$ with the
boundary operators $\Sigma^\alpha, \CO^\alpha$.  Note that $\CO$
will have zero $U(1)$ charge.  Let us deal
with each of these in turn; we will in fact
argue that this second contact term is taken care of by the
first.

The bulk-boundary OPE can be treated as a factorization
of the disc amplitude onto an intermediate open-string state.
The OPE coefficient $C^{I}_{i\alpha}$
will be proportional to the open-closed
disc amplitude $\langle \psi^i\CO^\alpha \rangle$.
There are in fact two classes of terms to
worry about in eq. \btobope: $\delta_{\CO^\alpha}<1$
and $\delta_{\CO^\alpha}=1$.  In the former case 
the intermediate state is a tachyon.
Either this is removed by the GSO projection,
or the perturbation by $\psi^i$ has changed
the acceptable boundary conditions -- for
example by changing the stability condition
on vector bundles -- so that the original
boundary is no longer a stable D-brane.
Such a divergence will have to be removed by
perturbing the boundary conditions.
The second case is a more genuine contact term;
it is a dimension-one operator which is
integrated over the boundary.  In this
case the 3-point correlator satisfies:
\eqn\correqn{
	\left(\p_i - C^{{\rm bound}}_{i\alpha}
	\p_{\alpha} \right)
	\langle \Sigma^{\alpha_1}_{IJ} \CO^{\alpha_2}_{JK} 
	\Sigma^{\alpha_3}_{KI} \rangle = 0\ .
}
If $\CO^\alpha$ is a topological operator then
the perturbation by $\psi^i$ has the fairly
simple effect of moving the vev slightly
along a flat direction.  It should
not affect the form of the superpotential,
in keeping with our claim.

A very similar formula to \correqn\
appears in \ooy.  In that case they
find that by defining a suitable
connection, the chiral primary part of the
boundary state of
the B-type brane is covariantly constant
with respect to deformations of the
Kahler moduli.  Our result should be the
open-string version of this fact.

The second contact term above is between the
bulk operator $\psi^i$ and boundary operator
$\CO^\beta$.  By using
the doubling trick this is described as
the coalescence of three operators
and, associativity allows us to
write this by taking the
bulk-boundary OPE \btobope\ of $\psi$ first,
and then taking the OPE of $\CO^\alpha$
and $\CO^\beta$.  But this will be included
in the limits of integration of 
the first contact term $\CO^\alpha$ over the
boundary.


Higher order
amplitudes are more subtle
since they have a moduli space of insertions of
vertex operators.  When applying the Ward identities,
we will find integrals of total derivatives
with respect to these moduli, leading to
contributions from the boundaries
of moduli space.\foot{This
is similar to the fact that
insertions of the stress tensor into
correlators on higher-genus Riemann surfaces
lead not only to transformations of the 
operators but to derivatives of the amplitude
with respect to the moduli of the surface
\eguchiooguri.  Indeed such
terms are important in deriving the one-loop
holomorphic anomaly for topological amplitudes \holomanom.}
We can already illustrate
this phenomenon by looking at the contribution
to the quartic term of the superpotential
from a single derivative with respect to the K\"ahler
moduli.  The resulting amplitude is:
\eqn\quarticder{
\eqalign{
	\lim_{\epsilon,\delta\to 0}
	& \langle \Sigma^{i,(-1/2)}(x_3) \int^{x_3-\epsilon}_{x_2 + \delta}
		dx \frac{1}{\sqrt{2}}\left\{G^- + \bar{G}^-,
		\CO^{j, (-1)}(x)\right\}\cr
		&\ \ \ \ \ \CO^{k,(-1)}(x_2)\Sigma^{\ell,(-1/2)}(x_1)
		\int d^2 w \oint_{z\to w}\oint_{\bar{z}\to\bar{w}}
		dzd\bar{z} G^+(z)\bar{G}^-(\bar{z})\psi^{(a,c)}\rangle
}}
plus a potential sum over orderings.  As
before we may fix the orderings of the boundary
operators via a judicious choice of Chan-Paton
factors.  Once again,
we pull the contour integral of $G^+$ off of
the bulk operator and apply the superconformal 
Ward identities.  In addition to the terms
which we have already argued to vanish, we get
a term coming from the contour integral of $G^+$
around the integrated NS operator.  Again,
let us look at the case where we may describe this
amplitude via the doubling trick.  Then
the above anticommutator can be replaced with a
contour integral of $G^-$ around a point on the real line,
and the contour integral of $G^+$ around this
leads simply to a derivative of $\psi^j$.  
The result is the difference of contact terms:
\eqn\fourpointcontact{
\eqalign{
	\lim_{\epsilon\to 0}
		&\langle \Sigma^i(x_3) \left(\CO^j(x_3-\epsilon) 
		- \CO^j(x_2+\epsilon)\right)\CO^k(x_2)
	\Sigma^{\ell}(x_1)\cr
	&\int d^2w \oint_{\bar{z}\to\bar{w}}d\bar{z}\bar{G}^-(\bar{z})
	\psi^{(a,c)}(w,\bar{w})\rangle
}}
In the twisted theory we might hope that factorization
and associativity means that this difference would
vanish.  This would be true if there was no
insertion of $\psi^{(a,c)}$.  With such an insertion, it is not
clear that the amplitude will factorize onto topological
intermediate states, so we cannot complete this argument at present.

The upshot of all of this is that there is a simple world-sheet mechanism which
could lead to decoupling.  It is very analogous to the known decoupling of bulk
K\"ahler and complex structure deformations: the decoupling operator is a descendant
with respect to an operator which annihilates the boundary chiral fields
(say for K\"ahler and B-type, the operator $G^+$).  The situation is better than that
for $(0,2)$ heterotic string models as there are still two $\CN=2$ algebras involved;
they are identified only on the boundary.  

Such world-sheet arguments are valid up to the possible contributions of contact 
terms and to make them precise, one needs to show that the contact terms either vanish or
have simple interpretations (e.g. as connection coefficients on the moduli space).
We have interpreted some but not all of these terms and thus can say that
we have found further evidence for decoupling but by no means a proof.

\newsec{Correlation functions in minimal models and Gepner models}

We now turn to the problem of computing correlation functions
in the Gepner model. 
To begin with, 
let us recall a few properties of
Gepner model boundary correlators, which  
are comparable to properties
of bulk correlators.
As with correlators in the bulk theory,
in the boundary theory there are restrictions 
due to ghost number conservation. This can easily be seen using the
doubling trick and  has been discussed in the previous section.
In addition, the boundary fields transform
under particular representations of the chiral algebra, similar to
chiral halves of bulk fields. The chiral algebra is the tensor
product of the chiral algebras of the minimal models involved.
The fields obey the same fusion rules.
Correlators forbidden by the fusion rules therefore vanish.

In this section we point out a number of
differences with bulk theory computations and interpret their
consequences.

\subsec{Ordering effects}

Correlation functions involving boundary operators 
require a specification of operator ordering
along
the boundary (which we will place on the real line):
$$
\vev{ \psi_1(x_1) \psi_2(x_2) \dots \psi_n(x_n) } \quad 
x_1 > x_2 > \dots > x_n
$$
This ordering corresponds directly to the 
ordering of the matrix fields in the
world-volume Lagrangian: for 
example terms $\tr \psi_1\psi_2\psi_3$ and
$\tr \psi_1\psi_3\psi_2$ come from 
these two orderings of the three-point function.

In some particularly simple models (for example, free field theory),
correlation functions of boundary 
operators can be analytically continued to the bulk.
In this case it is possible to determine the effect
of arbitrary permutations of the fields. This was formalized
by Recknagel and Schomerus \RSmarg\ in a discussion of
non-supersymmetric conformal field theories.
Two boundary operators $\psi_{1,2}$ were called
mutually local if
\eqn\ordering{
\psi_1(x_1) \psi_2(x_2) = \psi_2(x_2) \psi_1(x_1) 
}
inside correlators. Here, the left hand side implies
$x_1>x_2$ and the right hand side $x_1 , x_2$.
Recknagel and Schomerus then argued that 
self-local marginal boundary operators are truly marginal.
The argument is basically that an o.p.e. 
$$
\psi(x_1)\psi(x_2)\rightarrow {1\over x_1-x_2}\psi(x_1) + \ldots
$$
of the form which would spoil 
marginality is incompatible with \ordering.

Free fermion correlators can be continued into the bulk as well, and in section
6 we saw that the superpotentials governing these operators were completely
antisymmetric.  In particular they vanish in the theory of a single brane.
It seems quite plausible that this result applies to all operators which are
strictly marginal in the large 
volume limit; however since we do not know whether an
operator we find at the Gepner point is marginal in the large volume limit until
we compute the superpotential (and many interesting operators are never strictly
marginal), such considerations appear somewhat circular.

In general, one does not expect either that boundary correlation functions
have a continuation into the bulk or that the boundary operators have such simple
exchange relations.  By general principles
(which we review in the next subsection) boundary correlation functions in minimal
models and Gepner models are particular 
combinations of several chiral conformal blocks,
each of which has different exchange relations, chosen to be single-valued on the
boundary.  To make any statement about ordering effects, we must consider this
analysis.

\subsec{Sewing constraints}

In this section we will briefly discuss sewing constraints on
boundary fields.
Correlation functions in two-dimensional CFT with boundaries
have been studied for rational conformal field theories
in \refs{\lewellen, \cardylewellen}. In the bulk, the
n-point functions on the sphere are determined by
the three-point functions; the higher-point functions
can be computed by sewing. The result is independent
of the decomposition of the n-point function into three-point
functions, as guaranteed by crossing symmetry for the
four-point functions. Similar results hold for the case
with boundaries. Here, we have three types of sewing
constraints, those involving only boundary fields, those,
involving both bulk- and boundary fields and those involving
only bulk fields. The structure constants for boundary fields
depend on the boundary conditions. 

As discussed in section 4, for RCFTs the possible
boundary conditions preserving all the symmetries are labeled
by the primary fields and can be implemented by boundary states
carrying these labels, and we have written
analogous states for
Gepner models. 
The field content of the theory can be read off
from the partition function $Z_{\alpha \beta}$; thus
the propagating fields also carry the labels
$\alpha,\beta$. In the case that $\alpha\neq\beta$,
the field $\phi^{\alpha\beta}$ is a boundary
condition-changing operator. If $\alpha=\beta$,
it preserves the boundary condition. 

Let us concentrate on the correlation functions for
boundary fields. The boundary OPEs are:
\eqn\boundaryope{
\phi_i^{\alpha\beta}(x) \phi_j^{\beta\gamma}(y) = \sum_k
c_{ijk}^{\alpha\beta\gamma} \phi_k^{\alpha\gamma}(y)
(x-y)^{h_k-h_i-h_j} + \dots \quad y<x\ .
}
The structure constants $c_{ijk}^{\alpha\beta\gamma}$,
together with the vacuum amplitude, determine the
three-point functions:
\eqn\channelone{
\langle \phi_i^{\alpha\beta} (x_i)
\phi_j^{\beta\gamma}(x_j)\phi_k^{\alpha\gamma}(x_k)\rangle
= c_{ijk^{\dagger}}^{\alpha\beta\gamma} 
c_{k^{\dagger}k \one}^{\alpha\gamma\alpha} \langle \one
\rangle_{\alpha} (x_i - x_j)^{h_k-h_i-h_j} (x_j-x_k)^{-2h_k}\ .
}
We can also evaluate the correlator in the other channel:
\eqn\channeltwo{
\langle \phi_i^{\alpha\beta} (x_i)
\phi_j^{\beta\gamma}(x_j) \phi_k^{\alpha\gamma}(x_k) \rangle
= c_{i i^{\dagger} \one}^{\alpha\beta\alpha} 
c_{jki^{\dagger}}^{\beta\gamma\alpha} 
\langle \one \rangle_{\alpha} (x_j -x_k)^{h_i-h_j-h_k} (x_i-x_j)^{-2h_i}
}
The dependence on the coordinates is dictated by conformal
symmetry. As mentioned above, conformal symmetry does not
relate three-point functions with different orderings.
Comparison of \channelone,\channeltwo\ leads to a consistency condition 
on the structure constants. 

In addition to these conditions on the OPE coefficients,
we must demand the crossing symmetry of the four-point functions.
Nonvanishing correlation
functions for boundary fields are of the form
\eqn\boundcor{
\langle \phi_1^{\alpha\beta} \phi_2^{\beta\gamma}
\phi_3^{\gamma\delta} 
\phi_4^{\delta\alpha} \rangle,
}
\ifig\bc{Four-point function}{\epsfxsize2.0in\epsfbox{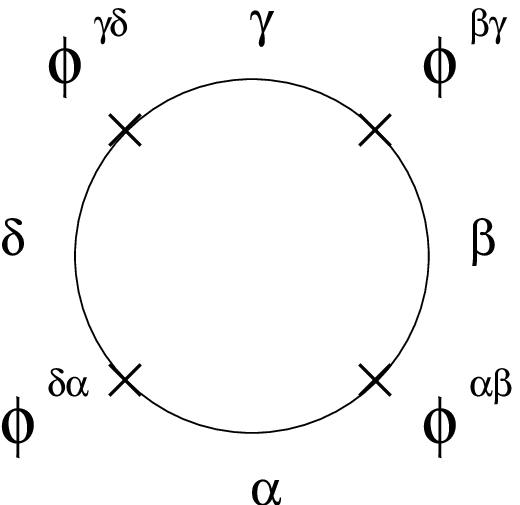}}
as illustrated in \bc.
In the case of rational symmetric models the factorization conditions can be
made explicit using the conformal blocks.
Note that for boundary correlators, the four point functions
are linear in the conformal blocks.

In \runkel\ an explicit solution was given for the
Virasoro minimal models:
$$
c_{ijk}^{\alpha\beta\gamma}=F_{k\beta} 
\left[ \matrix{\alpha&\gamma\cr i&j}\right]
$$
If we want to compare the four point functions for open
string operators with different orderings, we have to take
into account that the change in the ordering will in general
require different boundary conditions for the respective four
point function to be non-vanishing. Different boundary conditions
will in general change both the structure constants and the
expectation values of the identity so that we do not expect
the four point functions to agree. In the example of the Virasoro
minimal model, where we have an explicit solution, the boundary
structure constants satisfy
\eqn\opesymm{
c_{ijk}^{\alpha\beta\gamma} = c_{jik}^{\gamma\beta\alpha}.
}
In particular, they are completely symmetric in the case that there is only
one boundary condition involved. The symmetry of the structure
constant is of course a direct consequence of the symmetries of
the $F$-matrices, which are specific for minimal models. (In general,
there will be a phase involved \seibmoore.)

Another simple example is the $U(1)$ boson.
The primary fields are given by tachyon vertex operators $e^{ikX}$.
The vertex operator $e^{ikX}$ connects the boundary conditions
$|n\rangle\rangle$ to $|n+k\rangle\rangle$. Therefore, the
condition \boundcor\ is fulfilled, whenever momentum
conservation holds.

The sewing constraints determining the OPEs 
have not yet been solved for
$\CN=2$ minimal models or for Gepner models.  
We will return to this in future work.

\subsec{Boundary selection rules}

Given two boundary states, $\kket{\alpha}, \kket{\beta}$, the
partition function will contain the characters of a particular
set of marginal operators, whose insertion changes the boundary
conditions from $\alpha$ to $\beta$.
In general, this set of marginal operators will be a subset
of all possible weight one representations,
and are determined by the fusion rules \cardy. As a consequence,
certain correlators remain uncorrected, because the required
marginal operator does not propagate with the particular
boundary conditions. As a simple application, consider a boundary
condition-changing 
marginal operator $\phi^{\alpha\beta}$. 
All non-vanishing correlators
$\vev{\phi_1^{\gamma\delta} \phi_2^{\delta\epsilon}
\phi_3^{\epsilon\gamma}}$ containing only boundary changing operators
cannot be corrected by insertions of $\phi^{\alpha\beta}$.
On the other hand, one can generate a non-zero correlator
which vanishes at lower order.

\subsec{Three-point functions in the Gepner models}

A superpotential for massless fields is computed
from $n>2$-point functions as we have discussed.
Let us briefly discuss
the conditions under which a three point function can be
non-vanishing.
To compute a $\vev{\phi\phi F}$ term
we start by picking three vertex operators in the NS sector and
we apply spectral flow by one unit to one of them. This is done
by splitting the operators in a charged and an uncharged part as
in \chargedop\ and applying the spectral flow $e^{-iH\sqrt{3}}$.
This gives the following correlator:
\eqn\threebound{
\vev{ \CO_{0_1}^{\alpha\beta} e^{-iH {2\over{\sqrt{3}}}}
\CO_{0_2}^{\beta\gamma} e^{i{H\over{\sqrt{3}}}}
\CO_{0_3}^{\gamma\delta}e^{i{H\over{\sqrt{3}}}}}
}
Including the ghost contributions,
the result is the product of the OPE
coeffient $c^{\alpha\beta\gamma}_{123}$ for the uncharged operators,
with the vacuum expectation amplitude for the boundary condition
$\alpha$.  Thus, a cubic term in the superpotential is directly proportional
to a structure constant $c_{123}^{\alpha\beta\gamma}$.

\subsec{The $A_1$ model}

The power of boundary selection rules can be nicely illustrated
with the example of the $A_1$ model, where all correlators between
chiral fields are forbidden.
The model contains the chiral operators $\one$, $\phi_1^{(0)} =
e^{{i\over{\sqrt{3}}} \phi}$ and the antichiral operator
$\phi_{-1}^{(0)} =
e^{-{i\over{\sqrt{3}}} \phi}$. As discussed in section 6, from
these operators we can derive one-form operators. For the chiral
operator,
we get the one-form operator
$\phi_1^{(1)} = e^{{{-2i}\over{\sqrt{3}}}\phi}$. Alternatively,
this operation can be interpreted as picture changing, if the
$A_1$ model is part of a string theory compactification.
A candidate for a non vanishing correlator is
$\vev{(\phi_1^{(0)})^3 \phi_1^{(1)}}$ and we have to check whether
it is compatible with the boundary conditions.
We can determine, which boundary conditions allow for the field $\phi_1$
by using the fusion rules. The field $\phi_1^{\alpha\beta}$ exists
whenever $N_{\alpha\beta}^1$ does not vanish. This is the case
for $\alpha\beta = \one 1, -1 \one$ and $ 1-1$. As a consequence,
our candidate is suppressed by boundary selection rules.

\subsec{The $A_2$ model}

In the $A_2$ model, the boundary selection rules do
not forbid all correlators. We will give an example of an
allowed correlator, where permutation of the operators
requires different boundary conditions.

The $A_2$ minimal model can be seen as one real boson $\chi$
and one real fermion $\lambda$. 
The central charge is ${3\over{2}}$. Apart from the
identity there are two more chiral primary fields, 
$$
\phi_1 = \sigma e^{{i\chi\over{2\sqrt{2}}}}, \quad \phi_2 =
e^{{i\chi\over{\sqrt{2}}}}.
$$
There are two corresponding anti-chiral fields of opposite charges.
$$
\phi_{-1} = \mu e^{-{i\chi\over{2\sqrt{2}}}}, \quad \phi_{-2} =
e^{{-i\chi\over{\sqrt{2}}}}.
$$
There is also an uncharged field $\lambda$, which is the ordinary
fermion, or in minimal model language the field $l=2, m=0$.
The spectrum for various boundary conditions can now be 
determined by fusing the fields labeling the boundary conditions.
\eqn\diffseven{\vbox{
\offinterlineskip
\halign{\strut
\vrule$#$\hfill&\vrule$#$\hss\vrule\cr
\noalign{\hrule}
\qquad&\qquad\cr
\noalign{\hrule}
\one & \one^{\one,\one},\one^{1,1},\one^{-1,-1},\one^{2,2},\one^{-2,-2}
       \one^{\lambda,\lambda}\cr
\noalign{\hrule}
\phi_1 &\phi_1^{1,2}\phi_1^{-2,1}\phi_1^{1,-2}\phi_1^{2,-1}
         \phi_1^{-1,\lambda} \phi_1{\lambda,1}\cr
\noalign{\hrule}
\phi_2 & \phi_2^{1,-1}\phi_2^{-1,1} \phi_2^{2,\lambda}\phi_2^{\lambda,-2}\cr
\noalign{\hrule}
\lambda & \lambda^{1,1}\lambda^{-2,2}\lambda^{2,-2}\cr
\noalign{\hrule}
\noalign{\hrule}
}}}

The boundary conditions for the anti-chiral fields follow from
this table: For any chiral field $\phi_q^{\alpha\beta}$ we have an
antichiral field $\phi_{-q}^{\beta\alpha}$. A non-trivial four-point
function to compute is 
$$
\langle \phi_2 \phi_2 \phi_1 \phi_1 \rangle ,
$$
where one operator is a one-form and integrated over the
boundary. There exists another ordering
$$
\langle \phi_2 \phi_1 \phi_2 \phi_1 \rangle
$$
The first ordering requires the following boundary conditions:
$$
\langle \phi_2^{2,\lambda} \phi_2^{\lambda, -2} \phi_1^{-2,1}
\phi_1^{1,2} \rangle
$$

The other ordering requires the boundary conditions:
$$
\langle \phi_2^{1,-1} \phi_1^{-1,\lambda} \phi_2^{\lambda,-2}
\phi_1^{-2,1} \rangle.
$$
Evaluation of the two four-point functions leads to structure
constants with different boundary conditions (which will in general
be not equal). In the final result, the expectation value of
the identity is taken with two different boundary conditions.
Therefore, the two results are not expected to agree.

\subsec{The models $(k=2)^2$ and $(k=2)^4$, $(k=1)^3$ and $(k=1)^9$}

There are two Gepner models containing only the $(k=2)$
model: the model $(k=2)^2$, which corresponds to
a 2-torus;
and $(k=2)^4$, which corresponds to a K3 surface.
We will consider in this section
the case of a single boundary condition (\ie a single brane).
In this case it is known
that the superpotential vanishes, which we
can check using the Gepner model description.
The marginal operator in the $(k=2)^2$ model is
the operator $\phi_2^{(1)} \phi_2^{(2)}$. This operator
corresponds to the complex fermion $\psi$ in
sigma-model language.
We know that this operator is an
anticommuting variable.  Therefore,
all superpotential terms involving this operator vanish
in the absence of Chan-Paton factors, due to the
sum over operator ordering. 
Similarly, as a consequence of the fusion rules, in the
$(k=2)^4$ case all marginal operators are of the form
$\phi_2^{(i)} \phi_2^{(j)}$. We know from the torus that these
operators anticommute. Therefore, all superpotential terms
vanish after summing over all permutations. This verifies the
result discussed in section 4.4. for the case $N_g =1$.

Similar statements can be made for the models consisting only
of $(k=1)$ models, like the torus $(k=1)^3$ and the orbifolded
six-torus $(k=1)^9$. The fermion in the two-torus is given by
the field $\phi_{(1,1,0)}^{(1)} \phi_{(1,1,0)}^{(2)}
\phi_{(1,1,0)}^{(3)}$. Simlilarly, we can form three complex fermions
for the six-torus example. The marginal operators propagate
for the $L=1$ boundary conditions in these models. Again, we know
that the superpotential vanishes by antisymmetry.

\subsec{The quintic}

Let us now turn to the applications of the boundary selection
rules to the $(k=3)^5$ Gepner model.  We are particularly interested
in correlators of marginal operators.
For the B-boundary states discussed in section 5, we found
that if we impose the same boundary conditions on both
ends of the string there are
boundary conditions
with either 101, 50, 24, 11, 4 or 0 marginal operators
propagating.

The 101 marginal operators propagating between the boundary
states $L=\{ 1,1,1,1,1 \}$ are  of
the same form as the complex structure deformations in the
closed string case.  These are left-right symmetric fields
of charge $(1,1)$.  If we use
the doubling trick, the boundary marginal operators look
the holomorphic part of these operators. They are of
the form:
\eqn\diffseven{\vbox{
\offinterlineskip
\halign{\strut
\vrule\quad$#$\hss\quad&\vrule\quad$#$\hss\quad\vrule\cr
\noalign{\hrule}
(1,1,0)^5&1\cr
\noalign{\hrule}
(1,1,0)^3 (2,2,0) (0,0,0)&20\cr
\noalign{\hrule}
(1,1,0)^2 (3,3,0) (0,0,0)^2 &30\cr
\noalign{\hrule}
(1,1,0)^2 (2,2,0)^2 (0,0,0) &20\cr
\noalign{\hrule}
(2,2,0) (3,3,0) (0,0,0)^3 &20\cr
\noalign{\hrule}
}}}
For these boundary conditions there are no further
restrictions on possible correlators from the boundary selection
rules. All correlators allowed by $U(1)$ charge conservation
and fusion rules will also be allowed by the boundary selection
rules. The difference from closed string computations is
that these correlators depend on:
the one-point function of the identity in the
presence of particular boundary conditions; the values of
the fusion coefficients $c_{ijk}^{\alpha\alpha\alpha}$ 
for the boundary conditions;
and the different integration 
domain for the four- and higher point functions.

The 4,11, 24 and 50 marginal operators are particular
subsets of the 101 operators.  Here, restrictions from
boundary selection rules are possible: some of the
correlators which are present in the closed string case
cannot appear as the operators do not
propagate with the given boundary conditions. We expect
the strongest results for the case with 4 marginal
boundary operators. 

To compute superpotential terms, 
we start with three (or more) chiral primary fields in the NS
sector. One way to
relate this to a physical amplitude
is to apply a unit of spectral flow,
so that from the space time point of view we are computing
the $\langle F \phi\phi \rangle$ term in the
worldvolume Lagrangian. For 4- and higher-point
functions, charge conservation requires us to apply the operator
$G$ to the additional operators (which changes the label
$s$ of the Gepner model fields from $0$ to $2$). 
In our example, unit spectral flow
is implemented by the operator $\phi_{(3,-3,0)}^5$.

The four marginal operators for the boundary conditions
$L=\{ 1, 0,0,0,0 \}$ are given by 
$$
\psi_i=\phi^{(1)}_{(2,2,0)} \phi^{(i)}_{(3,3,0)}\ ,
$$ 
where the upper index
labels the minimal model, $i=2,3,4,5$. Spectral flow relates
these operators to $\phi^{(1)}_{(1,-1,0)} \phi^{(i)}_{(0,0,0)}
\phi^3_{(3,-3,0)}$. There is no  non-vanishing three-point
function
for these operators. However, there are some higher-order terms
which are allowed. The four point function
$$
\langle \psi_2 \psi_3 \psi_4 \int_{\partial\Sigma} G\psi_5\rangle
$$
is not suppressed by the selection rules. 
We can ask about possible
corrections to this correlator. The fusion rules tell us that all
higher order terms have to be of the form 
$$ 
\left(\int_{\partial\Sigma} G\psi_j \right)^5\ ,
$$
since the fifth powers of $\phi_{(2,2,2)}$ and $\phi_{(3,3,2)}$
contain the identity in their fusion.
Note that there are a lot more correction terms for the
corresponding bulk 4-point function. 

Let us move on to the next most complicated example, the example
with 11 marginal operators. The marginal operators are
of the form
\eqn\types{
\phi^{(1)}_{(2,2,0)}\phi^{(i)}_{(3,3,0)}, \quad \phi^{(2)}_{(2,2,0)}
\phi^{(i)}_{(3,3,0)},
\quad \phi^{(1)}_{(1,1,0)} \phi^{(2)}_{(1,1,0)} \phi^{(i)}_{(3,3,0)}.
}
Compared to the previous case, this example is already much less
restricted. For example, we have  in this case two
types of three-point functions: those containing each type
of operator listed in \types\ once; and that containing the
three operators of the third type listed in \types.
However, a lot of corrections which would be allowed in the
corresponding bulk case are absent for these boundary conditions,
because in the factors $3,4,5$ of the minimal models, 
$\phi_{(3,3,0)}$ is the only chiral primary which propagates.

Likewise, we potentially get more three point functions for the
cases with 24 and 50 operators, and less corrections are
suppressed.  Finally, for the case with 101 marginal
operators, all allowed bulk correlators have a boundary
equivalent.

To conclude this section, let us briefly commment on the
A-type boundary conditions. Here, the $L=1$ states have
one marginal operator, which is the operator
$\phi_{(1,1,0)}^{(1)}\dots \phi_{(1,1,0)}^{(5)}$. 
The three-point function between
three operators of this type is allowed by the selection
rules. Also, higher-order correlators containing
$\left( \int G\left(
\phi_{(1,1,0)}^{(1)} \dots \phi_{(1,1,0)}^{(5)} \right)\right)^5$ are
allowed. Taking the 5th power is required by the fusion rules:
$\phi_{(1,1,0)}^5 $ contains the identity. For the A-type
boundary conditions we have also argued for a coupling of this
operator to a bulk field. The closed string observables in the
A-model are the $(a,c)$ fields. On the quintic, this is the
K\"ahler deformation 
$(\prod_{i=1}^5 \phi_{(1,1,0)}, \prod_{i=1}^5 \phi_{(1,-1,0)})$.
Taking this operator to the boundary, we get boundary fields contained
in the OPE of 
$\prod_{i=1}^5 \phi_{(1,1,0)} \times \prod_{i=1}^5 \phi_{(1,1,0)}$.
This certainly makes a bulk-boundary coupling of the desired
form possible.

\subsec{Consequences of the selection rules}

Perhaps the simplest conclusion we can draw from these selection rules
is that the B branes with $L\ge 1$ have non-trivial moduli spaces.
Consider the example of $\ket{10000}_B$:
the superpotential must take the form
$$
W = \psi_2\psi_3\psi_4\psi_5 f(\psi_2^5,\psi_3^5,\psi_4^5,\psi_5^5).
$$
No matter what $f$ is, the subspace $\psi_2=\psi_3=0$ (or any two $\psi$'s
zero) solves $W'=0$.

On the other hand, we found that the branes $\ket{11111}_A$, which
we identified with the large volume $\BR P^3$s, admitted a superpotential
$W=\psi^3 f(\psi^5) + \phi \psi g(\psi^5)$, 
which would resolve the potential contradiction with the
lack of moduli in the large volume limit.  A non-trivial $f$ and $g$
would break the $\psi\rightarrow -\psi$ R symmetry of the leading order
superpotential, which has no reason to exist in the large volume limit.
However we cannot test the prediction for the number of minima of $W$ 
at this point.

\newsec{Conclusions and further directions}

In this work we began a systematic study of D-branes in the stringy
regime of the quintic Calabi-Yau.  Our main result was the determination
of the charges (in the usual large volume conventions) of the explicit
Gepner model boundary states constructed by Recknagel and Schomerus.
Our tools were the intersection form, and the monodromy and continuation
formulas for the CY periods.  The techniques clearly generalize to any 
Calabi-Yau given this data.

The primary question we hope to address is whether the spectra and
low energy world-volume theories of branes in the stringy regime
are the same (up to renormalizations of couplings)
as in the large volume limit or not.  
We will refer to this as the ``geometric hypothesis.''
Unlike the previously studied cases, supersymmetry is not sufficient
to answer this question.  From the bulk point of view, $\CN=2$, $d=4$
supersymmetry allows lines of marginal stability for BPS states, while
from the brane point of view $\CN=1$, $d=4$ supersymmetry allows
transitions from Higgs to Coulomb branch which are essentially unpredictable
from the large volume point of view.

There are a number of considerations which would lead us to expect
non-geometric phenomena.  
Perhaps the simplest is that 
the B monodromies relate branes of different dimension.
Another essentially non-geometric
phenomenon is the topology change
seen in various Calabi-Yaus; both phenomena make the
the geometric interpretation of brane probes ambiguous.
The mere existence of these phenomena however does not really contradict
the geometric hypothesis as we have stated it,
if the different geometric objects related by monodromy 
and topology change
lead to the same low energy theories.  What we would be saying is that
the same brane theory can have multiple large volume limits, a familiar
phenomenon in duality.

Small instantons and the $\BC^3/\BZ_3$ orbifold provide
examples which do contradict the 
geometric hypothesis in its simplest form.
At a special point (more generally, in complex codimension one)
in moduli space, enhanced gauge symmetry and
additional states appear.
This might be considered a relatively mild failure as it is associated
with a singularity of the Riemannian geometry or gauge bundle.  
If all failures of the
geometric hypothesis were associated with singularities, conversely
it would be true under the mild condition that the geometry stayed
non-singular.  As we mentioned in the introduction, we can imagine
much more drastic failures -- a priori, the spectrum of branes at the
Gepner point might have satisfied none of the relations we expected
from geometry and gauge theory.

The results we have presented here are
not (yet) inconsistent with the geometric hypothesis.
Most of the branes we find could certainly correspond to
the appropriate geometric constructions -- holomorphic
vector bundles and special Lagrangian submanifolds.
For example, all of the branes we found satisfied the (mathematical)
stability condition on vector bundles.
The lack of any classification of these makes it difficult for
us to assert that branes which we did not identify actually do not have
geometric constructions.  The elliptically fibered case may be more
promising in this regard.  The most problematic case was a brane which
would correspond to a rank $1$ bundle with $c_1=0$ but $c_2\ne 0$.
Although such things do not exist in conventional gauge theory, they
are known to exist in modified gauge theories 
(such as noncommutative gauge theory), so one can imagine that this
object has a description in the large volume limit.

We also presented a general argument that B-type branes should be
described by geometric considerations -- namely, that their world-volume
potentials are determined by quantities in B-twisted topological models,
which are equal to their classical values in the B-twisted topological
sigma models, up to hopefully
minor effects of K\"ahler deformations.  
Besides a formal world-sheet argument we showed that many
known cases fit with this idea.  By contrast,
the superpotentials in A brane theories can depend directly on K\"ahler
moduli and a priori it would seem much more likely that the geometric
hypothesis fails.

Finally, we made some first steps towards explicit computation of the
superpotentials on these branes.  These superpotentials appear
to be quite non-trivial and it appears that
such computations are doable with existing techniques; we will return
to this in future work.
An exciting possibility is that topological open string theory can
be developed to the point where exact superpotentials can be obtained,
perhaps with some analog
of the special geometry determining the bulk prepotentials.

One important direction to develop is to find more direct ways to
get the geometric interpretation of these states.  The results here
suggest that this will be simpler in the A picture -- the simplest
picture is that each component minimal model has a specific boundary
condition for its LG superfield.  If we had the D$0$-brane boundary
state, we could apply a probe construction to get the geometrical
picture for the B boundary states (indeed we could derive the corresponding
$(0,2)$ models); perhaps larger classes of boundary states containing
the D$0$ can be found.

A study of curves of marginal stability is in progress, to decide
whether the large volume and Gepner D-branes should be expected to
match up, and whether new phenomena appear near the conifold point.  

Let us close with a brief
discussion the physical relevance of our primary question.
To the extent that branes in the stringy regime are qualitatively
different than geometric branes, all of the work on compactification
using branes will have to be reconsidered.
On the other hand, to the extent that they are qualitatively the same, 
these techniques
will provide new ways of deriving geometric results, such as the
existence and moduli space dimension for vector bundles.

Questions of existence of branes are also directly relevant for
non-perturbative constructions of M theory.  For example, Matrix
theory constructions to date 
use D$0$-branes as their starting point.
Compactifications on some manifold $M$
are believed to be described by D$0$-branes in a certain scaling
limit of type IIA string theory on $M$ \seibergsen.
The authors of ref. \kls\ argued that 
for Calabi-Yau compactifications
this limit was the mirror of the conifold point.
If it were to turn out that the D$0$-brane did
not exist in the stringy regime, this construction
would have to be reconsidered.

In any case we believe there is much to be discovered in this direction.

\smallskip
We would like to thank Paul Aspinwall, Tom Banks, 
Michael Bershadsky, Ralph Blumenhagen, Robert Bryant, Emanuel Diaconescu, 
Robbert Dijkgraaf, Dan Freed, Bartomeu Fiol,
Jaume Gomis, Rajesh Gopakumar, Sheldon Katz, Albrecht Klemm,
Maxim Kontsevich, Juan Maldacena, John Morgan, Hirosi Ooguri,
Arvind Rajaraman, Andreas Recknagel, Moshe Rozali, 
Volker Schomerus, Shishir Sinha,
Andy Strominger, Gang Tian, Cumrun Vafa,
Edward Witten and Sasha Zamolodchikov for useful discussions and correspondence.

This research was supported by DOE grant DE-FG02-96ER40959.
A.L. is also supported in part by NSF grant PHY-9802709
and DOE grant DE-FG02-91ER40654.
\smallskip

\appendix{A}{An explicit calculation for A-type boundary states}

This appendix shows the explicit calculation of the intersection
number of two A-type boundary states. The 
Witten index $\tr_R(-1)^F$ in the open
string sector is obtained from the transition amplitude between the
(internal) RR parts of the boundary states with $(-1)^{F_L}$
inserted\foot{One has to be careful with the definition of
$(-1)^{F_L}$ in the RR sector. It should be defined by
$(-1)^{F_L}=(-1)^{J_0+d/2}$ because the charge might be half-
integer moded.}. The first part of this calculation is very close to
that in \reckone.

For the A-type boundary states 
the $\delta_A$ constraint is trivial, as we have discussed.
\eqn\stepone{
\eqalign{
\tr_R(-1)^Fq^{H}=&{1\over{\kappa_\alpha^A\kappa_{\tilde\alpha}^A}}
\sum_{\lambda,\mu}^{\beta,R}
\sum_{\tilde\lambda,\tilde\mu}^{\beta,R}
B^{\tilde\lambda,\tilde\mu}_{\tilde\alpha} 
B^{\lambda,\mu}_{\alpha}\,_A
\bbra{\tilde\lambda,-\tilde\mu}(-1)^{F_L}
\tilde q^{L_0-{c\over 24}}\kket{\lambda,\mu}_A= \cr
=&{1\over{\kappa_\alpha^A\kappa_{\tilde\alpha}^A}}
\sum^{\beta,R}_{\lambda,\mu}\sum^{ev}_{\lambda',\mu'}
B^{\lambda,-\mu}_{\tilde\alpha}B^{\lambda,\mu}_{\alpha}
(-1)^{Q(\mu)+d/2}
S_{(\lambda,\mu),(\lambda',\mu')}\chi^{\lambda'}_{\mu'}(q),}
}
where $S_{(\lambda,\mu),(\lambda',\mu')}$ is the modular transformation
matrix and $ev$ means $l_j+m_j+s_j=\even$.
The $\beta$-constrained sum together with the charge projection
operator can be rewritten as
\eqn\steptwo{
\eqalign{
\sum^{\beta,R}_{\lambda,\mu}(-1)^{Q(\mu)+d/2}=&\sum^R_{\lambda,\mu}
{1\over K}\sum_{\nu_0=0}^{K-1}
e^{i\pi(2\nu_0+1)(-\sum{s_j\over 2}+
\sum{m_j\over{k_j+2}}+{d\over 2})}=\cr
=&{1\over K}\sum^R_{\lambda,\mu}
\sum_{\nu_0}e^{i\pi{d\over 2}(2\nu_0+1)}\prod_{j=1}^r
e^{i\pi{m_j\over{k_j+2}}(2\nu_0+1)}e^{-i\pi{s_j\over 2}(2\nu_0+1)}}
}
Putting all these equations together and collecting terms we get:
\eqn\stepthree{
\eqalign{
\tr_R(-1)^Fq^{H}=&
{1\over{\kappa_\alpha^A\kappa_{\tilde\alpha}^A}}
{1\over K}\sum^{ev}_{\lambda',\mu'}
\sum_{\nu_0}e^{i\pi{d\over 2}(2\nu_0+1)}
\prod_{j=1}^r\Big({1\over{2(k_j+2)^2}}\times \cr
\times&\sum_{l_j,m_j,s_j}^R
{{\sin(l_j,L_j)\sin(l_j,\tilde
L_j)\sin(l_j,l_j')}\over{\sin(l_j,0)}}\times \cr
\times&e^{i\pi{m_j\over{k_j+2}}(2\nu_0+1+M_j-{\widetilde M}_j+m_j')}
e^{i\pi{s_j\over 2}(-2\nu_0-1-S_j+{\widetilde S}_j-s_j')}\Big)
\chi^{\lambda'}_{\mu'}(q).}
}
The sums over $l_j,m_j,s_j$ can be evaluated as follows:
\eqn\stepfour{
\eqalign{
&{1\over{2(k+2)^2}}\sum_{l,m,s}^R
{{\sin(l,L)\sin(l,\tilde L)\sin(l,l')}\over{\sin(l,0)}}
e^{i\pi{m\over{k+2}}M}e^{i\pi{s\over 2}S}=
\cr
=&{1\over{(k+2)^2}}\sum_{l,m}^R
{{\sin(l,L)\sin(l,\tilde L)\sin(l,l')}\over{\sin(l,0)}}
e^{i\pi{m\over{k+2}}M}(-1)^{S\over 2}\delta^{(2)}_S=
\cr
=&{1\over{k+2}}\sum_l
{{\sin(l,L)\sin(l,\tilde L)\sin(l,l')}\over{\sin(l,0)}}
(-1)^{{M\over{k+2}}(l+1)}\delta^{(k+2)}_M
(-1)^{S\over 2}\delta^{(2)}_S=\cr
=&\half N^{l'}_{L,\tilde L}\delta^{(2k+4)}_M
\delta^{(2)}_S(-1)^{S\over 2}\ .
}}
Inserting this result into \stepthree\ gives
\eqn\stepfive{
\eqalign{
\tr_R(-1)^Fq^{H}=&{1\over{\kappa_\alpha^A\kappa_{\tilde\alpha}^A}}
{1\over K}{1\over 2^r}(-1)^{{S-\widetilde S}\over 2}\sum^R_{\lambda',\mu'}
\sum_{\nu_0}(-1)^{{d\over 2}(2\nu_0+1)}\prod_{j=1}^rN^{l_j'}_{L_j,\tilde L_j}
\delta^{(2k_j+4)}_{2\nu_0+1+M_j-{\widetilde M}_j+m_j'}\times\cr
\times&(-1)^{\nu_0+{{1+s_j'}\over 2}}
\chi^{\lambda'}_{\mu'}(q)\ .}
}
This fits with the normalization constant being
$\kappa^A_{\alpha}=\sqrt{C\over{K2^r}}$, where $C$ is an extra
integer constant depending on the specific model. It can be understood
from the $\beta$ constraints. Imposing the same spin structure on all
subtheories reduces the number of states by a factor of
${1\over 2^r}$, the $U(1)$ constraint gives another factor of
${1\over K}$.

To simplify this result we have to use the fact that the R ground
states are given by $\phi^l_{l+1,1}$ which are identified with
$\phi^{k-l}_{-k+l-1,-1}$; only these
states will contribute to the Witten index.
We continue the upper index of fusion 
rule coefficients $N^l_{L,\tilde L}$ with a period of $2k+4$;
we
identify $N^{-l-2}_{L,\tilde L}=-N^l_{L,\tilde L}$; and we set
$N^{-1}_{L,\tilde L}=N^{k+1}_{L,\tilde L}=0$. This continuation is
natural from the point of view of the Verlinde formula. Neglecting an
overall factor of $(-1)^{d\over 2}$ we find that:
\eqn\stepsix{
\eqalign{
\tr_R(-1)^Fq^{H}=&{1\over C}(-1)^{{S-\widetilde S}\over 2}
\sum_{\nu_0}(-1)^{(d+r)\nu_0}\prod_{j=1}^r
\sum_{m_j=0}^{2k_j+3}N^{-m_j'-1}_{L_j,\tilde L_j}
\delta^{(2k_j+4)}_{2\nu_0+1+M_j-{\widetilde M}_j+m_j'}=\cr
=&{1\over C}(-1)^{{S-{\widetilde S}}\over 2}
\sum_{\nu_0}(-1)^{(d+r)\nu_0}\prod_{j=1}^r
N^{2\nu_0+M_j-{\widetilde M}_j}_{L_j,\tilde L_j}.}
}

\bigskip

\listrefs

\end